%% file: Klochkova_atlas.tex
\documentclass[
 aps, pra,
 amsmath,amssymb,
 12pt,
 final,
tightenlines,
 nofootinbib,
 superscriptaddress,
 ]
{revtex4}

\usepackage[utf8x]{inputenc}
\usepackage[english]{babel}
\usepackage{graphicx}
\usepackage{longtable}
\usepackage{appendix}

\input{sao_cmd_author.tex}

\input{maik.rty}

\begin{document}

\title{B[e]-star CI\,Cam: a high-resolution spectrum atlas in the range of 395--780\,nm}

\author{\firstname{V.G.}~\surname{Klochkova}}
\email{Valentina.R11@yandex.ru}
\affiliation{Special Astrophysical Observatory, Nizhnii Arkhyz, 369167 Russia}
\author{\firstname{A.S.}~\surname{Miroshnichenko}}
\affiliation{University of North Carolina, Greensboro, NC~27402, USA}
\affiliation{Fesenkov Astrophysical Institute,  Almaty 050020, Kazakhstan}
\author{V.N.~Komarova}
\affiliation{Special Astrophysical Observatory,  Nizhnii Arkhyz, 369167 Russia}
\author{N.S.~Tavolzhanskaya}
\affiliation{Special Astrophysical Observatory, Nizhnii Arkhyz, 369167 Russia}

\begin{abstract}

The  atlas of the spectrum of the B[e] star CI\,Cam obtained with the 6-meter  BTA telescope
in combination with the NES spectrograph in the wavelength interval of 395--780~nm with a resolution
of  $\lambda/\Delta\lambda\ge60\,000$  is presented.
The atlas contains about 400 spectral features and illustrates the diversity of spectral
features of the unique star, forming in a complex circumstellar environment. The spectrum
is dominated by forceful single-peaked H\,I, He\,I emissions and numerous double-peaked
permitted and forbidden emissions of ions of chemical elements starting from the CNO triad
up to metals (Mg, Al, Ti, V, Cr, Fe) with practically ``rectangular'' profiles. The FeII and
[FeII] emissions predominate in the spectrum. However, several other double-peaked forbidden ion
emissions were also detected: [VII], [Cr II], [Ni II]. The atlas is presented  graphically, with a
separate figure corresponding to each echelle order. A list of identified lines, including
a number of known interstellar features is presented in table form.  CI\,Cam  supergiant status
is confirmed  due to the richness of its spectrum by different details of nitrogen.
\end{abstract}

\maketitle

\section{INTRODUCTION}
A hot star CI\,Cam is the optical component of an X-ray binary system XTE~J0421+560. Astronomers'
perceptive  interest in this object emerged after the detection of a powerful X-ray flare in the spring
of 1998, which was accompanied  by an increase in the visble flux by two stellar magnitudes (\citet{Clark2000}).
After the detection of the  short-lived flare, an intensive study of this
object commenced in all wavelength ranges, and the results of subsequent  studies allowed to
classify  CI\,Cam  as an ultraluminous X-ray source (\citet{Bartlett}), exhibiting
moderate variability in both X-ray and optical fluxes. The first post-outburst optical spectra
of CI\,Cam revealed  by~\citet{Mirosh2002} the key features: powerful H\,I and He\,I emissions and
numerous permitted and forbidden Fe\,II emissions with double-peak profiles.
  Data of high-resolution spectroscopy performed by~\citet{Robinson}
two weeks after the 1998 outburst with the McDonald Observatory 2.7-m telescope are crucial for
studying the behavior of the optical spectrum of CI\,Cam.  These authors identified  spectral features,
estimated their parameters, and specified the formation regions of various emission-line types.
The results of \citet{Mar2017, Maravelias}
are important for understanding the spectral features of stars with the B[e] phenomenon.
Having obtained high-resolution spectra for a sample of B[e]--stars, these authors examined
various emission types and concluded that specific double-peak emissions arise from a collection
of shell structures in the circumstellar medium, most often in the form  of an optically thin
compact disk or a collection of arcs. \citet{Maravelias} examined the possible
formation scenarios of such a circumstellar medium. One of the fundamental results of this study is the
conclusion that there is no consensus on whether binarity is the true cause  of ring structure
formation, since not all the studied B[e]--phenomenon objects have a companion. Therefore,
alternative scenarios can also be considered, such as (quasi-periodic) mass ejections caused
by pulsations or other instabilities of a central star.

Over the past three decades, high-resolution spectroscopy of selected hot stars of various masses
(LBV candidates, Be and  B[e] stars, including CI\,Cam) has been performed using several echelle
spectrographs of the 6-m BTA telescope of the Special Astrophysical Observatory of the Russian
Academy of Sciences  under the programs of E.L.\,Chentsov and A.S.\,Miroshnichenko. The first
results for CI\,Cam were published in 2002 by~\citet{Mirosh2002}, revealing
the main  features of the spectrum, namely, extremely intense H\,I and He\,I emissions and numerous emissions
of ions  of several chemical elements, primarily iron, with ``rectangular'', nearly
vertical profiles. Experience from the first years of working with the optical spectrum of CI\,Cam
demonstrated the necessity and feasibility of spectral  monitoring of the star to track the
variability of spectral feature profiles and the radial velocity pattern. This time-consuming
task requires repeated observations with high spectral resolution over a wide wavelength range.
The results of the analysis of the optical spectra for 2002--2023 were published by~\citet{KMP}.
All the spectra for  these dates contain the so-called ``rectangular'' profiles
of two-peak, low-excitation ion emissions.

The conclusion about the quiescent kinematic state of the CI\,Cam system was unexpected: the average
velocity found from the emissions for all observation dates is V$_r({\rm aver})=-53.1\pm0.5$\,km\,s$^{-1}$.
The half-amplitude of the variation (root-mean-square deviation) is  $\Delta V_r=2.5$\,km\,s$^{-1}$.
The behavior of the forbidden  [N\,II]\,5755\,\AA{} emission over time is no different from other emissions
with vertical profiles. The radial velocity corresponding to its position is stable over the course of observational
sets from 2002 and on and is taken as the systemic velocity: V$_r=-55.4\pm0.6$\,km\,s$^{-1}$.

Interest in studying CI\,Cam never yields. The results of its long-term multicolor photometry demonstrate
a weak photometric instability of the star, suggesting that unusual events may recur in the CI\,Cam system
in the future.
Therefore, we considered it important to prepare a detailed spectral atlas, which could significantly
facilitate the further study of the star optical spectrum. Section~2 of this article briefly describes
the observational methods and data processing. A list of all identified spectral features, a description
of the atlas, and comments on its individual sections are presented in Section~3. In Section~4, we briefly
summarize the main results. The complete atlas is  provided in the Appendix.

\section{OBSERVATIONS AND SPECTRUM REDUCTION}\label{obs}

To create the atlas, we used two high-resolution spectra from a collection
of spectra obtained during the long-term spectral monitoring of this star with the 6-m BTA telescope in combination with the NES echelle spectrograph~\cite{NES}, permanently mounted at the Nasmyth focus.  The NES provides a spectral resolution of \mbox{$R=\lambda/\Delta\lambda\ge60\,000$}. To compile the atlas, we have selected from our collection a spectrum obtained on September 3, 2015 ($\rm JD=2457269.47$). The advantage of this spectrum is that it was registereed over a broad wavelength range of 395--698\,nm. To extend the range to longer wavelengths,
\mbox{$\lambda=778$\,nm}, we used the spectrum from February 9, 2023 ($\rm JD=2459985.28$). Both spectra were obtained with a $4608\times2048$ CCD, with an element sized 0.0135$\times$0.0135\,mm; readout noise of
1.8e$^-$. To reduce the flux loss at the entrance slit, the NES spectrograph is equipped with a stellar image slicer. Using the slicer, each spectral order is repeated thrice.

Extraction of one-dimensional data from the two-dimensional echelle images was performed using the MIDAS
package ECHELLE  context, modified by~\citet{MIDAS} to account for the geometry of the NES spectrograph echelle frame. The cosmic particle traces were removed applying the standard technique of median averaging of  pairs of spectra obtained sequentially. A Th-Ar lamp was used to calibrate the wavelength scale.
All subsequent steps in the reduction of the one-dimensional spectra were performed using the current version of the DECH20t package by~\citet{DECH}. The systematic error in measuring the heliocentric radial velocity $V_r$ from telluric features and from the interstellar lines of the Na\,I doublet does not exceed
0.25\,km\,s$^{-1}$ for a single line; the $V_r$ measurement error for broad features does not
exceed 0.5~km s$^{-1}$. It should be noted that the high accuracy of the heliocentric velocity
finding served as an additional criterion for the reliability of details identification.

\section{THE ATLAS AND A LIST OF IDENTIFIED LINES}\label{atlas}

The atlas is a collection of 59~figures, each corresponding to a separate echelle order of
the CI\,Cam spectrum.  The overlap of the orders decreases with wavelength  increasing, gaps
start appearing between them  at $\lambda \approx 5600$\,\AA{}. The relative intensity
$I/I_{\rm cont}$ is plotted on the ordinate axis, and the observed wavelength ---
on the abscissa axis. The figures mark the identified spectral features with the laboratory
wavelengths rounded to 0.1\,\AA{}. The positions of the identified features in the figures
correspond to the observed wavelengths. The peaks of the strong H\,I and He\,I emissions are
cropped in the atlas figures, but their full profiles can be seen in Figs.\,1 and 2 in~\cite{KMP}.

\begin{figure}[hbtp]
\includegraphics[angle=0,width=0.5\textwidth,bb=0 0 750 1250,clip]{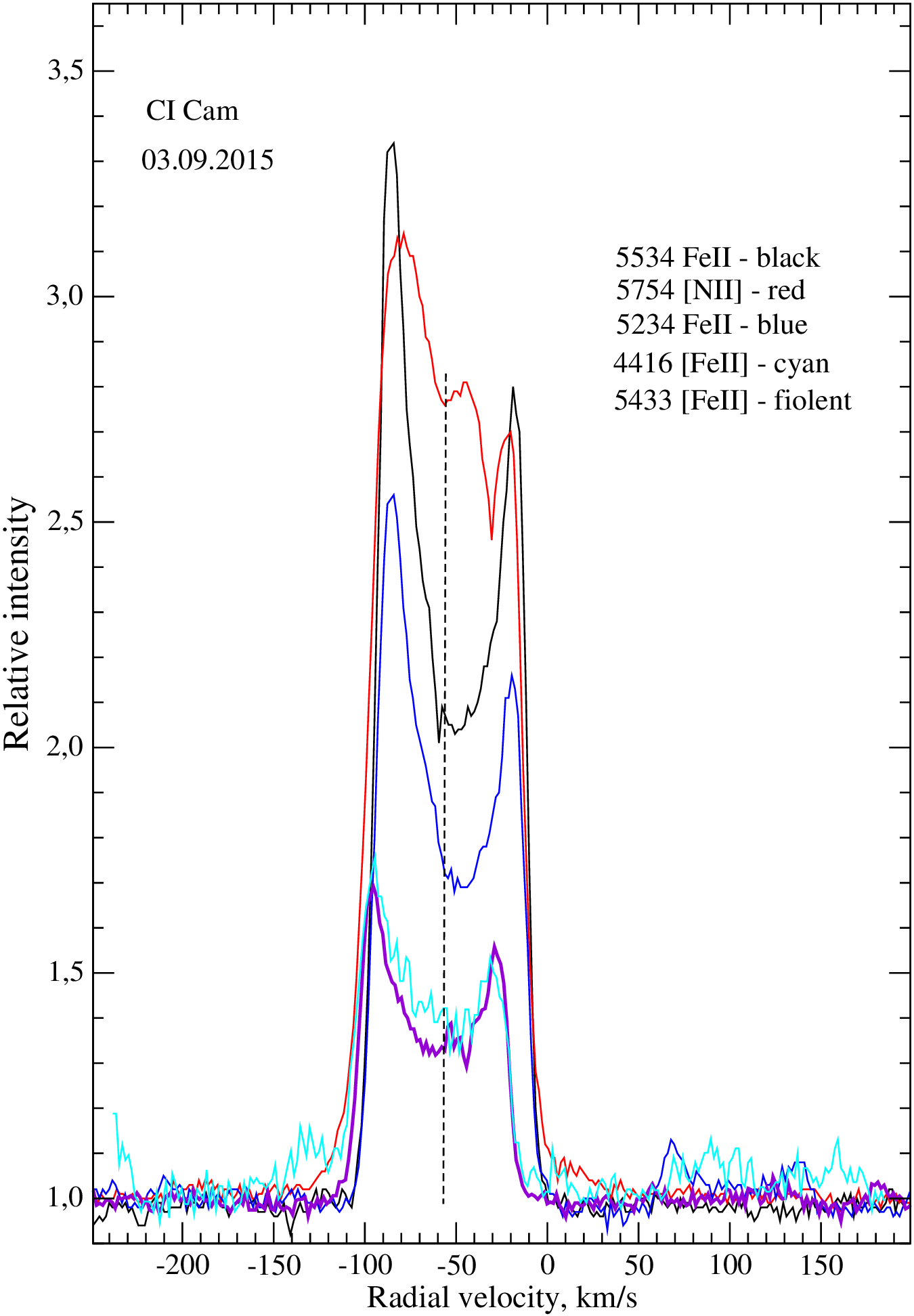}
\caption{The profiles of selected  emissions in the spectrum obtained on September 3, 2015.
The position of the vertical  corresponds to a systemic velocity of $V_{\rm sys}=-55.4$\,km\,s$^{-1}$.}
\label{Profiles}
\end{figure}

All the spectral features are listed in Table~1. The laboratory wavelengths are indicated for
each of them, and for most lines, the multiplet is given according to Moore (1945). Some
known interstellar absorptions are also included in the atlas and Table~1. The atlas, below
the continuum level, lists several DIBs, interstellar features of Ca\,II, K\,I, and interstellar
components of the Na\,I D-lines doublet. The interstellar spectral features and interstellar
extinction  for CI\,Cam were previously studied in more detail in~\cite{KMP}.
Several atlas fragments contain telluric absorptions of O$_2$ and H$_2$O molecules; their
positions are indicated by dots in the figures. Larger dots indicate the positions of telluric
emissions of  [O\,I]\,5577 and 6300\,\AA{}.

\begin{figure}[hbtp]
\hbox{
\includegraphics[width=0.5\textwidth, bb=10 10 1500 1000,clip]{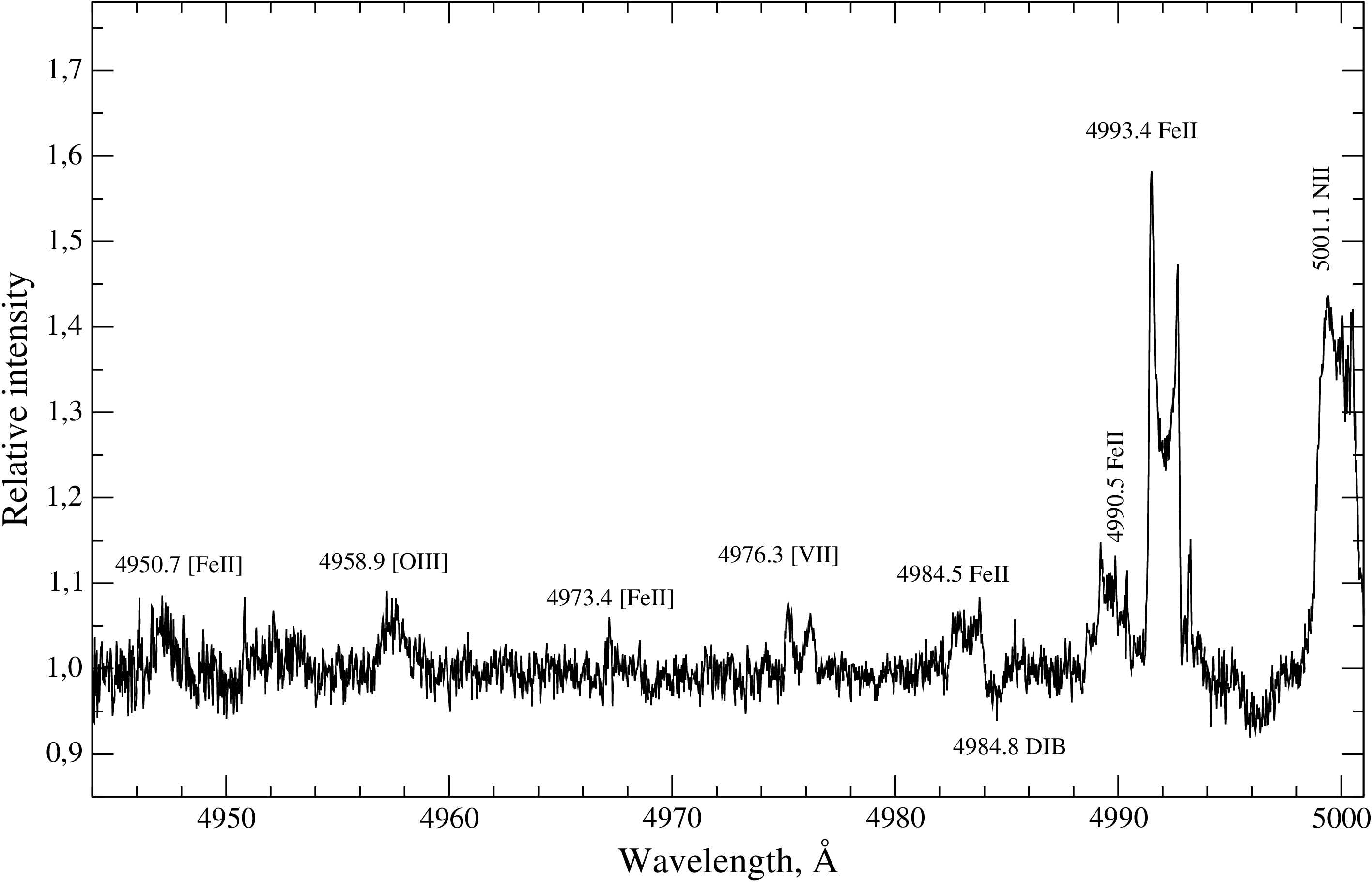}
\includegraphics[width=0.5\textwidth, bb=10 10 1500 1000,clip]{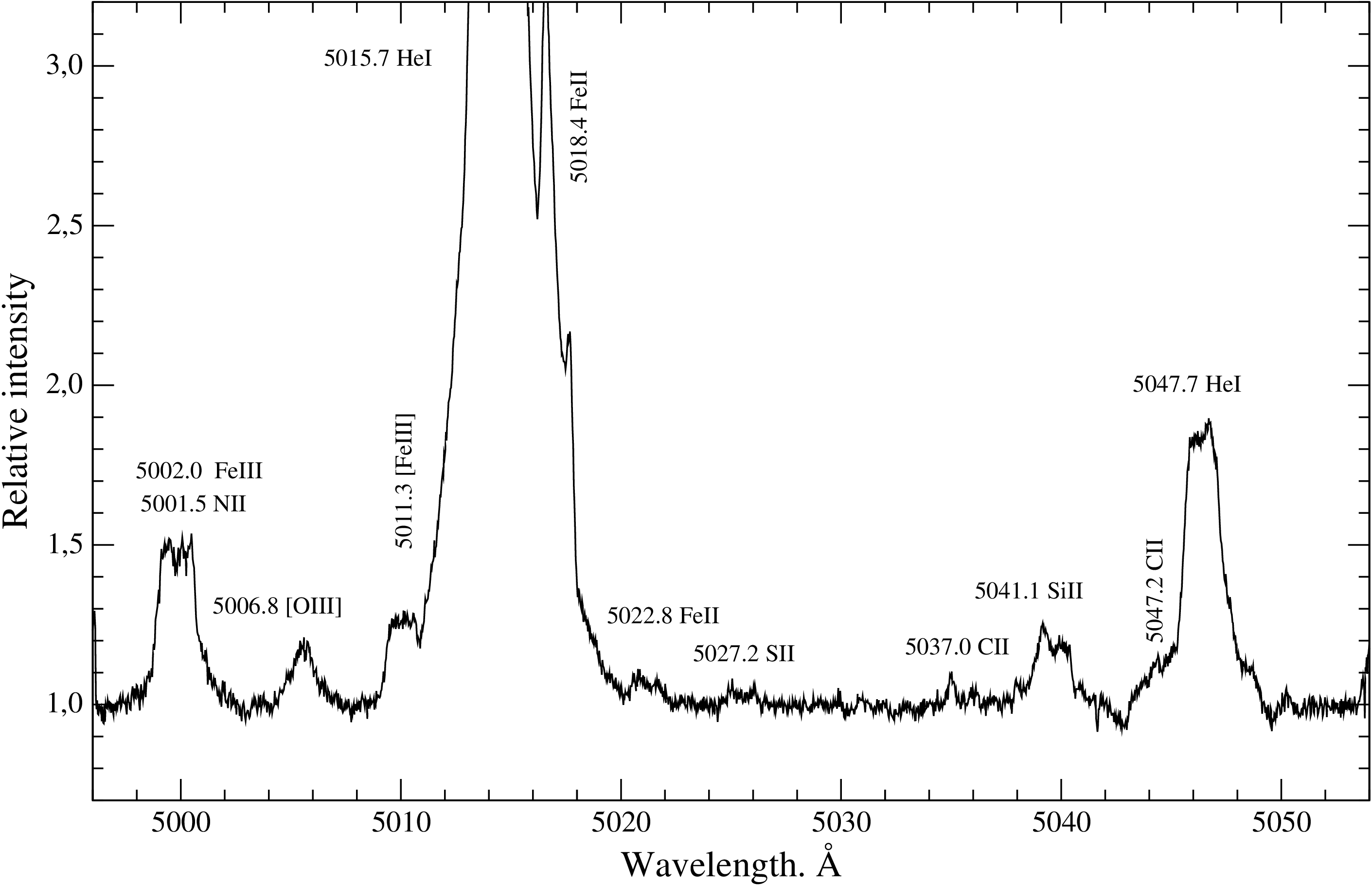} 
}
\caption{The Bowen emissions of [O\,III]\,4959 and 5007\,\AA{} lines, left  and right panels,
   respectively.}
\label{frag26}
\end{figure}

We have identified the features in the spectra of CI\,Cam using the data from papers~\cite{3Pup, Chen2016}  based on the BTA+NES spectroscopy of related stars with the B[e]--phenomenon,
as well as the information from the VALD database~\cite{VALD}  and the Multiplet Tables in
the well-known work by~\citet{Moore}.

Figure\,2 in the article~\cite{KMP} demonstrates the main difference between the two types of
emissions in the CI\,Cam spectrum, namely, the manyfold difference in the widths of intense
H\,I, He\,I emissions and emissions with ``rectangular'' profiles. The H\,I and He\,I emissions
are the dominant features in the CI\,Cam spectrum, with the peak intensities of their profiles
exceeding the local continuum level by a factor of 25--45. Moreover, in contrast to the spectra
of most stars with the B[e] phenomenon, the H$\alpha$ profile in the CI\,Cam spectrum is single-peaked  (see Fig.\,1 in~\cite{KMP}. For comparison, see the H$\alpha$ profiles in the studies~\cite{Aret2016,Maravelias,Mirosh2023}.

While the H$\alpha$ emission profile in CI\,Cam spectra  is stationary, minor variations in the intensity
of the peak of  this line profile  in~\cite{KMP, Zickgraf}  and the presence of a small hump on its long-wavelength wing  in~\cite{KMP} were registered, respectively.
Note also the absence   of P\,Cyg-type profiles in the spectrum of CI\,Cam in all H\,I and He\,I lines, indicating the absence
of wind and the associated mass loss  (see Figs.~1 and 2 in~\cite{KMP}).  The lack of
P\,Cyg-type profiles was also underlined in the earliest study of the CI\,Cam high-resolution spectra
obtained immediately after the outburst in April 1998~\cite{Robinson}.

The next characteristic peculiarity of the  CI\,Cam spectrum is a multitude of narrow two-peak
emissions of ions (Fe\,II, [Fe\,II], Cr\,II, [Cr\,II],  S\,II, [S\,II], N\,II,  [N\,II], etc.).
In terms of the number of narrow emissions found and their intensity, Fe\,II ions lead, there are
110 of them in the presented wavelength range. Forbidden [Fe\,II] emissions are four times fewer and
have low intensities, generally no higher than 10\% above the local continuum. For the illustration
purposes, Fig.\,\ref{Profiles}  shows typical permitted Fe\,II emissions and the pair of the most
intense forbidden [Fe\,II] emissions in our spectrum. All the profiles of forbidden and permitted
emissions in the figure have a peak intensity ratio of $\rm V/R>1$, and all of them have  a deep
central depression. The forbidden emissions are somewhat narrower than the permitted ones. A comparison
of similar figures for different observation dates, given in the paper~\cite{KMP}, indicates
synchronicity in the intensity variations of permitted and forbidden emissions.

 The presence of forbidden lines among these narrow emissions indicates their formation in an optically
 thin and weakly expanding medium. Previously, \citet{Zickgraf85} proposed a model for
 the formation of the hybrid spectrum of a hot, massive hypergiant in a structured circumstellar medium.
 In subsequent years, this two-component wind model obtained the  observational and theoretical support, and the paper~(\citet{Zickgraf85})  is now widely cited.
After studying the spectra of CI\,Cam obtained with the NES/BTA in early 2002  authors~\cite{Mirosh2002}
concluded that the narrow spectral features form in an inclined circumstellar disk, seen edge-on.

\begin{figure}[hbtp]
\hbox{
\includegraphics[width=0.5\textwidth, bb=10 10 1500 1000,clip]{631_25k.eps} 
\includegraphics[width=0.5\textwidth, bb=10 10 1500 1000,clip]{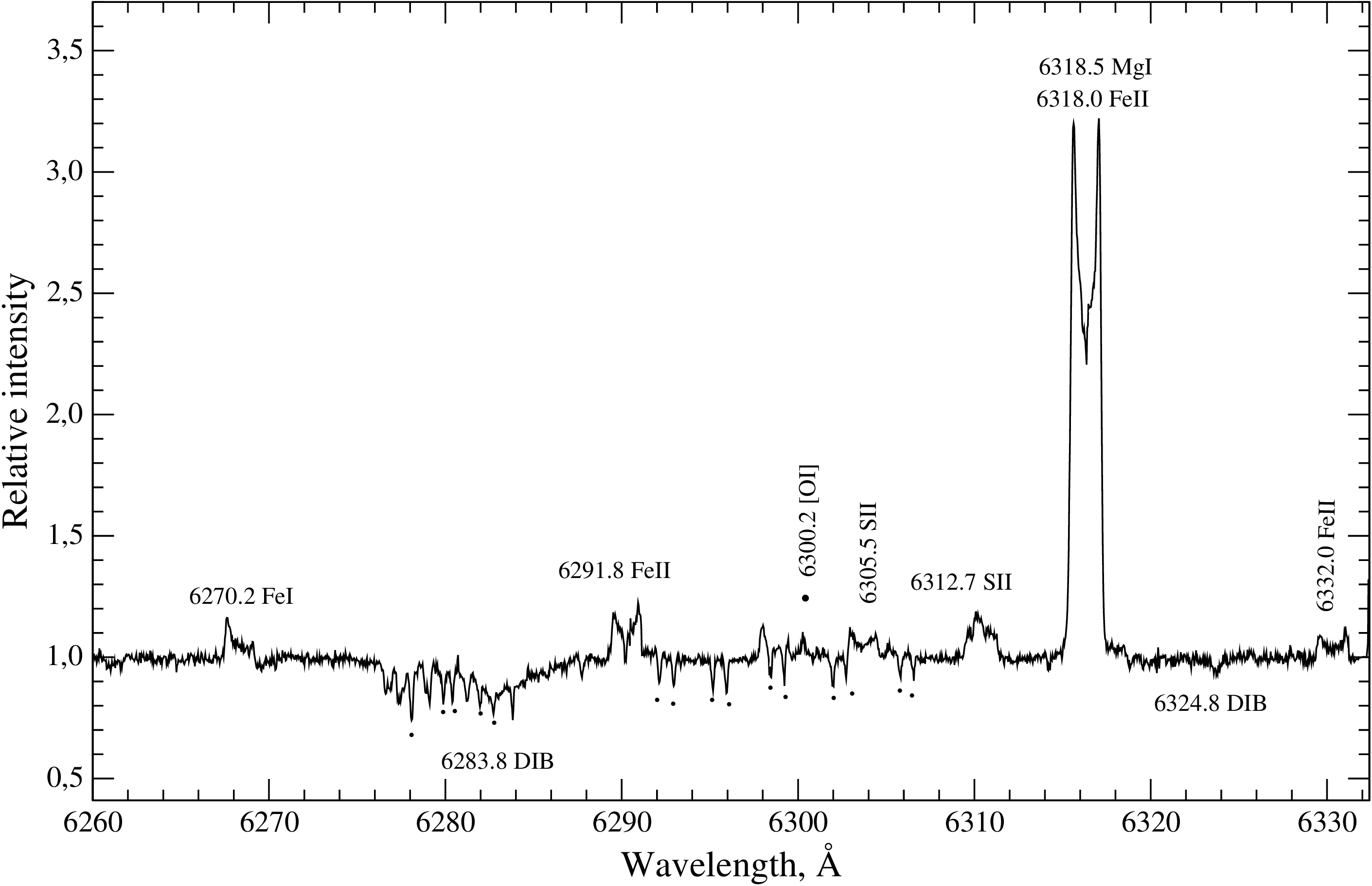}
}
\caption{Fragments with telluric emissions of the [O\,I]\,5577 and 6300\,\AA{}  lines.}
\label{frag45}
\end{figure}

The specific profiles of these emissions have almost vertical slopes and concave peaks, confirming their
formation in an optically thin compact disk located close to the star.  The presence of a structure at
the peaks of permitted  and forbidden emissions indicates the stratification of the medium in which they formed.
It is worth noting here the absence of such strictly rectangular profiles in the spectra of stars with the  B[e] phenomenon, members of the sample in the aforementioned paper by~\citet{Maravelias}.
According to the results in~\cite{KMP}, the intensities of the short-wavelength peaks
of narrow emissions in the CI\,Cam spectrum are higher than the long-wavelength ones for all dates of our observations,
with the exception of the spectrum of January~13, 2011 with an inverse $\rm V/R$  ratio,
and the spectrum of November~18, 2008 with equal peak intensities. For these emissions, a significant
temporal intensity decrease was found the longer the time passed since the 1998 flare.

As follows from the atlas fragments and Table\,1, the spectrum of CI\,Cam contains
several  weak symmetric emissions  of multiply ionized oxygen and metals ([O\,III], Si\,III, Al\,III, Fe\,III,  [Fe\,III]).
Emissions of different types are formed under different physical conditions of the complex envelope
and have widths differing by manyfold, which is well illustrated by Fig.\,2 in the paper~\cite{KMP}.
Low-intensity [O\,III]\,4959 and 5007\,\AA{} emissions with symmetric profiles (see here Fig.\,\ref{frag26})
are reliably  identified; their positions, as follows from Table~1 in the paper~\cite{KMP},
correspond to the general radial velocity pattern.

The fragments of  the  Fig.\,\ref{frag45}  with telluric forbidden emissions of [O\,I]\,5577\,\AA{}
do not contain any obvious emissions at all  that could belong to the CI\,Cam system.
In Fig.\,\ref{frag45}  one can barely discern a faint trace of a detail with an intensity about
5\% near the telluric feature of [O\,I]\,5577\,\AA{}. Yet right is the intensity of the feature, which can be considered as
an emission of [O\,I]\,6300\,\AA{} with its reduced excitation potential. A similar pattern of weak
[O\,I]\,5577 and 6300\,\AA{} emissions was detected for a larger proportion of the sample stars,
including the bright B[e] star 3\,Pup~\cite{3Pup}. The spectrum of the cooler star 3\,Pup differs significantly from the spectrum of CI\,Cam. Their spectra show different types of H\,I emission profiles and metal emissions. Furthermore, photospheric absorptions with a reliably detected position variability were detected in the spectrum of 3\,Pup.

\begin{figure}[hbtp]
\hbox{
\includegraphics[width=0.5\textwidth, bb=10 10 1500 1000,clip]{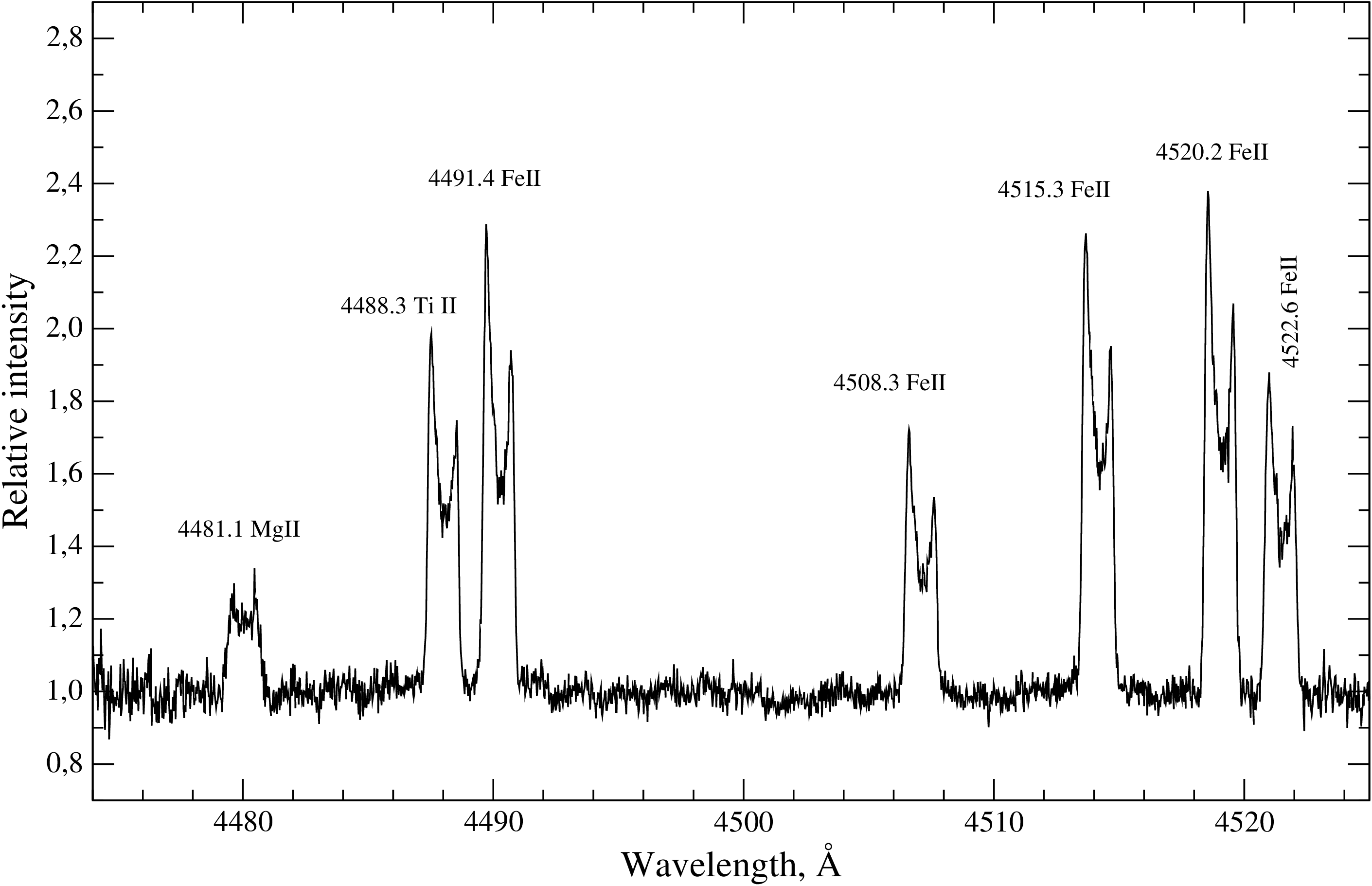}   
\includegraphics[width=0.5\textwidth, bb=10 10 1500 1000,clip]{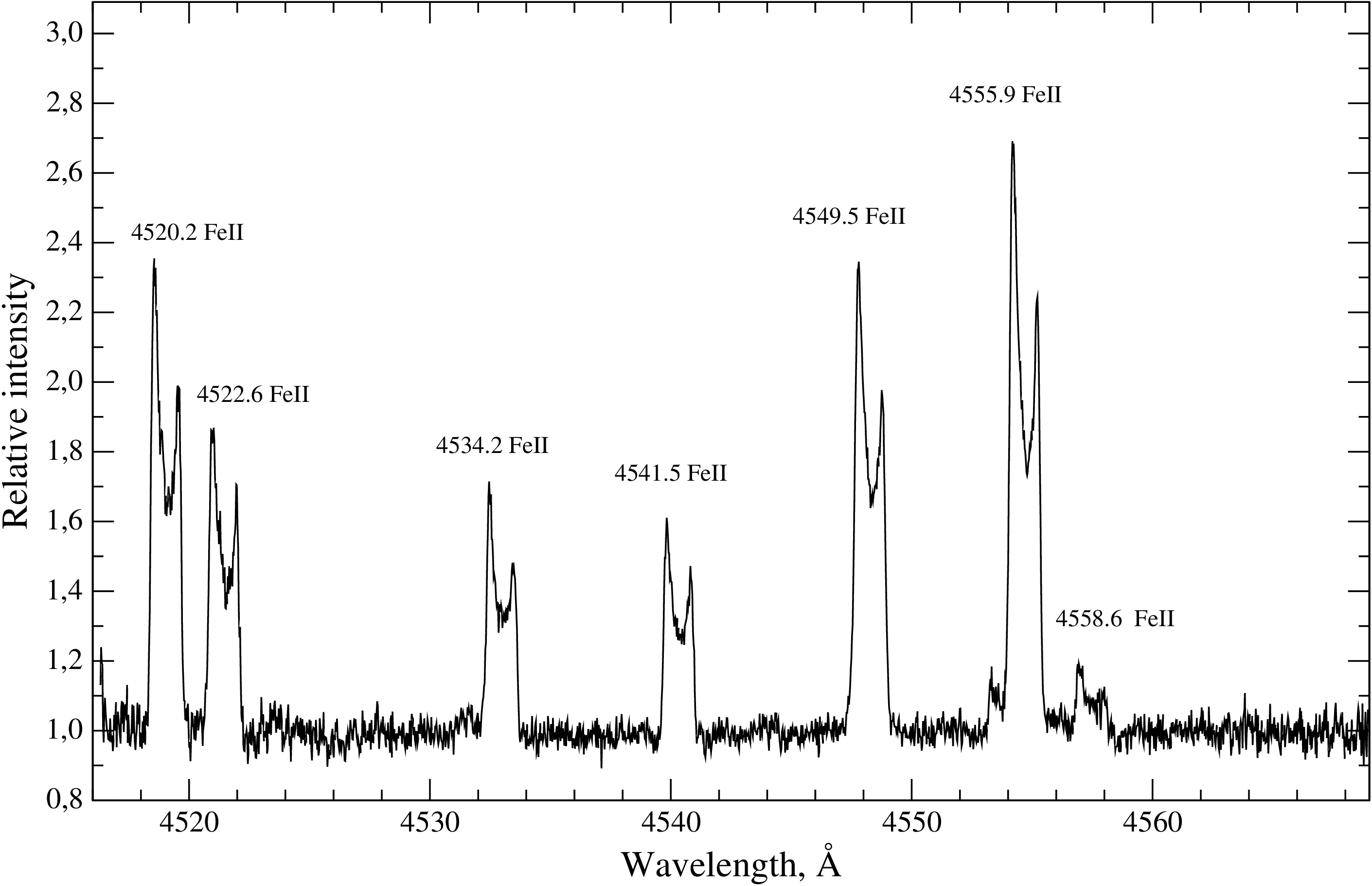}
}
\caption{Fragments with  emissions of the Mg\,II, Ti\,II,  Fe\,II ions.}
\label{frag15}
\end{figure}

\begin{figure}[hbtp]
\hbox{
\includegraphics[width=0.5\textwidth,bb=10 10 1500 1000,clip]{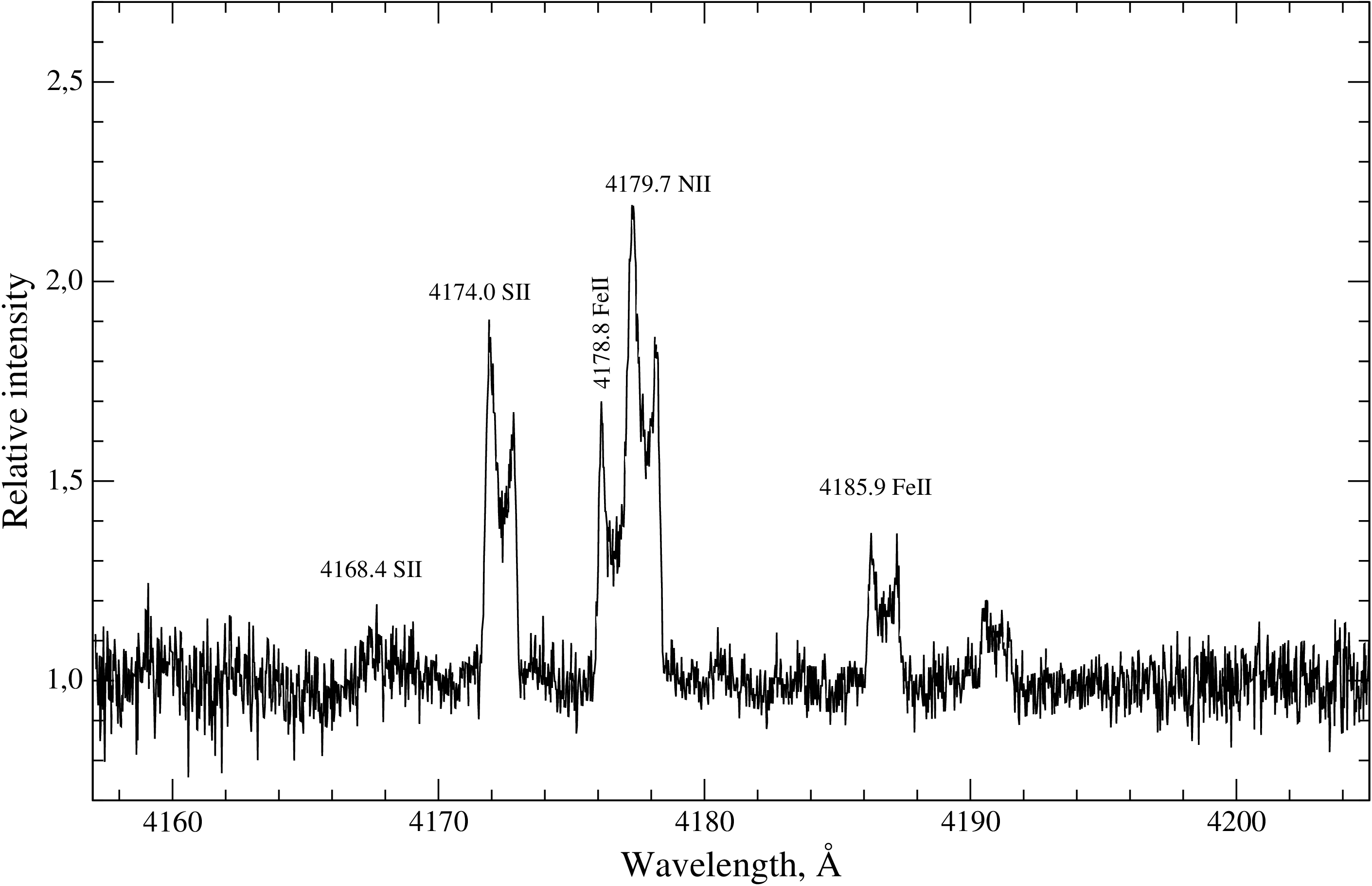}  
\includegraphics[width=0.5\textwidth,bb=10 10 1500 1000,clip]{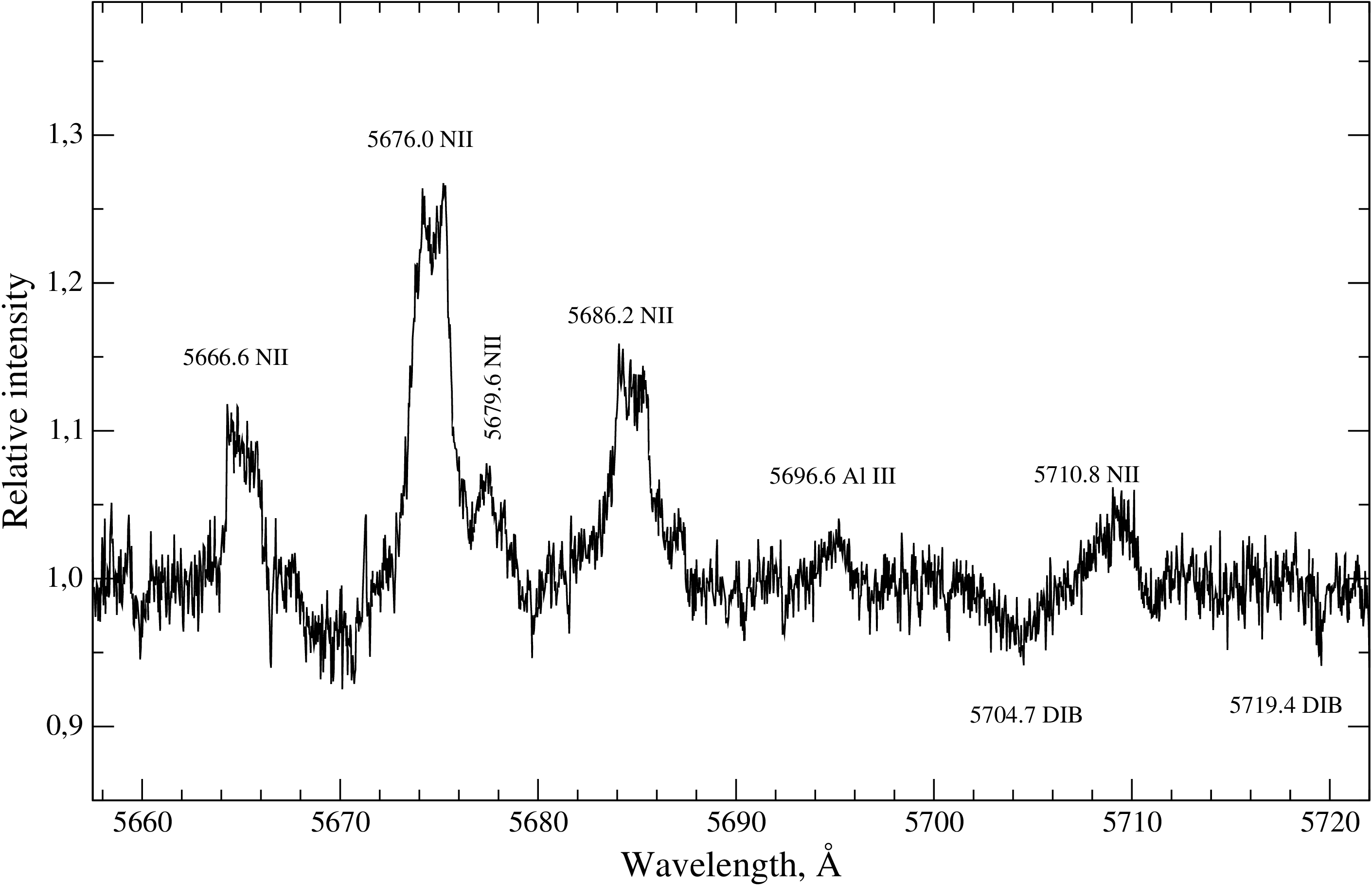}
}
\caption{The fragments of the spectrum with permitted emission lines of the N\,II ions.}
\label{frag07}
\end{figure}

\begin{figure}[hbtp]
\hbox{
\includegraphics[width=0.5\textwidth,bb=10 10 1500 1000,clip]{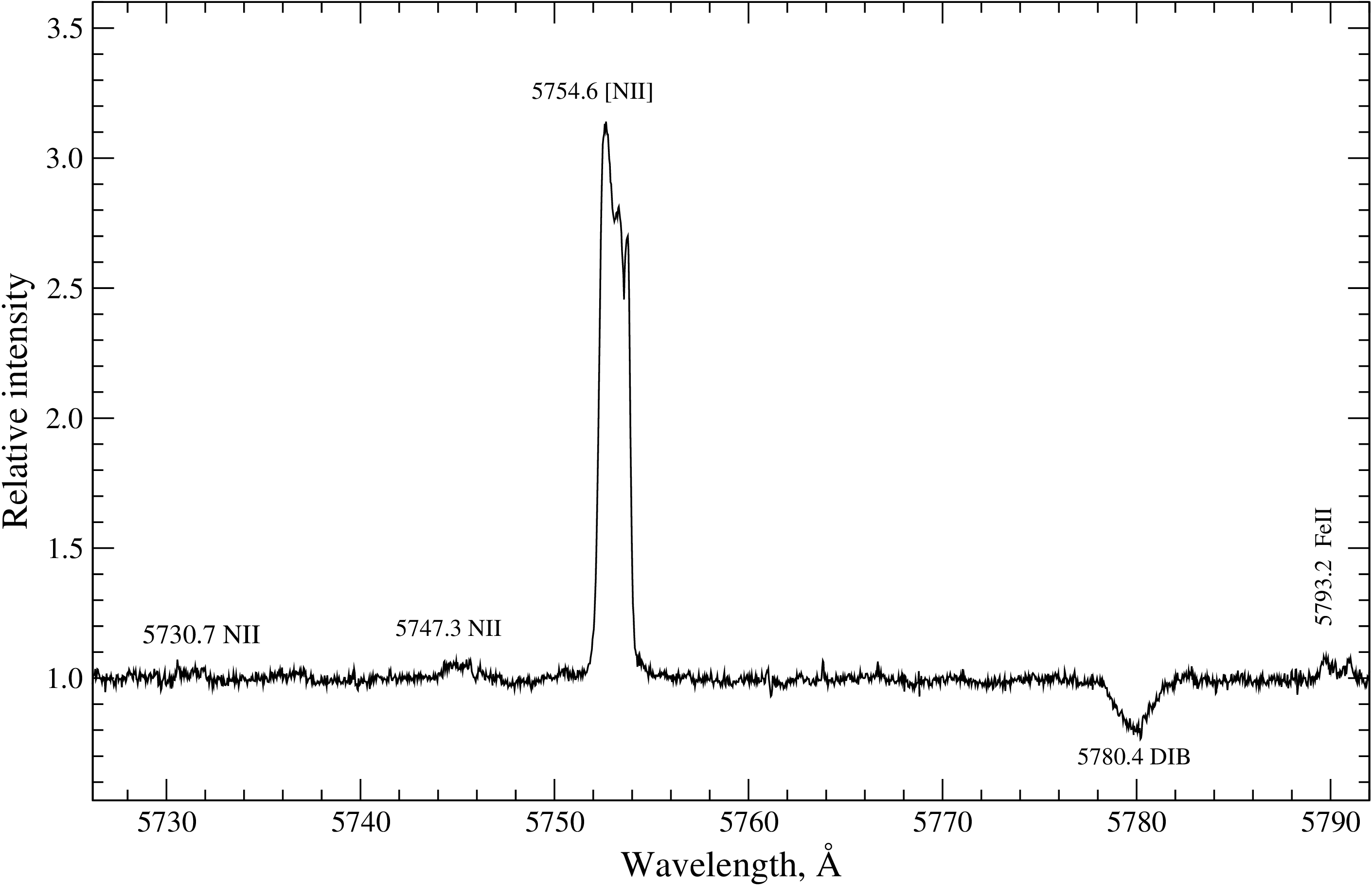}   
\includegraphics[width=0.5\textwidth,bb=10 10 1500 1000,clip]{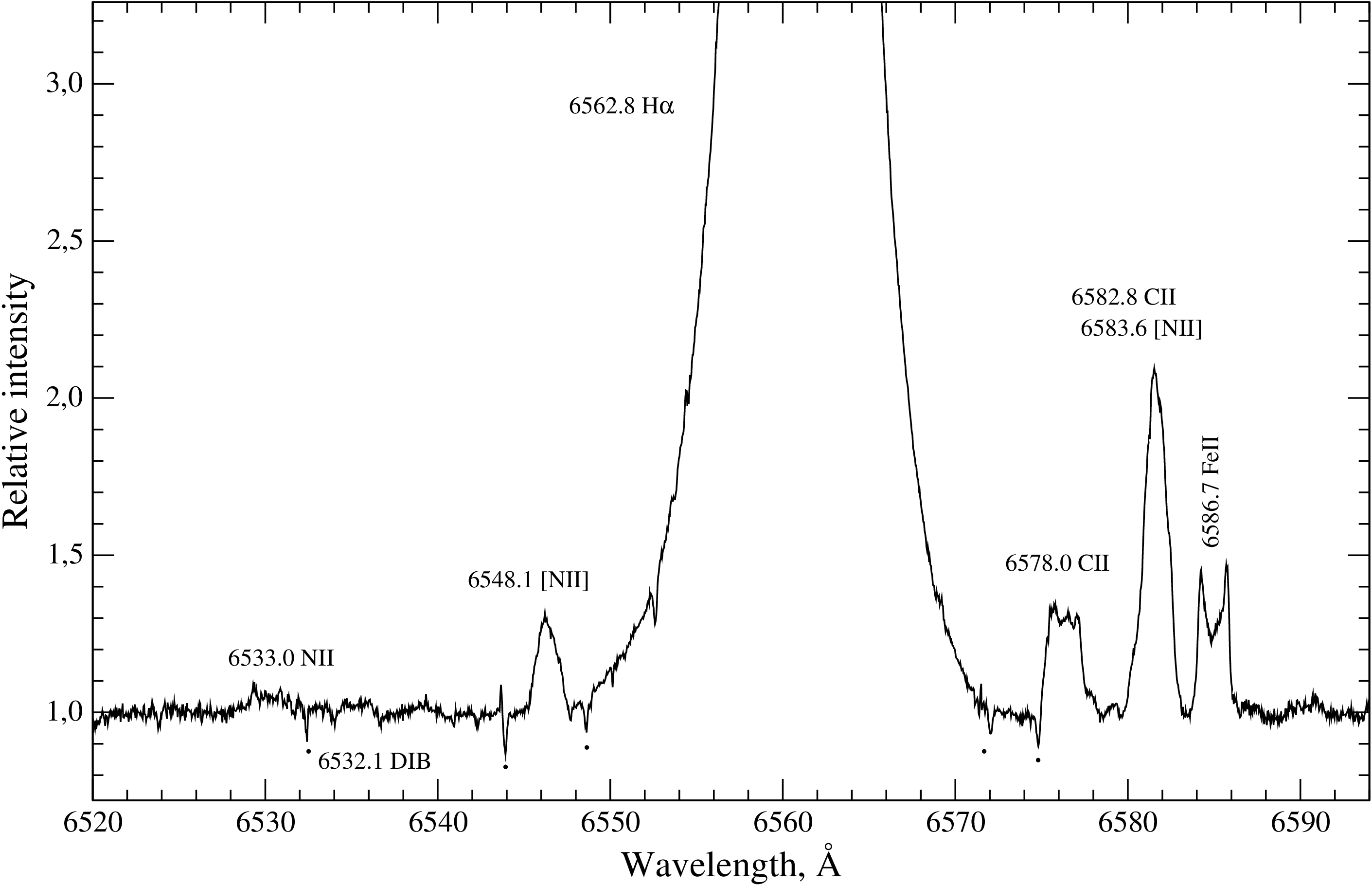}
}
\caption{The fragments of the spectrum with permitted and forbidden emissions of ionized nitrogen.}
\label{frag38}
\end{figure}

\begin{figure}[hbtp]
\hbox{
\includegraphics[width=0.5\textwidth,bb=10 10 1500 1000,clip]{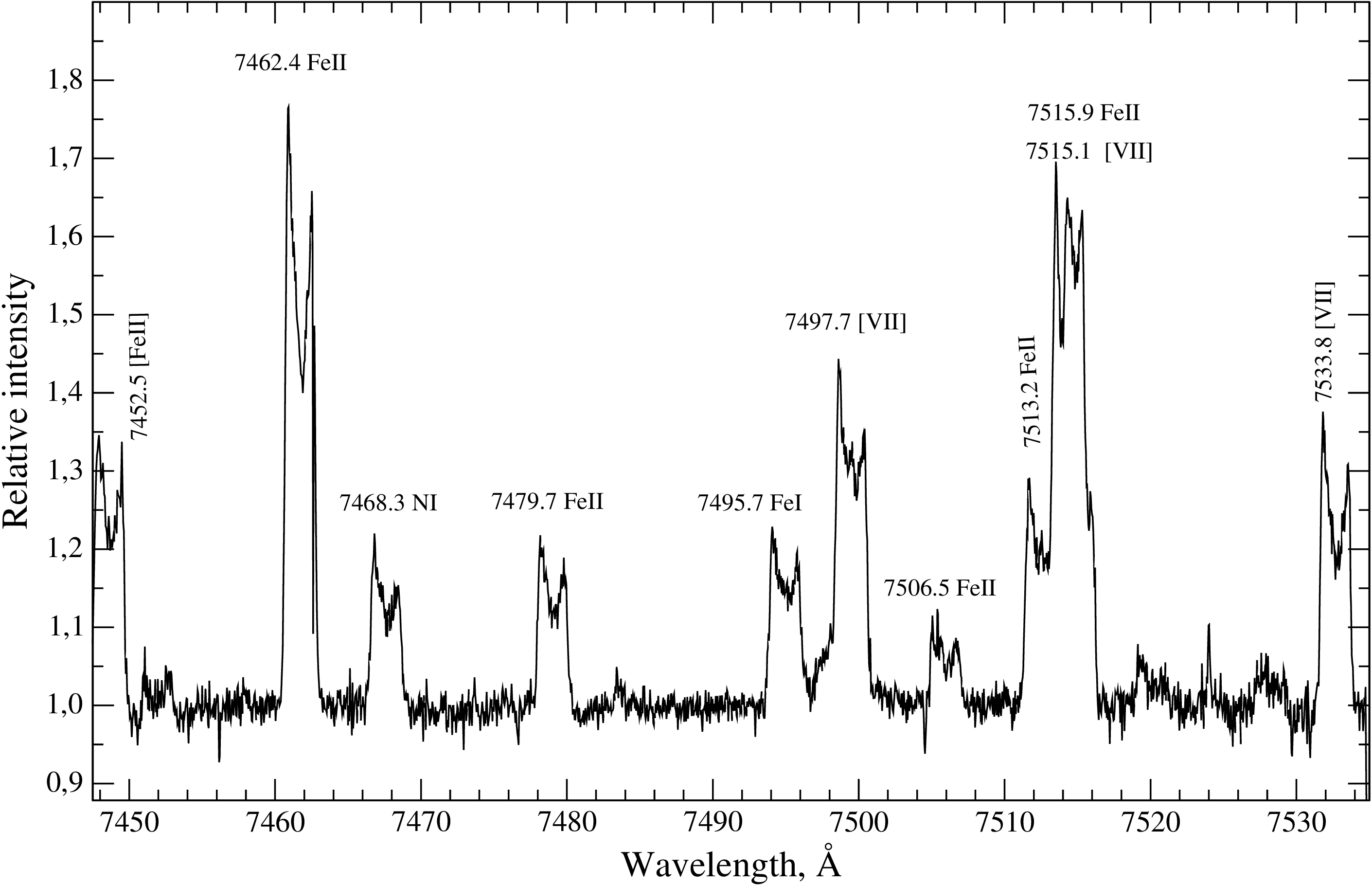} 
\includegraphics[width=0.5\textwidth,bb=10 10 1500 1000,clip]{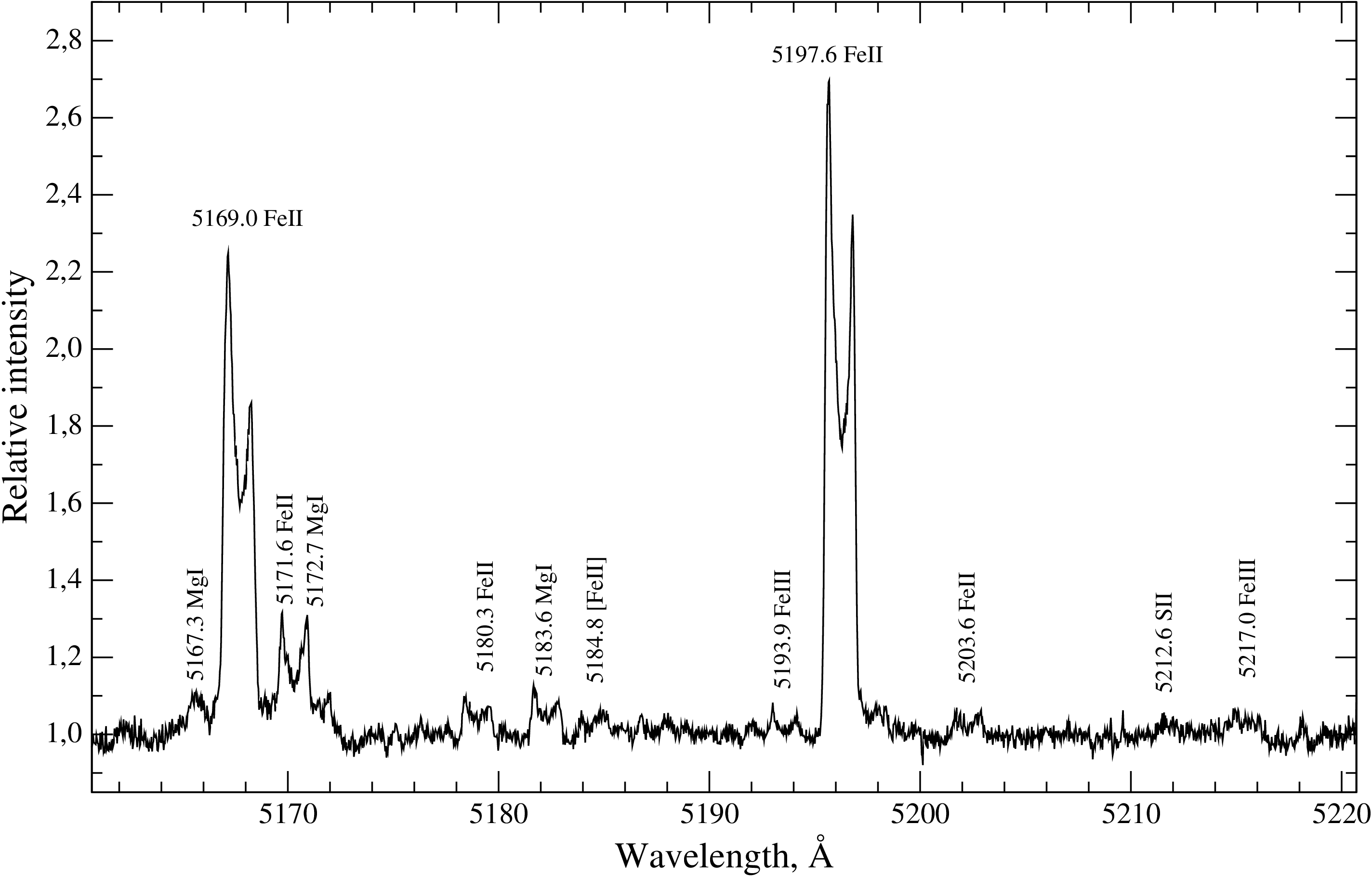}
}
\caption{Emissions of the N\,I atom (left panel)  and the Mg\,I triplet (right one).}
\label{frag759_38}
\end{figure}

Note that~\citet{Robinson} revealed the presence of the [O\,I]\,5577~\AA{}
emission in the  most early high-resolution spectra (R=60\,000) of CI\,Cam after the outburst.
High quality of the spectroscopic data  in our atlas and in the study~\cite{Robinson} allows us
to discuss the good chances for the
presence of  physical condition differences in the region of formation of these emissions immediately
after the outburst and 17 years on.
In addition to the situation with the [O\,I]\;5577 emission, the differences in the emission profiles of
metals can also be noted. For instance, in Fig.~7 of~\cite{Robinson}, the emission profiles of
the Ti\,II, Cr\,II, Fe\,II  ions had rounded tops, while the profiles of the same metal emissions
in the fragments of our atlas (see Fig.\,\ref{frag15} in this text) are two-peaked, which very
well illustrates the changes that have occurred  over the years in the physical conditions of the
circumstellar medium of CI\,Cam.

Let's emphasize  the richness of the CI\,Cam spectrum with nitrogen features: this is a multitude
of resolved emissions of the N\,II ions, which is well illustrated by Fig.\,\ref{frag07}, as well
as intense forbidden emissions of [N\,II] (Fig.\,\ref{frag38}). Panels  of Fig.\,\ref{frag759_38}
also contains a fairly
strong emission from the nitrogen atom N\,I\,7468.3\,\AA{}.  This  richness of nitrogen details
of a far evolved star can naturally be
explained by the production of nitrogen during the previous phases of the massive star's evolution
and the subsequent releasing  of the newly produced  element into the circumstellar medium. Therefore,
there is an indication on the high initial mass of the star and a reason to classify CI\,Cam as a supergiant.
The main criteria for classifying a star as a B[e] supergiant are presented in the survey by~\citet{Kraus2019}.
The B[e] phenomenon consists of the presence in the stellar  spectrum of a number of peculiar features:
strong H\,I  (and He\,I emissions in the spectra of the hottest objects with T$_{\rm eff}>20\,000$\,K),
as well as emissions of permitted metal ion lines and low-excited forbidden lines.
The second  significant characteristic of stars exhibiting the B[e] phenomenon is a large excess
of infrared flux, caused by the presence of hot dust in the stellar  envelope. The known
features of CI\,Cam satisfy these requirements. However, since the discovery of the B[e] phenomenon
by~\citet{Allen}, it has been known that stars exhibiting these two key characteristics
form a group of highly diverse objects.

As follows from Table\,1, in addition to numerous ion emissions, the spectrum of CI\,Cam
contains identified emissions of neutral  N\,I, O\,I, Mg\,I atoms, and about 20 Fe\,I lines.
Left panel in~Fig.\,\ref{frag759_38}, along with intense emissions of Fe\,II and [VII] ions,
contains emissions of neutral  N\,I and Fe\,I atoms. Right panel in the same Figure\,\ref{frag759_38}
shows the Mg\,I triplet.  All the components of the IR triplet of the oxygen atom O\,I\;7773\,\AA{}
have high intensity.
Forbidden lines of the [Ca\,II]\;7291 and 7323\,\AA{} doublet, typical of stars with the
B[e] phenomenon (see, e.g.,~\cite{3Pup,Chen2016,Maravelias}) are missing in the spectrum
of CI\,Cam. The absence of the [Ca\,II] doublet lines and the [O\,I]\,6300\,\AA{} line in
the spectrum of CI\,Cam was alreay noted  by~\citet{Aret2016}.

\begin{figure}[hbtp]
\hbox{
\includegraphics[width=0.5\textwidth,bb=10 10 1500 1000,clip]{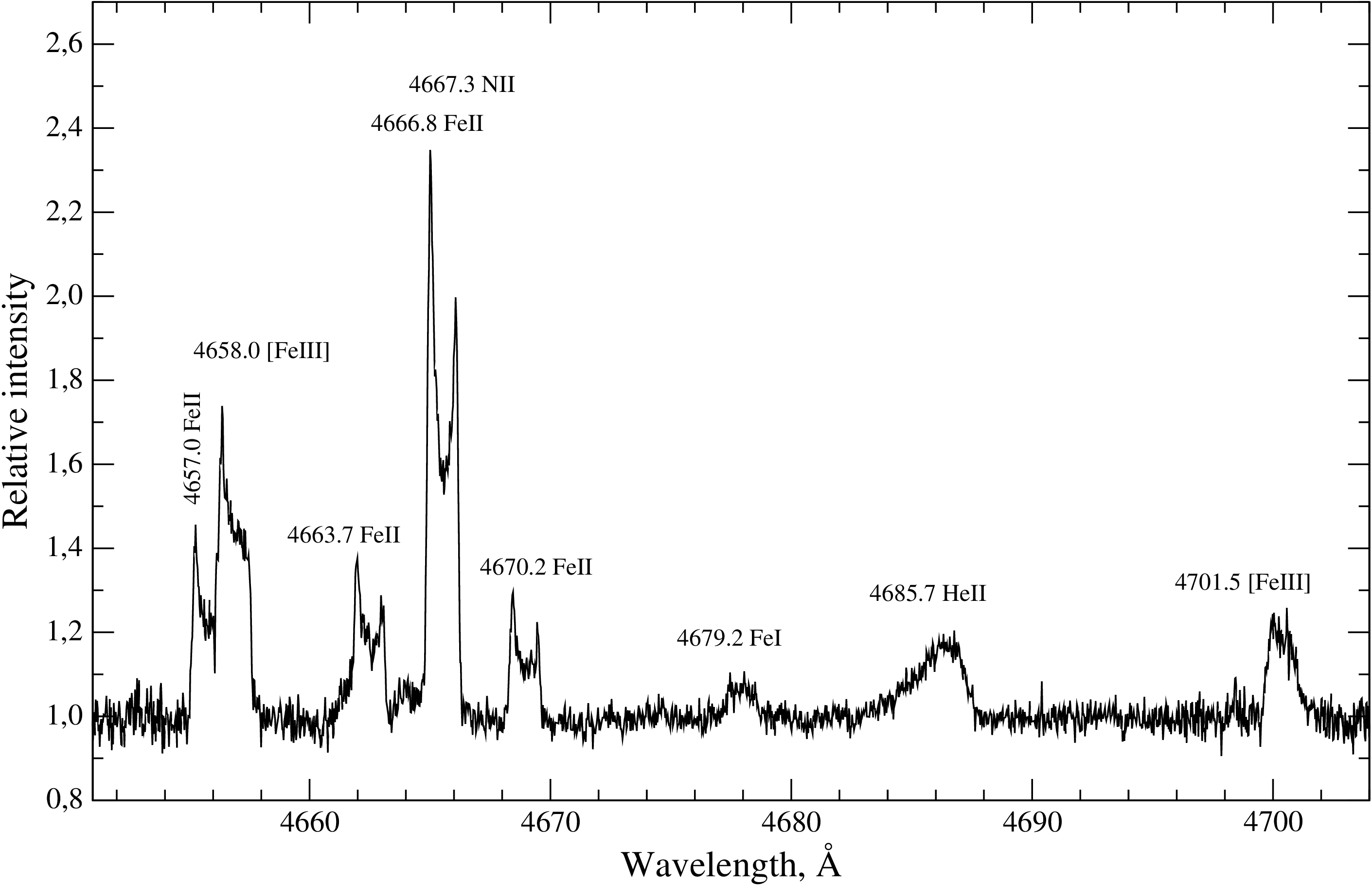} 
\includegraphics[width=0.5\textwidth,bb=10 10 1509 1000,clip]{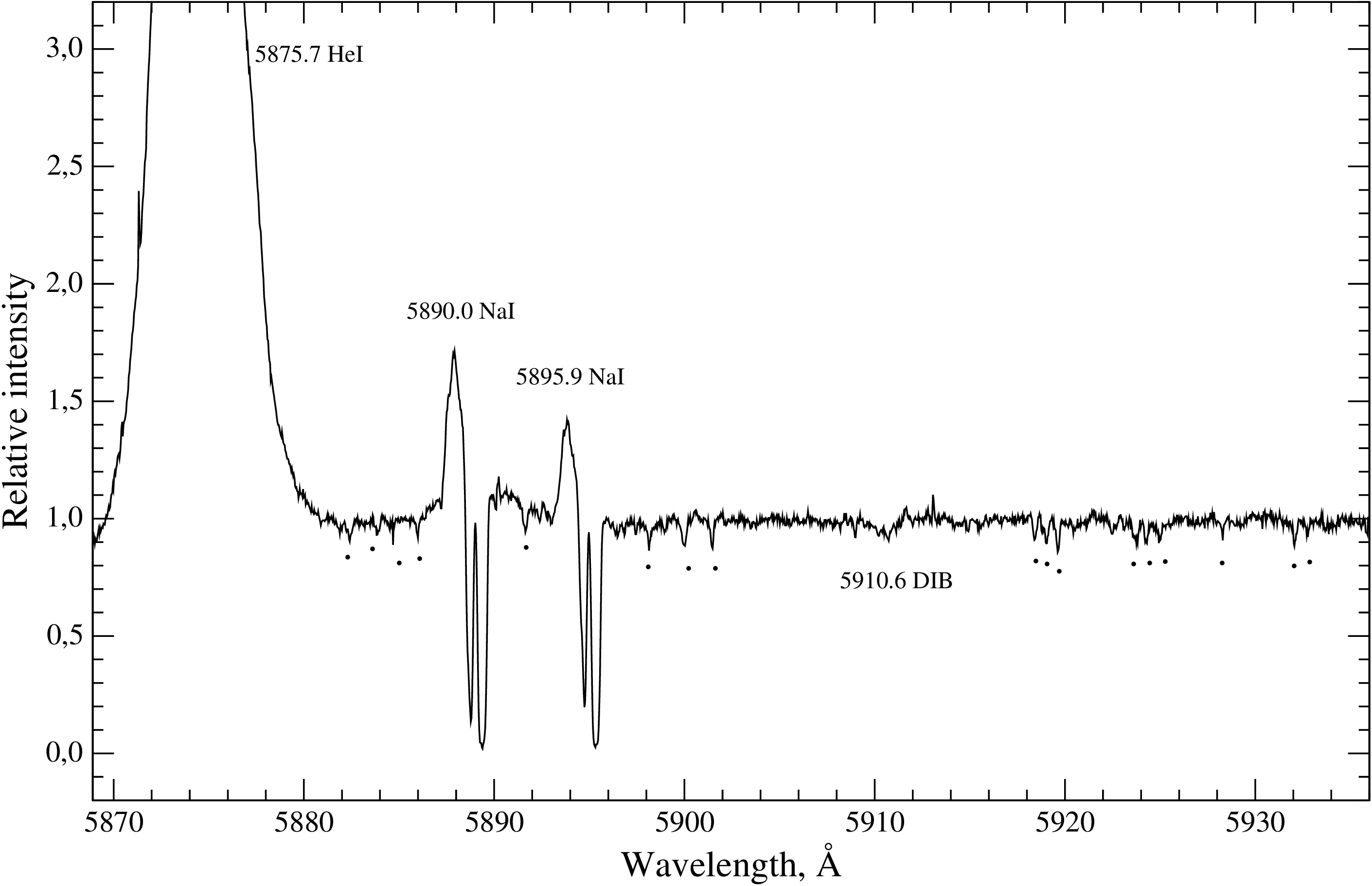}
}
\caption{Fragments of the spectrum of CI\,Cam with an emission near $\lambda$\,4686\,\AA{} and
        with a doublet of Na\,I D-lines,  left  and right panels, respectively.}
\label{frag19}
\end{figure}

As authors~\cite{KMP} have shown,  the intensity and position of the emission  near
$\lambda$\,4686 \AA{} in the stellar spectrum vary significantly. Figure\,6
in the paper~\cite{KMP} shows fragments  for six observation dates.
This line is missing in the spectra number  ``1'', ``3'', ``4'' and ``5'' in the spectrum
number ``2'' the emission intensity is low (less than 5\% of the local continuum).
A sufficient  intensity of this line, approximately 16\% of the local continuum level (see left panel
of  Fig.\,\ref{frag19}), was registered only in the spectrum for September~2015, which was used to create this atlas.
Besides, the profile of this line  has an atypical shape, and the radial velocity corresponding
to the position of this feature, V$_r(\lambda\,4686)=63.2\,\text{km\;s}^{-1}$,
differs significantly  from the velocity of other emissions in the CI\,Cam spectrum.

\clearpage
\onecolumngrid
\renewcommand{\baselinestretch}{1.0}
\setlength{\tabcolsep}{2.0pt} 
\begin{longtable*}[c]{ l | l ||l | l ||l | l ||l | l }
\caption{The features identified in the spectrum of CI\,Cam}  \\   
 \hline
\multirow{2}{*}{ $\lambda_{\rm lab}$, \AA}&  \multicolumn{1}{c||}{Element}&\multirow{2}{*}{$\lambda_{\rm lab}$, \AA}&  \multicolumn{1}{c||}{Element}&\multirow{2}{*}{$\lambda_{\rm lab}$, \AA}& \multicolumn{1}{c||}{Element}& \multirow{2}{*}{$\lambda_{\rm lab}$, \AA}&  \multicolumn{1}{c}{Element}\\
&\multicolumn{1}{c||}{(multiplet)}&& \multicolumn{1}{c||}{(multiplet)}&& \multicolumn{1}{c||}{(multiplet)}&& \multicolumn{1}{c}{(multiplet)}\\
 \hline
 \endfirsthead
 \caption{(Contd.) \\}\\
 \hline
 \multirow{2}{*}{ $\lambda_{\rm lab}$, \AA}&  Eleme\ref{frag15}nt&\multirow{2}{*}{$\lambda_{\rm lab}$, \AA}&  Element&\multirow{2}{*}{$\lambda_{\rm lab}$, \AA}& Element& \multirow{2}{*}{$\lambda_{\rm lab}$, \AA}&  Element\\
&(multiplet)&& (multiplet)&&  (multiplet)&& (multiplet) \\
 \hline
 \endhead
 \hline
 \endfoot
 \hline
 \endlastfoot
3964.73 &He\,I~(5)      &4178.85 &Fe\,II~(28)    &4369.28 &O\,II~(26)    &4520.22 &Fe\,II~(37)   \\
3970.07 &H$\epsilon$    &4179.67 &N\,II~(50)     &4384.32 & Fe\,II~(32)  &4522.63 &Fe\,II~(38)   \\
3973.26 &O\,II~(6)      &4185.92 &Fe\,II         &4385.38 & Fe\,II~(27)  &4534.17 &Fe\,II~(37)   \\
3982.72 &O\,II~(6)      &4185.95 &S\,II          &4387.93 &He\,I~(51)    &4541.52 &Fe\,II~(38)   \\
3995.00 & N\,II~(12)    &4233.17 &Fe\,II~(27)    &4410.06 &C\,II~(40)    &4549.47 &Fe\,II~(38)   \\
4002.55 & Fe\,II~(190)  &4243.98 &[Fe\,II]~(21F) &4413.60 &Fe\,II~(32)   &4555.89 &Fe\,II~(37)   \\
4009.27 & He\,I~(55)    &4257.38 &[Fe\,II]       &4413.78 & [Fe\,II]~(7F)&4558.58 &Fe\,II~(20)   \\
4023.97 & He\,I~(54)    &4257.42 & S\,II~(66)    &4416.27 &[Fe\,II]~6F   &4576.33 &Fe\,II~(38)   \\
4026.19 & He\,I~(18)    &4263.70 &Fe\,II         &4416.37 & S\,II~(53)   &4580.07 &Fe\,II~(26)   \\
4068.62 & [S\,II]~(1F)  &4267.80 &S\,II~(49)     &4419.60 &Fe\,III~(4)   &4582.84 &Fe\,II~(37)   \\
4069.64 & O\,II~(10)    &4276.8  &[Fe\,II]       &4427.99 &Mg\,II~(9)    &4583.83 &Fe\,II~(38)   \\
4076.22 & [S\,II]~(1F)  &4273.42 &Fe\,III~(121)  &4433.99 &Mg\,II~(9)    &4588.22 &Cr\,II~(44)   \\
4101.74 & H$\delta$     &4278.54 &S\,II~(49)     &4437.55 &He\,I~(50)    &4596.18 &O\,II~(15)    \\
4112.03 & O\,II~(21)    &4287.40 &[Fe\,II]~(7F)  &4452.11 &[Fe\,II]~(7F) &4601.48 &N\,II~(5)     \\
4119.22 & O\,II~(30)    &4296.85 &Fe\,III~(121)  &4457.96 &[Fe\,II]~(6F) &4607.15 &N\,II~(5)     \\
4120.82 &He\,I~(16)     &4303.17 &Fe\,II~(27)    &4461.43 &Fe\,II~(26)   &4620.51 &Fe\,II~(38)   \\
4122.98 &Fe\,III~(118)  &4313.43 &O\,II~(78)     &4471.48 &He\,I~(14)    &4629.34 &Fe\,II~(37)   \\
4128.07 &Si\,II~(3)     &4319.63 &O\,II~(2)      &4474.91 &[Fe\,II]~(7F) &4635.33 &Fe\,II~(186)  \\
4130.89 &Si\,II~(3)     &4325.77 &O\,II~(2)      &4481.13 &Mg\,II~(4)    &4643.09 &N\,II~(5)     \\
4137.93 &Fe\,III~(118)  &4340.46 &H$\gamma$      &4488.33 &Ti\,II~(115)  &4649.14 &O\,II~(1)     \\
4143.76 &He\,I~(53)     &4351.76 &Fe\,II~(27)    &4491.40 &Fe\,II~(37)   &4656.97 &Fe\,II~(43)   \\
4168.41 &S\,II~(44)     &4359.33 &[Fe\,II]~(7F)  &4508.28 &Fe\,II~(37)   &4658.03 &[Fe\,III]~(3F)\\
4174.04 & S\,II~(65)    &4363.21 &[O\,III] (2F)  &4515.33 &Fe\,II~(37)   &4663.7  &Fe\,II~(44)   \\
\hline
\hline
4665.80 &Fe\,II~(26)    &4876.41 & Cr\,II~(30)   &5027.19 &S\,II~(1)      &5306.6  & Fe\,III~(113) \\
4666.78 &Fe\,II~(37)    &4889.63 &[Fe\,II]~(4F)  &5037.0  & C\,II~(17)    &5316.61 & Fe\,II~(49)   \\
4667.28 & N\,II~(11)    &4893.78 & Fe\,II~(36)   &5041.06 &Si\,II~(5)     &5322.78 &Cr\,II~(24)    \\
4670.17 &Fe\,II~(25)    &4905.35 &[Fe\,II]~(20F) &5047.29 &C\,II (15)     &5325.56 & Fe\,II~(19)   \\
4679.23 &Fe\,I~(688)    &4906.88 &O\,II~(28)     &5047.74 &He\,I~(47)     &5337.71 &Fe\,II~(48)    \\
4685.68 & He\,II~(1)    &4921.93 &He\,I~(48)     &5061.79 &  Fe\,II       &5330.6  &Fe\,II         \\
4701.5  &[Fe\,III]~(3F) &4923.92 &Fe\,II~(42)    &5065.42 & [Ti\,II]~(19F)&5339.65 & [Cr\,II]~(12F)\\
4713.14 &He\,I~(12)     &4930.5  &[Fe\,III]~(1F) &5073.59 &N\,II~(10)     &5346.56 & Fe\,II~(49)   \\
4728.07 &[Fe\,II]~(4F)  &4950.74 &[Fe\,II]~(20F) &5075.83 &Fe\,II         &5354.15 & [Cr\,II]~(12F)\\
4731.44 &Fe\,II~(43)    &4958.92 &[O\,III]~(1F)  &5081.92 &Fe\,II~(221)   &5362.86 &   Fe\,II~(48) \\
4733.9  &[Fe\,III]~(3F) &4973.39 &[Fe\,II]~(20F) &5086.69 &Fe\,III (5)    &5376.47 & [Fe\,II]~(19F)\\
4754.7  &[Fe\,III]~(3F) &4976.33 & [V\,II]~(7F)  &5089.28 &Fe\,II         &5381.02 & Ti\,II~(69)   \\
4762.47 & DIB           &4984.5  & Fe\,II        &5093.47 &Fe\,II~(205)   &5395.86 &Fe\,II         \\
4769.4  &[Fe\,III]~(3F) &4990.5  & Fe\,II        &5100.65 &Fe\,II~(35)    &5408.84 & Fe\,II~(184)  \\
4814.55 &[Fe\,II]~(20F) &4993.35 &Fe\,II~(36)    &5106.11 &Fe\,II         &5412.64 & [Fe\,II]~(17F)\\
4824.14 &Cr\,II~(30)    &5001.47 &N\,II~(19)     &5117.11 &Fe\,II         &5414.09 & Fe\,II~(49)   \\
4825.72 &Fe\,II~(30)    &5002.02 &Fe\,III~(19)   &5120.34 &Fe\,II~(35)    &5420.90 &Cr\,II~(23)    \\
4833.20 &Fe\,II~(30)    &5006.84 &[O\,III]~(1F)  &5127.32 &Fe\,III~(5)    &5425.27 &Fe\,II~(49)    \\
4840.00 &Fe\,II~(30)    &5011.3  &[Fe\,III]~(1F) &5132.66 &Fe\,II~(35)    &5427.83 &Fe\,II         \\
4848.25 & Cr\,II~(30)   &5014.03 &S\,II~(15)     &5276.00 &Fe\,II~(48)    &5432.98 &Fe\,II~(55)    \\
4855.54 & Fe\,II~(25)   &5015.68 &He\,I~(4)      &5284.09 &Fe\,II~(41)    &5433.15 &[Fe\,II]~(18F) \\
4861.33 & H$\beta$      &5018.44 &Fe\,II~(42)    &5291.67 &Fe\,II         &5445.97 &Fe\,II~(53)    \\
4871.27 & Fe\,II~(25)   &5022.79 & Fe\,II        &5298.79 & Fe\,I~(875)   &5457.72 &Fe\,II         \\
\hline
\hline
5466.55 &S\,II~(11)           &5640.32 &S\,II~(11)     &5952.53 & Fe\,II~(182)  &6160.75 & Fe\,II~(161)      \\
5472.63 &Cr\,II~(50)          &5648.90 &Fe\,I~625)     &5957.61 &Si\,II~(4)     &6195.98 & DIB               \\
5478.36 &Cr\,II ~(50)         &5666.64 &N\,II~(3)      &5978.97 &Si\,II~(4)     &6203.05 & DIB               \\
5481.17 & [Fe\,I]~(20F)       &5676.02 &N\,II~(3)      &5979.90 &Fe\,III~(117)  &6229.34 &Fe\,II~(34)        \\
5487.52 &[V\,II]~(5F)         &5679.56 &N\,II~(3)      &5991.38 &Fe\,II~(46)    &6231.75 &Al\,II~(10)        \\
5492.82 &Ti\,II~(68)          &5686.21 &N\,II~(3)      &5999.30 &Fe\,III~(117)  &6233.52 &Fe\,II             \\
5503.18 &Cr\,II~(50)          &5696.60 &Al\,III~(2)    &6007.34 &[NiII]~(8F)    &6238.38 &Fe\,II~(74)        \\
5506.27 &Fe\,II               &5710.76 &N\,II~(3)      &6032.30 &Fe\,III~(117)  &6240.66 &Fe\,I~(64)         \\
5510.68 &Cr\,II~(23)          &5730.67 &N\,II~(3)      &6041.90 & S\,I~(10)     &6242.52 & N\,II~(57)        \\
5525.14 &Fe\,II~(56)          &5747.29 &N\,II~(9)      &6045.50 & Fe\,II~(200)  &6247.56 &Fe\,II~(74)        \\
5529.94 &Ti\,II~(68)          &5754.64 &N\,II]~(3)     &6061.50 & [Cr\,I]~(12F) &6248.92 &Fe\,II             \\
5534.86 &Fe\,II~(55)          &5780.48 &DIB            &6067.88 & [Cr\,I]~(12F) &6270.24 &Fe\,I~(342)        \\
5544.20 &Fe\,II               &5793.16 &Fe\,II~(47)    &6077.80 & [Ti II]~(26F) &6283.86 & DIB               \\
5548.21 &Fe\,II               &5795.16 &DIB            &6084.11 & Fe\,II~(46)   &6291.83 &Fe\,II             \\
5554.68 &[V\,II]~(20F)        &5797.06 &DIB            &6103.54 & Fe\,II~(200)  &6300.23 & [O\,I]~(1F) tellur\\
5564.94 &S\,II~(6)            &5813.67 &Fe\,II~(163)   &6113.38 & Fe\,II~(46)   &6305.51 & S\,II~(19)        \\
5567.82 &Fe\,II               &5823.17 &Fe\,II~(164)   &6116.05 & Fe\,II~(46)   &6312.68 & S\,II~(26)        \\
5577.34 & [O\,I]~(3F) tellur  &5835.43 &Fe\,II~(58)    &6124.57 & [Ti\,II]~(22F)&6317.99 & Fe\,II            \\
5588.15 & [Fe\,II]~(39F)      &5849.81 & DIB           &6129.71 & Fe\,II~(46)   &6318.23 & Mg\,I~(23)        \\
5606.11 &S\,II~(11)           &5875.66 & He\,I~(11)    &6147.74 & Fe\,II~(74)   &6318.75 & Mg\,I~(23)        \\
5609.73 &DIB                  &5889.95 &Na\,I~(1)      &6149.24 & Fe\,II~(74)   &6324.80  & DIB              \\
5627.49 &Fe\,II~(57)          &5895.92 &Na\,I~(1)      &6155.98 & O\,I~(10)     &6331.96 & Fe\,II~(199)      \\
5639.96 & S\,II~(14)          &5910.57 &  DIB          &6158.19 & O\,I~(10)     &6347.09 &Si\,II~(2)         \\
\hline
\hline
6369.46 &Fe\,II~(40)     &6583.6  & [N\,II]~(1F)  &6860.29 &Fe\,I~(205)     & 7357.20 &  DIB           \\
6371.36 &Si\,II~(2)      &6586.69 &Fe\,II         &6861.93 & Fe\,I~(109)    & 7363.96 & Fe\,I~(1274)   \\
6379.32 & DIB            &6613.62 & DIB           &7042.06 &Al\,II~(3)      & 7377.83 & [NiII]~(2F)    \\
6383.75 &Fe\,II          &6626.75 & [Ti\,II]~(37F)&7056.60 &Al\,II~(3)      & 7379.57 & [NiII]~(2F)    \\
6385.47 & Fe\,II         &6630.5  & N\,II~(41)    &7065.72 &He\,I~(10)      & 7388.16 & [Fe\,II]~(14F) \\
6402.42 &Fe\,I~(1344)    &6638.36 &Fe\,I~(1279)   &7115.13 &C\,II~(20)      & 7411.90 & [V\,II]~(4F)   \\
6407.30 &Fe\,II~(74)     &6660.71 &DIB            &7119.45 &C\,II~(20)      & 7413.33 & [NiII]~(2F)    \\
6416.90 &Fe\,II~(74)     &6665.15 & DIB           &7125.49 &C\,II~(20)      & 7452.50 & [Fe\,II]~(14F) \\
6432.65 & Fe\,II~(40)    &6678.15 &He\,I~(46)     &7134.99 & Fe\,II~(197)   & 7462.38 & Fe\,II~(73)    \\
6442.97 &Fe\,II          &6721.36 &O\,II~(4)      &7155.14 & [Fe\,II]~(14F) & 7468.29 & N\,I~(3)       \\
6446.43 &Fe\,II~(199)    &6726.84 &C\,II~(21)     &7181.02 & Fe\,I~(33)     & 7479.69 & Fe\,II~(72)    \\
6456.38 &Fe\,II~(74)     &6770.05 & DIB           &7191.66 & Fe\,I~(1274)   & 7495.67 & Fe\,I~(1275)   \\
6482.07 &N\,II~(8)       &6779.74 &C\,II~(14)     &7222.39 &Fe\,II~(73)     & 7497.68 & [V\,II]~(3F)   \\
6482.20 &Fe\,II~(199)    &6780.27 &C\,II~(14)     &7231.12 &C\,II~(3)       & 7506.54 & Fe\,II         \\
6491.61 &Ti\,II~(91)     &6783.75 &C\,II~(14)     &7236.19 &C\,II~(3)       & 7513.17 & Fe\,II         \\
6493.06 &Fe\,II          &6811.44 & DIB           &7237.17 &C\,II~(3)       & 7515.13 & [V\,II]~(4F)   \\
6504.9  & N\,II~(45)     &6823.48 &Al\,II~(9)     &7255.28 &Si\,I~(59)      & 7515.88 & Fe\,II~(73)    \\
6506.33 &Fe\,II          &6829.01 & [Fe\,II]~(31F)&7278.48 &Fe\,I~(1274)    & 7533.84 & [V\,II]~(3F)   \\
6533.0  & N\,II~(45)     &6829.61 & FeI~(1195)    &7281.35 &He\,I~(45)      & 7571.69 & [V\,II]~(3F)   \\
6548.1  &[N\,II] (1F)    &6837.00 & Fe\,I~(1225)  &7289.05 &Fe\,II~(72)     & 7578.96 & S\,I           \\
6562.82 &H$\alpha$       &6837.14 & Al\,II~(9)    &7301.57 & Fe\,II~(72)    & 7698.98 & K\,I~(1)       \\
6578.03 &C\,II~(2)       &6852.67 &  DIB          &7341.78 & Fe\,I~(1307)   & 7711.71 & Fe\,II~(73)    \\
6582.85 &C\,II~(2)       &6857.25 & Fe\,I~(1006)  &7353.53 & Fe\,I~(1251)   & 7724.7  & [S\,I]~(3F)    \\
\hline
\hline
7731.7  & Fe\,II          & 7771.96 & O\,I~(1)     & 7774.18 & O\,I~(1)      & 7775.40 & O\,I~(1)       \\
7752.86 & [Cr\,II]~(11F)  &         &              &         &               &                          \\
\end{longtable*}
\renewcommand{\baselinestretch}{1.0}
\onecolumngrid

The currently available high-resolution spectra do not allow us to resolve the question of
such a feature  $\lambda$\,4686\,\AA{}  for estimating the orbital period or to localize the
formation of this  emission in the CI\,Cam system.
Note that in our spectrum for September~3, 2015, in which the emission near $\lambda\approx4686$\,\AA{}
was confidently detected, another important feature is observed: reduced intensity of the  emissions
with ``rectangular'' profiles~\cite{KMP}. This coincidence suggests the appearance
of the $\lambda$\,4686\,\AA{} emission during some phases of the as-yet-undetermined orbital period.
The possible binarity of CI\,Cam has already been  suggested   before the outburst, in an early paper
by~\citet{Mirosh1995}. This was later confirmed in the paper~\cite{KMP} by the detection of a
small sample of photospheric type absorptions of Fe\,III, N\,II, Ti\,II, and S\,II with variable positions
in the 2019--2023 spectra. However, to accept this result, a significant increase in the number of high-quality  spectra is clearly required. By the way, no photospheric absorptions  were found in the 2015 spectrum that we have used to create the atlas.

Let us briefly examine the complex profile of the Na\,I D-lines (right panel of Fig.\,\ref{frag19}).
The complex profile of each line in the doublet includes two interstellar absorptions: with the velocities of $-7.1\pm 0.5$ and $-5.8\pm 0.3$\,km\,s$^{-1}$, as well as the circumstellar emission.
The exact position of the emission is impossible to determine because its profile is distorted by the absorption components. According to the data in the article~\cite{KMP}, the positions of all
components of the doublet are stable, and the positions of the two interstellar absorptions are consistent with the positions of components of  interstellar potassium K\,I\,7696\,\AA{}.

Along with the above mentioned star 3\,Pup, the distant hot star MWC\,17, exhibiting the B[e]
phenomenon (\citet{Chen2016}), can be considered a close analogue of CI\,Cam.  Its stable spectrum contains approximately the same set of emissions: H\,I, He\,I, as well as permitted and forbidden ion emissions.   No photospheric  absorptions were detected by~\cite{Chen2016}
in the spectrum of MWC\,17. The absence of photospheric type lines may be a consequence of the star's rapid rotation coupled with a strong circumstellar continuum, veiling the absorptions.
The extreme H$\alpha$ intensity in its spectrum has been known since the first spectroscopic data were registered by~\citet{Zickgraf}. Moreover, according to the results by~\citet{Chen2016}, there are no emissions with vertical slopes in the spectrum of  MWC\,17.
These authors noted the separation  of the H$\alpha$ profile by an absorption,
but the variability  of its position on the profile was not detected. The intensity of the profiles (Fe\,II, [Fe\,II], [O\,III], etc.) is many times greater than their intensity in the spectrum of CI\,Cam.
The spectrum of MWC\,17 does not have the same richness of nitrogen emissions found in the spectrum of CI\,Cam; only three well known forbidden [N\,II] features are present, while the intensity of the
[N\,II]\,5755\,\AA{} emission is 5.5 times greater than the intensity of this line in the spectrum of CI\,Cam.  The main problem that needs to be solved for this pair and other related stars with complex envelopes is determining their distance and luminosity.

\section{MAIN RESULTS}\label{conclus}

We present here a prepared atlas of the optical spectrum of the B[e] star CI\,Cam
and the results of an identification of its details in the wavelength range of 395$\div$780\,nm.
This atlas is an essential supplement to the study of the spectrum of CI\,Cam performed
by~\citet{KMP}.
All 59 fragmrents of the complete atlas are provided in the Appendix.
The set of features observed in the optical spectrum of CI\,Cam
allows us to consider it as a stripped core of a star that has lost its atmosphere during the
evolution and is now surrounded by a complex gas and dust envelope.
The radiation from this hot core provides the continuous spectrum of CI\,Cam. All the emissions
in the observed optical spectrum are formed within a complex optically thin envelope.
The atlas identifies the vast majority (about 400) of features with peculiar
profiles of various types. In particular, we signalize the discovery of two-peak forbidden
emissions of the [V\,II] and [Cr\,II] ions from several multiplets.

The high quality of the observational data gives us reason to hope for the future use of the results
of this work in the spectroscopy of CI\,Cam and other hot stars with complex gaseous-dusty
envelopes and the B[e] phenomenon in their spectra (see, for example, stars studied
in~\cite{Maravelias,Mirosh2020}), as well as a number of hot stars with unclear evolutionary
status (e.g. articles~\cite{CygOB2,VES723,Schulte12,AS314}).

The results of detailed spectroscopy shall be also clearly useful in the studies of the
circumstellar spatial  structure imaging (see, e.g.,~\cite{Liimets}). The atlas may also be
inightful for studying the
spectra of a rare  type of stars---extremely hydrogen-deficient supergiants in common-envelope binaries. Examples include the hot  stars $\upsilon$\,Sgr (\citet{Kipper2008})  and KS\,Per~(\citet{Kipper2012}) with accretion disks  in close  hydrogen-deficient binaries (HdBs), which are thought to be precursors to type SN\,Ia supernovae.
The spectra of low-mass post-AGB supergiants at the end of their transition to the planetary nebula stage  also contain many similar features.
An excellent example is the hot star in the  IR-source system  IRAS\,01005+7910,
in whose optical spectrum the forbidden  [N\,I], [N\,II], [O\,I], [S\,II] and [Fe\,II] emissions
have been  identified~\cite{IRAS01005}. The presence of forbidden  [N\,II] and [S\,II] emissions indicates the onset  of ionization of the circumstellar envelope and the proximity of the planetary nebula phase.

A similar combination of features was also detected in the spectrum of a distant B-supergiant
LS\,III\,+52$\degr$24 (\citet{IRAS22023}), whose status is not entirely clear.
This star holds the record for the intensity  of H$\alpha$ emission, reaching the values
of $I/I_{\rm cont}\ge 70$.
The H$\alpha$ and H$\beta$ lines have a P\,Cyg-type profile, their wind absorption component
varies in position in the range of V$_{\odot}=-(270\div290)$\,km\,s$^{-1}$.
The spectrum contains numerous permitted (O\,I, Si\,III, Al\,III, C\,II, Fe\,I, Fe\,II,
Fe\,III) and forbidden low-excitation emissions ([N\,II], [O\,I], [S\,II]).
Observations with the BTA+NES on arbitrary  dates from 2010 to 2021 revealed signs
of wind variability and velocity stratification in the extended atmosphere of this supergiant.
Based on the positions of the N\,II and O\,II absorptions, a temporal radial velocity variability
was found in the range of V$_{\odot}=-(127.2\div178.3)$\,km\,s$^{-1}$, indicating the presence
of a companion and/or pulsations in the atmosphere.

Thanks to the high quality of the spectroscopic data in the atlas and in the work
by~\citet{Robinson},  we were able to make a correct comparison of the profiles of
narrow disk emissions for two separate  stages of  the  CI\,Cam  spectroscopy.
A decrease in the intensity of the [O\,I]\;5577\;\AA{} emission and a change in the shape
of the Ti\,II, Cr\,II, and Fe\,II ion emission peaks were detected; their profiles, with
tops rounded  immediately  after the outburst, acquired a two-peaked shape, indicating the
reality of differences in the physical conditions in the region of disk emission formation
immediately after the outburst and 17 years onward.

The behavior of the high-excitation emission near $\lambda$\,4686\,\AA{}  remains unclear.
In this regard, the [O\,III] and H$\alpha$   images of the emission envelope of the supernova
remnant G150.3+4.5, centered on CI\,Cam, recently obtained by~\citet{Fesen}, are
of keen interest. Unfortunately, the lack of information on the distance of RSN~G150.3+4.5 does not
yet allow us to discuss a physical relationship between the supernova envelope and CI\,Cam.

\section*{FUNDING}

This work was carried out within the framework of a state assignment to SAO RAS,
approved by the Ministry of Science and Higher  Education of the Russian Federation.

\newpage

\section*{\large APPENDIX}

Below are presented  all 59 fragments of the CI Cam spectrum obtained on the BTA telescope using
the NES spectrograph in the wavelength range of 395$\div$780\,nm with a resolution of
$\lambda/\Delta\lambda\ge$ 60\,000. The relative intensity $I/I_{\rm cont}$ is plotted on the
ordinate axis, and the observed wavelength --- on the abscissa axis.

\setcounter{figure}{0}
\begin{figure*}[h]
\includegraphics[angle=0,width=0.8\textwidth,bb=40 40 1450 950,clip]{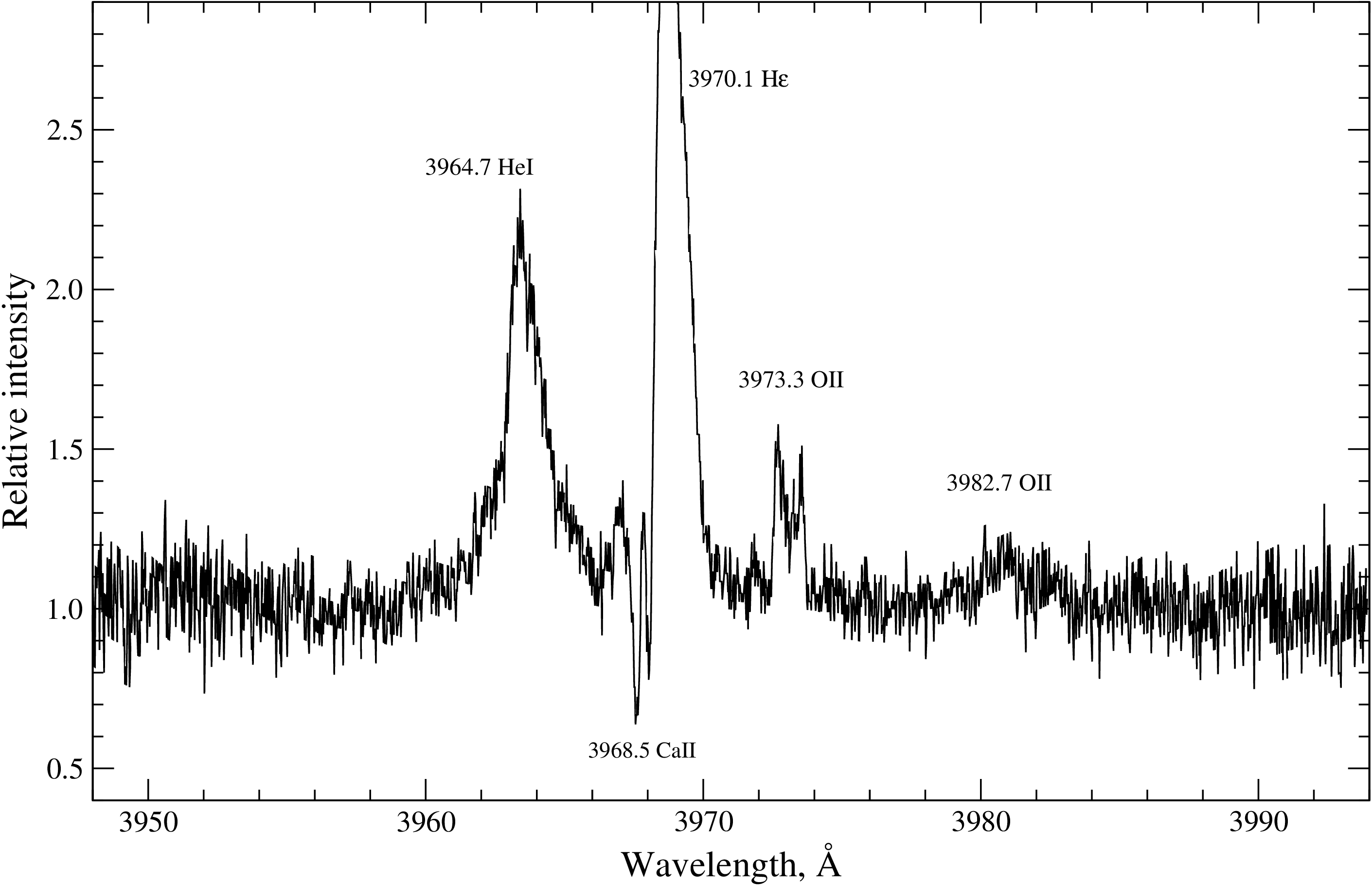}
\includegraphics[angle=0,width=0.8\textwidth,bb=40 40 1450 950,clip]{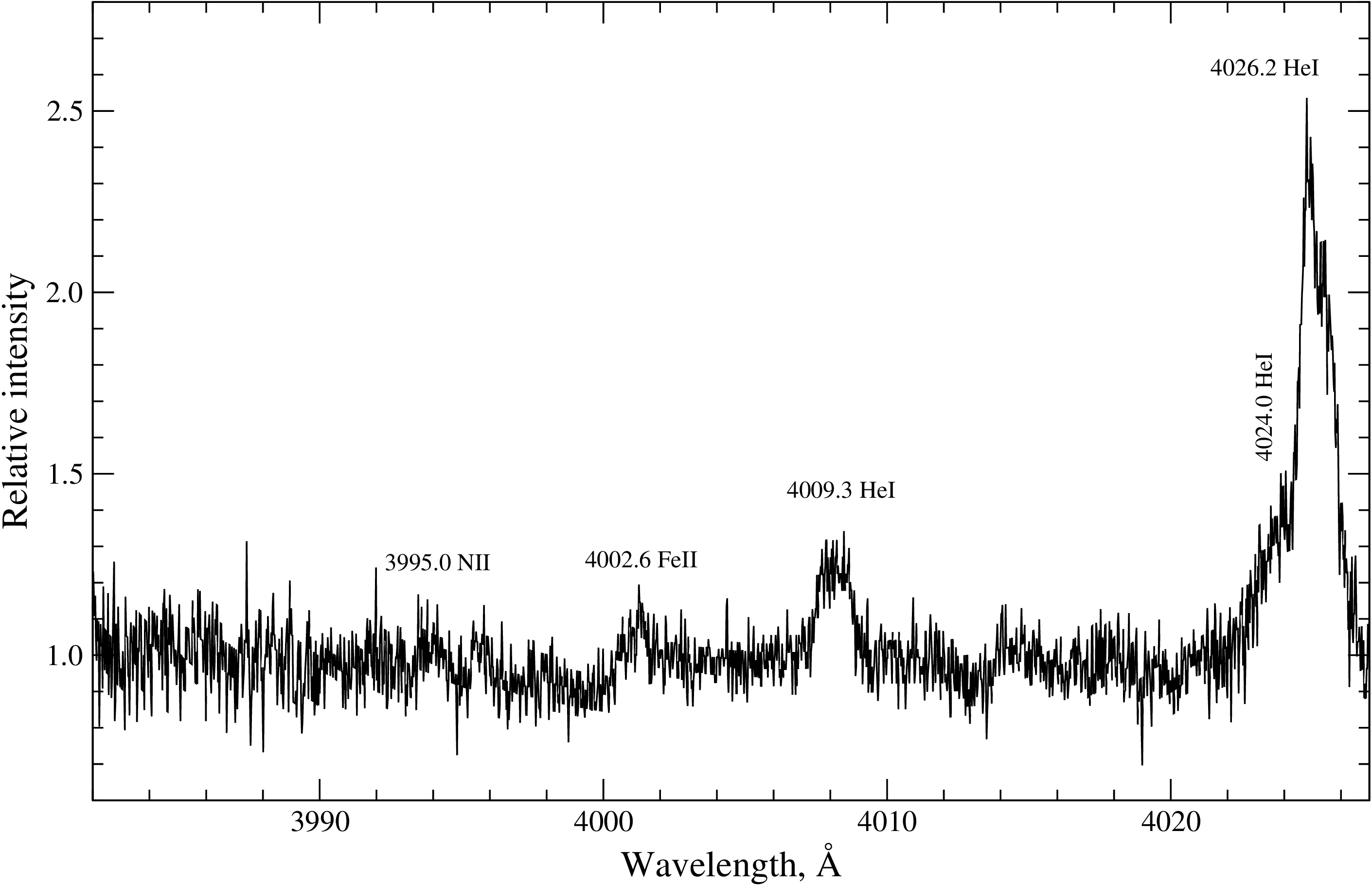}
\caption{CI\,Cam spectrum: fragments 1--2.}
\end{figure*}

\setcounter{figure}{0}
\begin{figure*}
\includegraphics[angle=0,width=0.8\textwidth,bb=0  40 1450 950,clip]{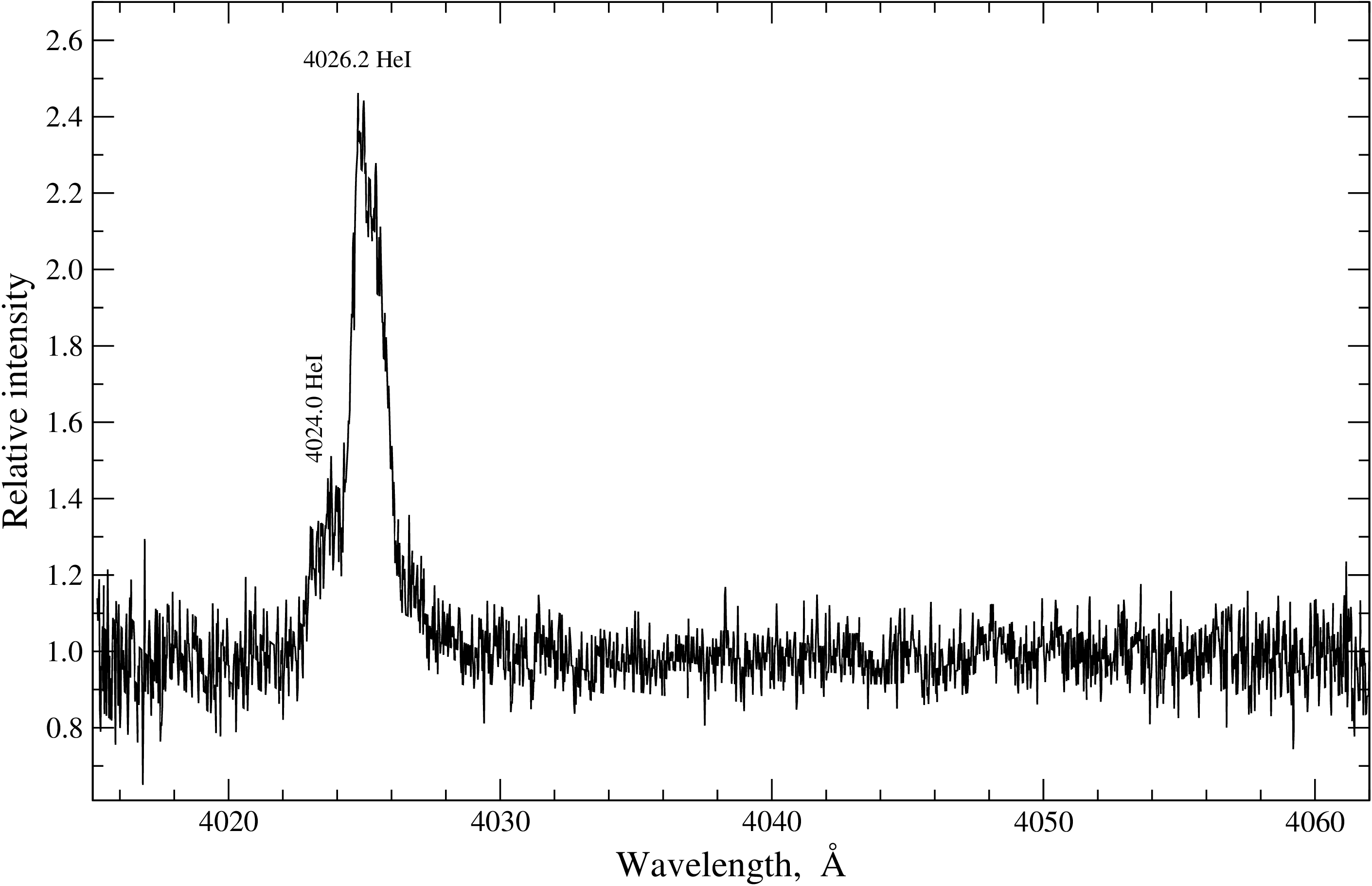}
\includegraphics[angle=0,width=0.8\textwidth,bb=0 -10 1450 950,clip]{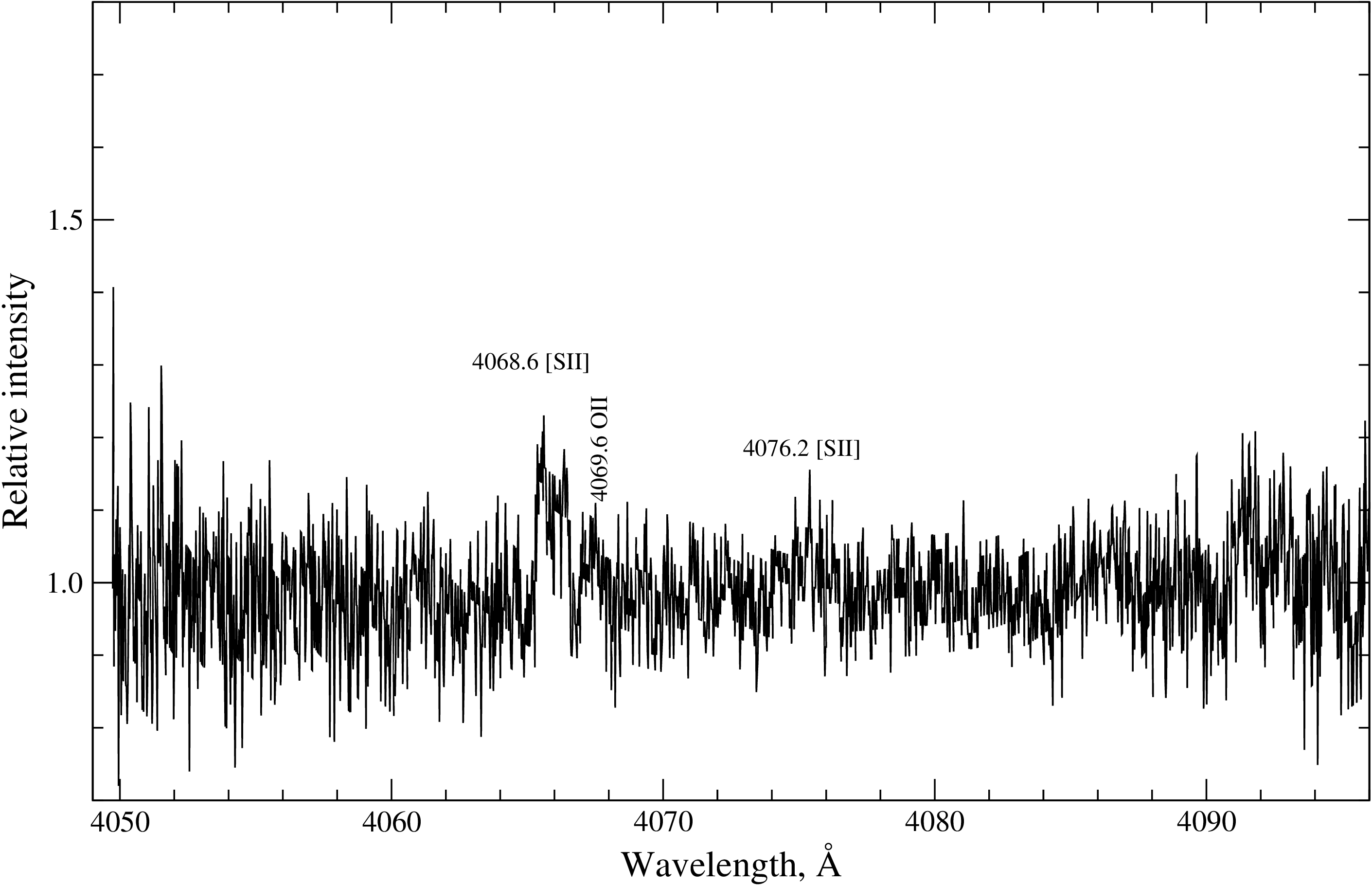}
\includegraphics[angle=0,width=0.8\textwidth,bb=0  40 1450 950 clip]{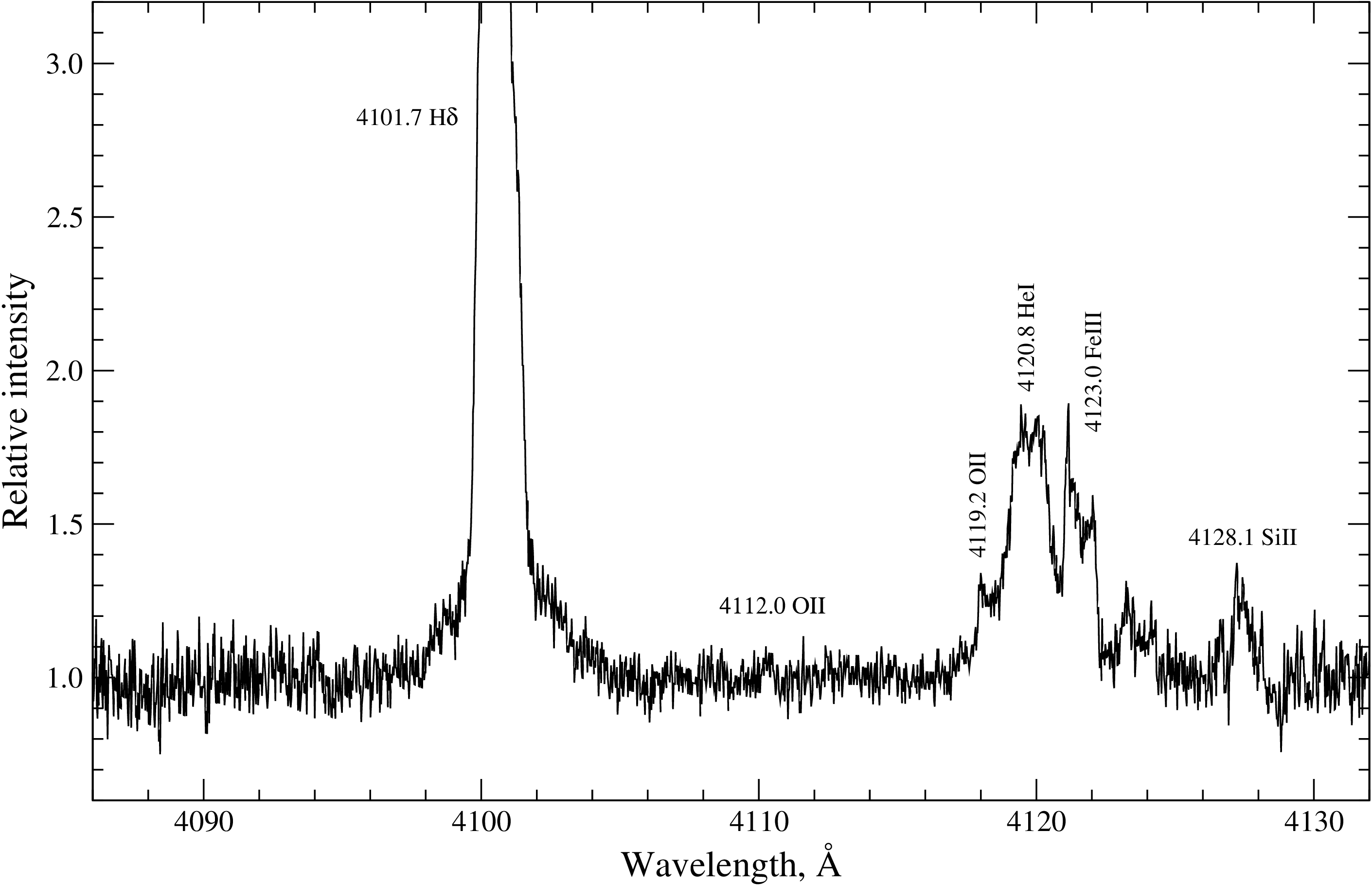}
\caption{CI\,Cam spectrum: fragments 3--5.}
\end{figure*}

\setcounter{figure}{0}
\begin{figure*}
\includegraphics[angle=0,width=0.8\textwidth,bb=0 40 1450 950,clip]{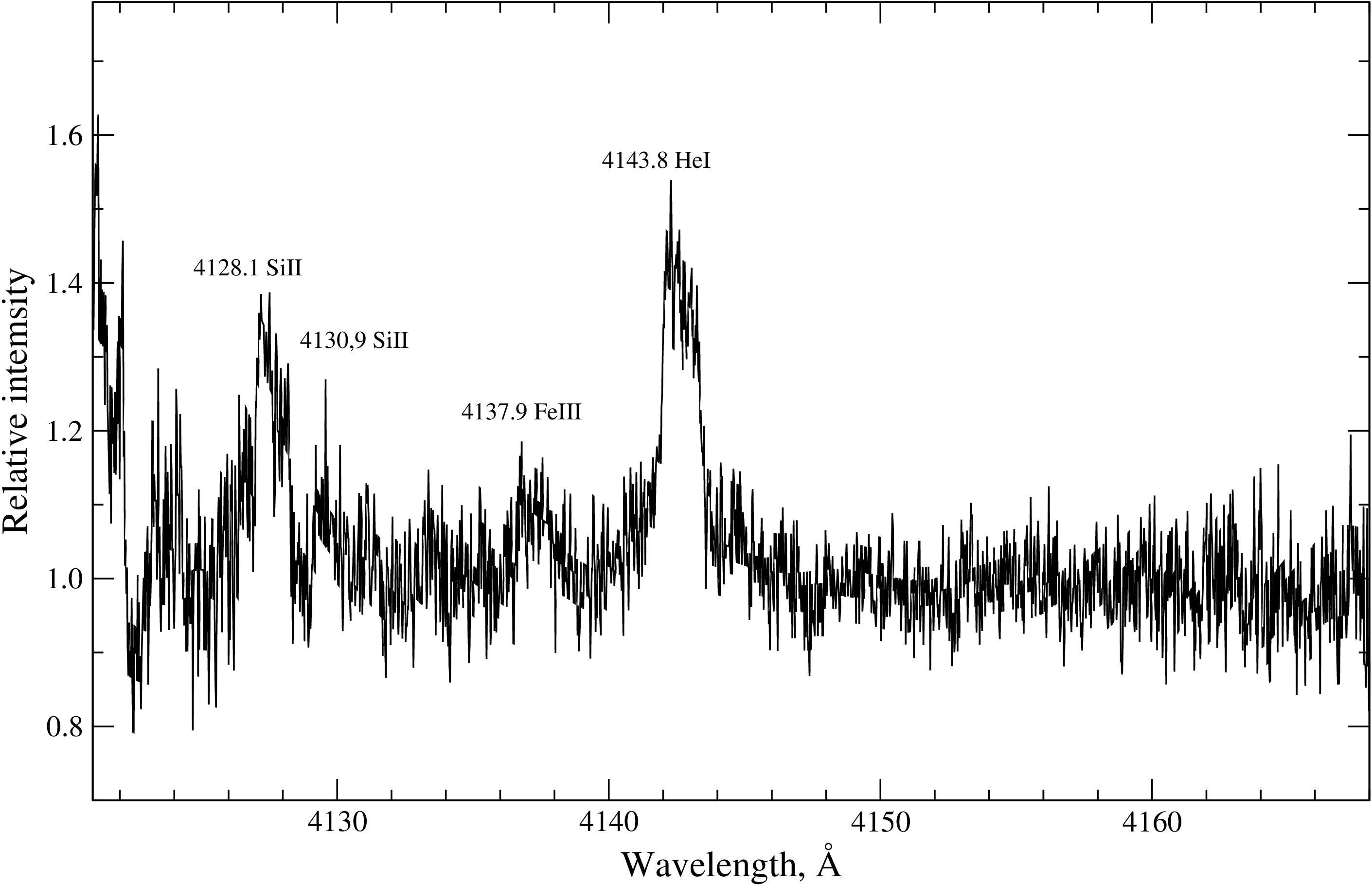}
\includegraphics[angle=0,width=0.8\textwidth,bb=0 40 1450 950,clip]{631_07k.eps}
\includegraphics[angle=0,width=0.8\textwidth,bb=0 40 1450 950,clip]{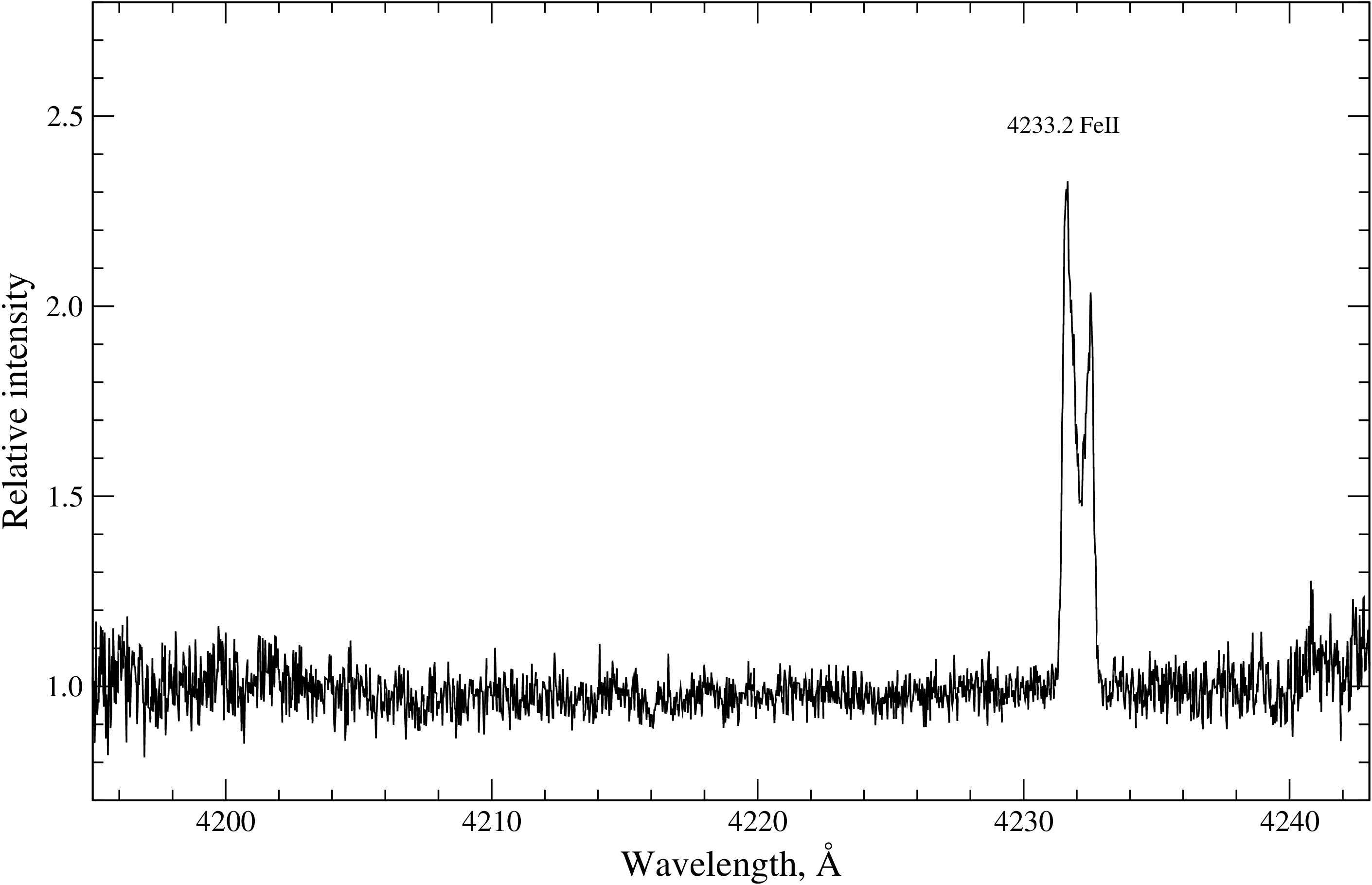}
\caption{CI\,Cam spectrum: fragments 6--8.}
\end{figure*}

\setcounter{figure}{0}
\begin{figure*}
\includegraphics[angle=0,width=0.8\textwidth,bb=0  40 1450 950,clip]{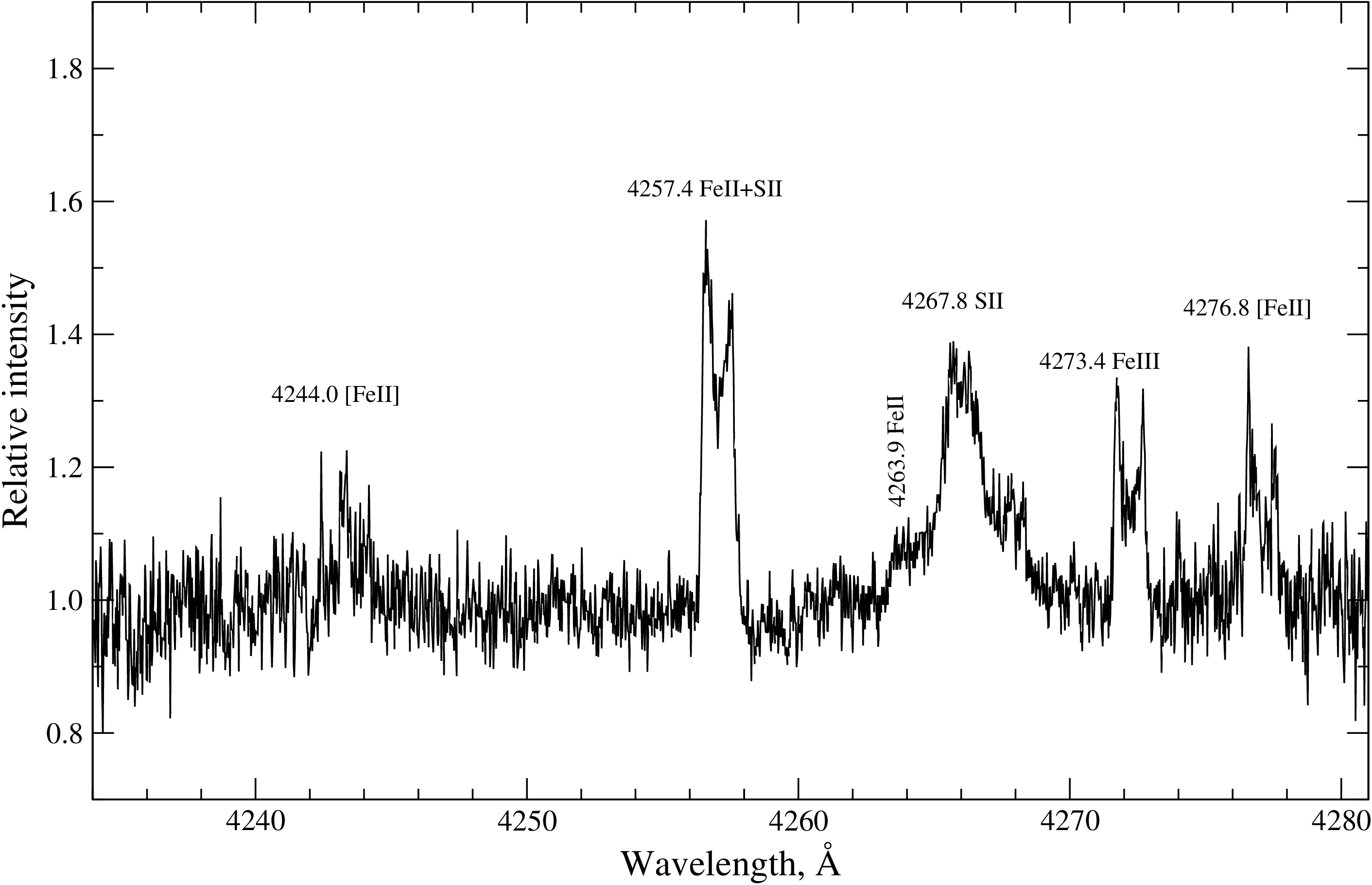}
\includegraphics[angle=0,width=0.8\textwidth,bb=0  -10 1450 950,clip]{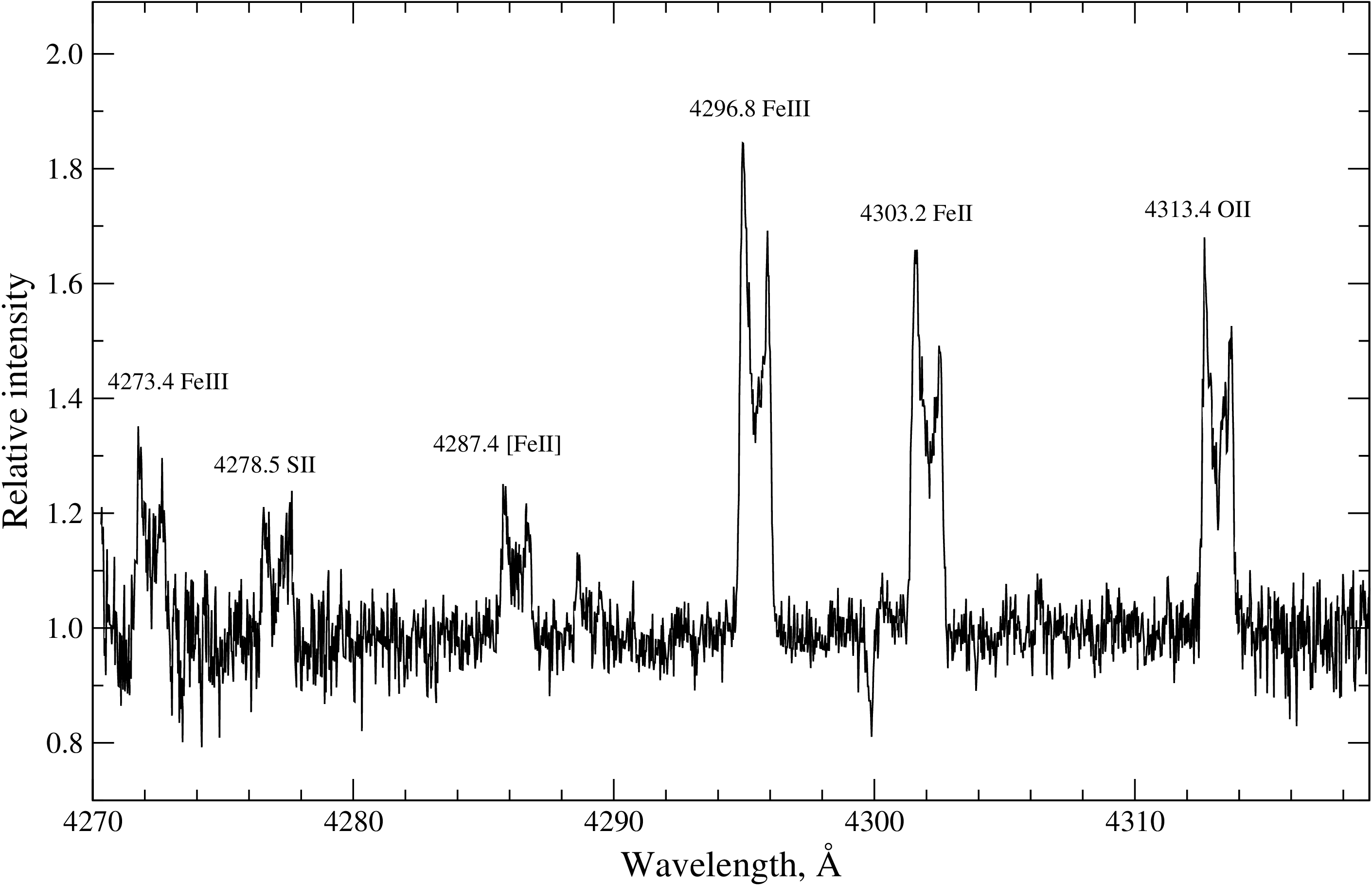}
\includegraphics[angle=0,width=0.8\textwidth,bb=0  40 1450 950 clip]{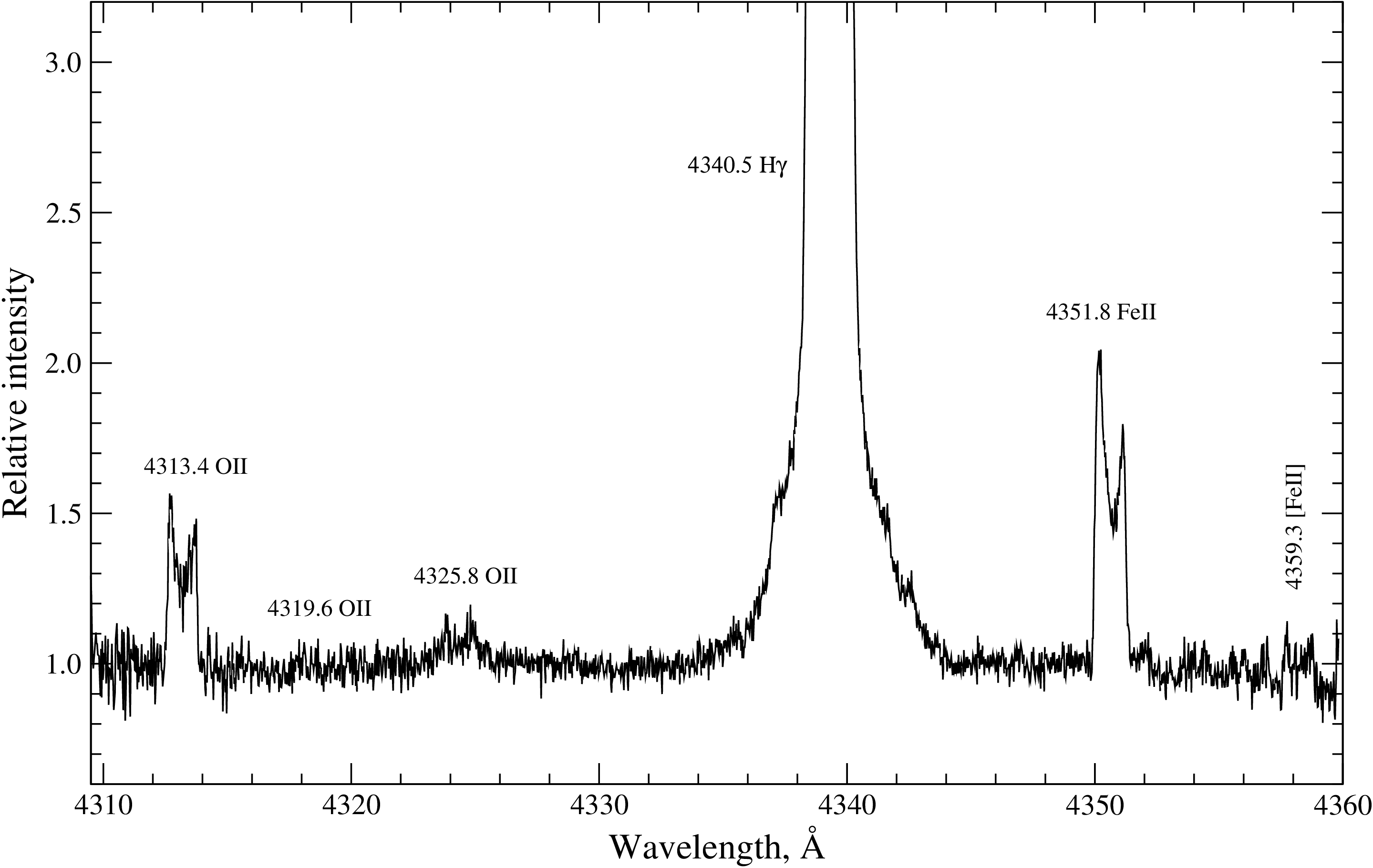}
\caption{CI\,Cam spectrum: fragments 9--11.}
\end{figure*}

\setcounter{figure}{0}
\begin{figure*}
\includegraphics[angle=0,width=0.8\textwidth,bb=0 40 1450 950,clip]{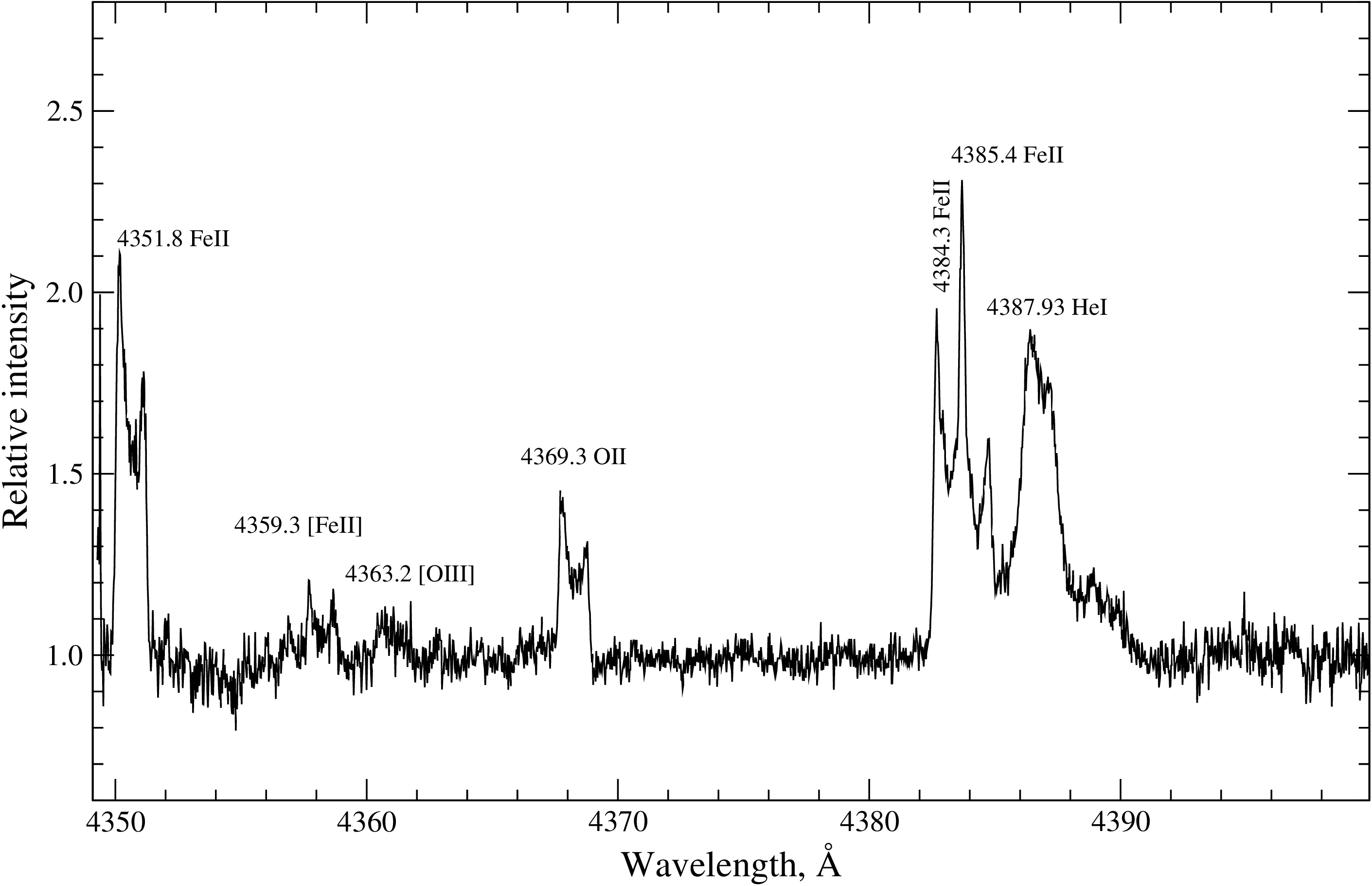}
\includegraphics[angle=0,width=0.8\textwidth,bb=0 40 1450 950,clip]{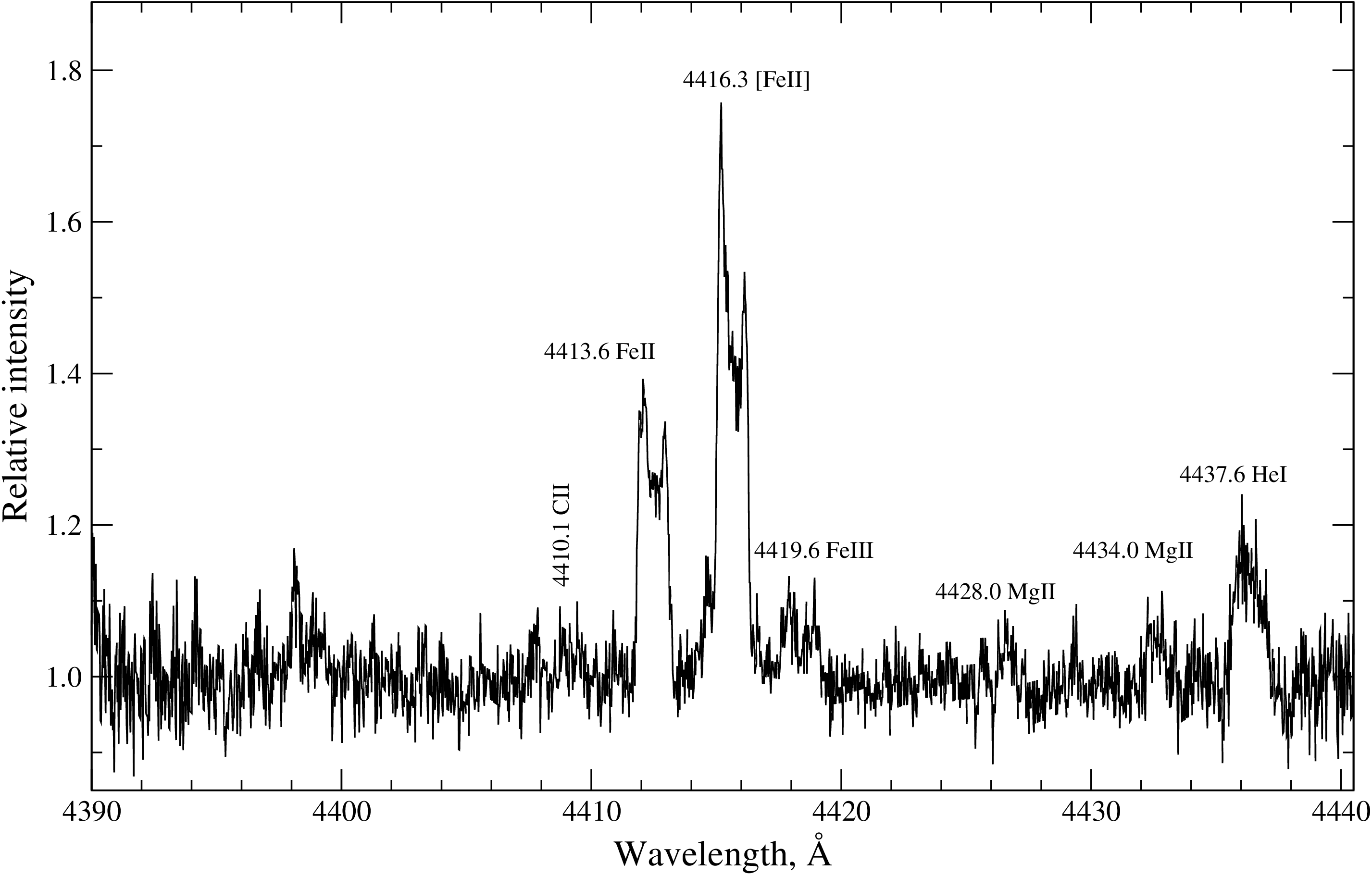}
\includegraphics[angle=0,width=0.8\textwidth,bb=0 40 1450 950,clip]{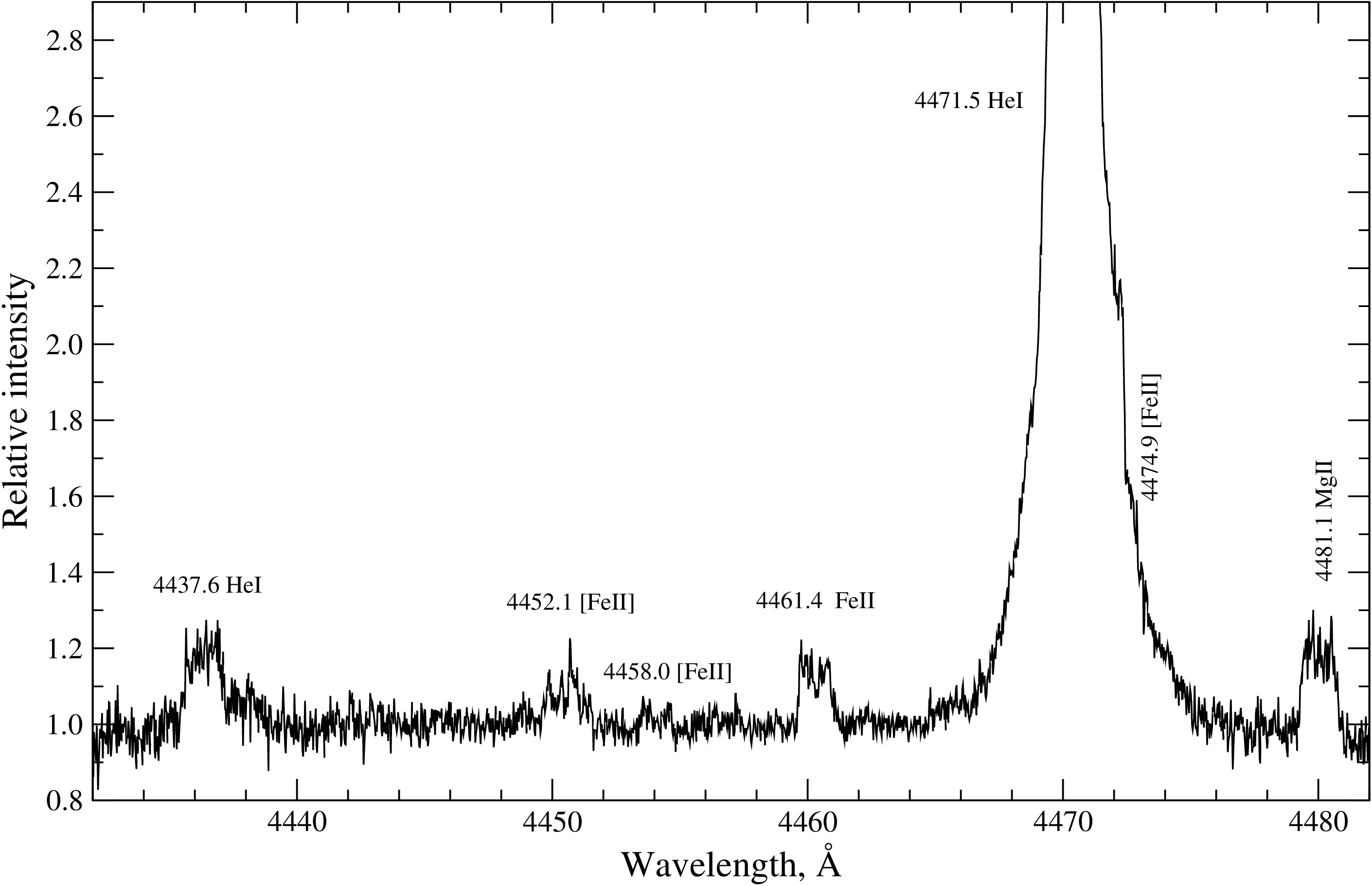}
\caption{CI\,Cam spectrum: fragments 12--14.}
\end{figure*}

\setcounter{figure}{0}
\begin{figure*}
\includegraphics[angle=0,width=0.8\textwidth,bb=0 40 1450 950,clip]{631_15k.eps}
\includegraphics[angle=0,width=0.8\textwidth,bb=0 40 1450 950,clip]{631_16k.eps}
\includegraphics[angle=0,width=0.8\textwidth,bb=0 40 1450 950,clip]{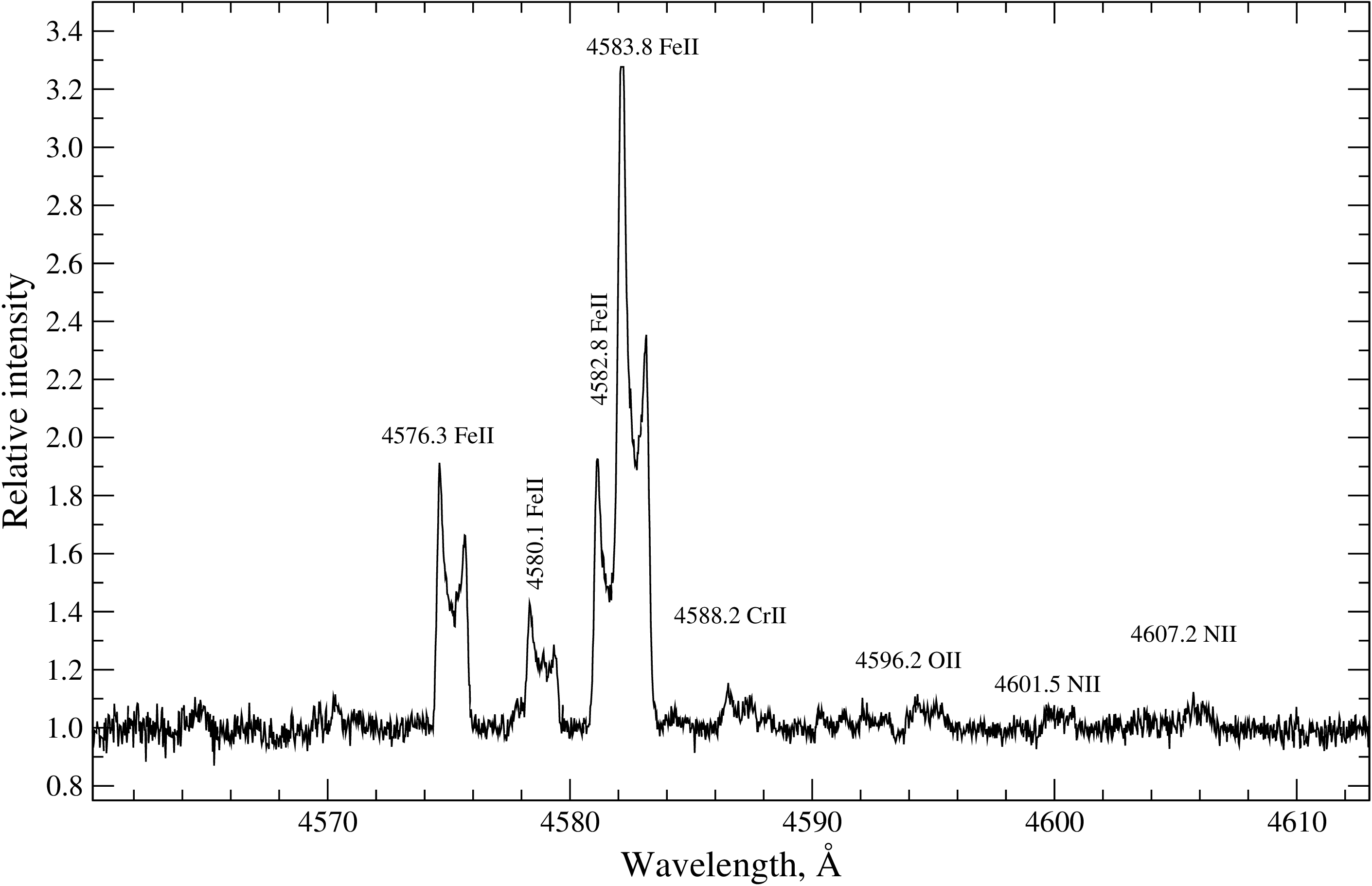}
\caption{CI\,Cam spectrum: fragments 15--17.}
\end{figure*}

\begin{figure*}
\includegraphics[angle=0,width=0.8\textwidth,bb=0 40 1450 950,clip]{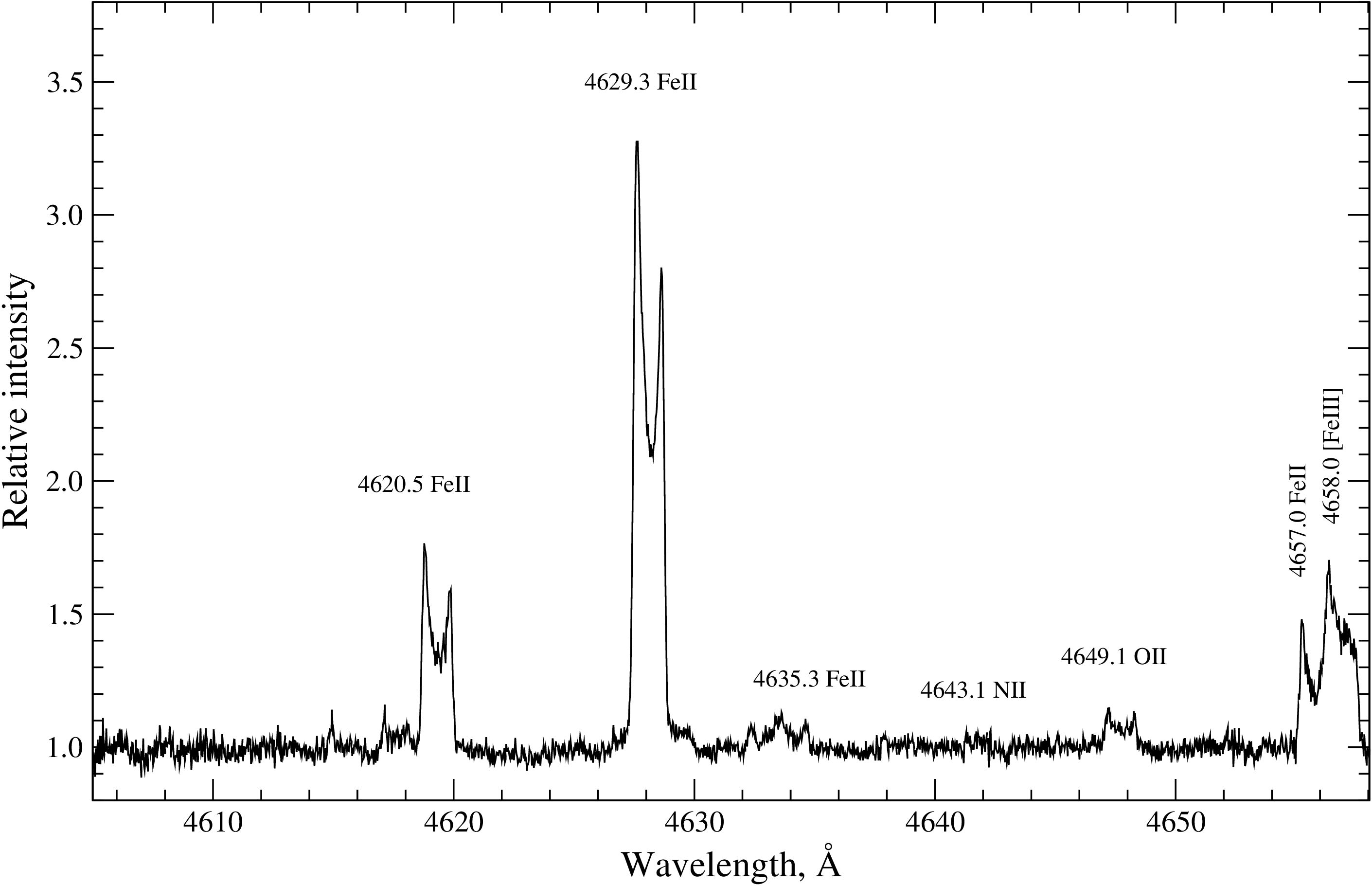}
\includegraphics[angle=0,width=0.8\textwidth,bb=0 40 1450 950,clip]{631_19k.eps}
\includegraphics[angle=0,width=0.8\textwidth,bb=0 40 1450 950,clip]{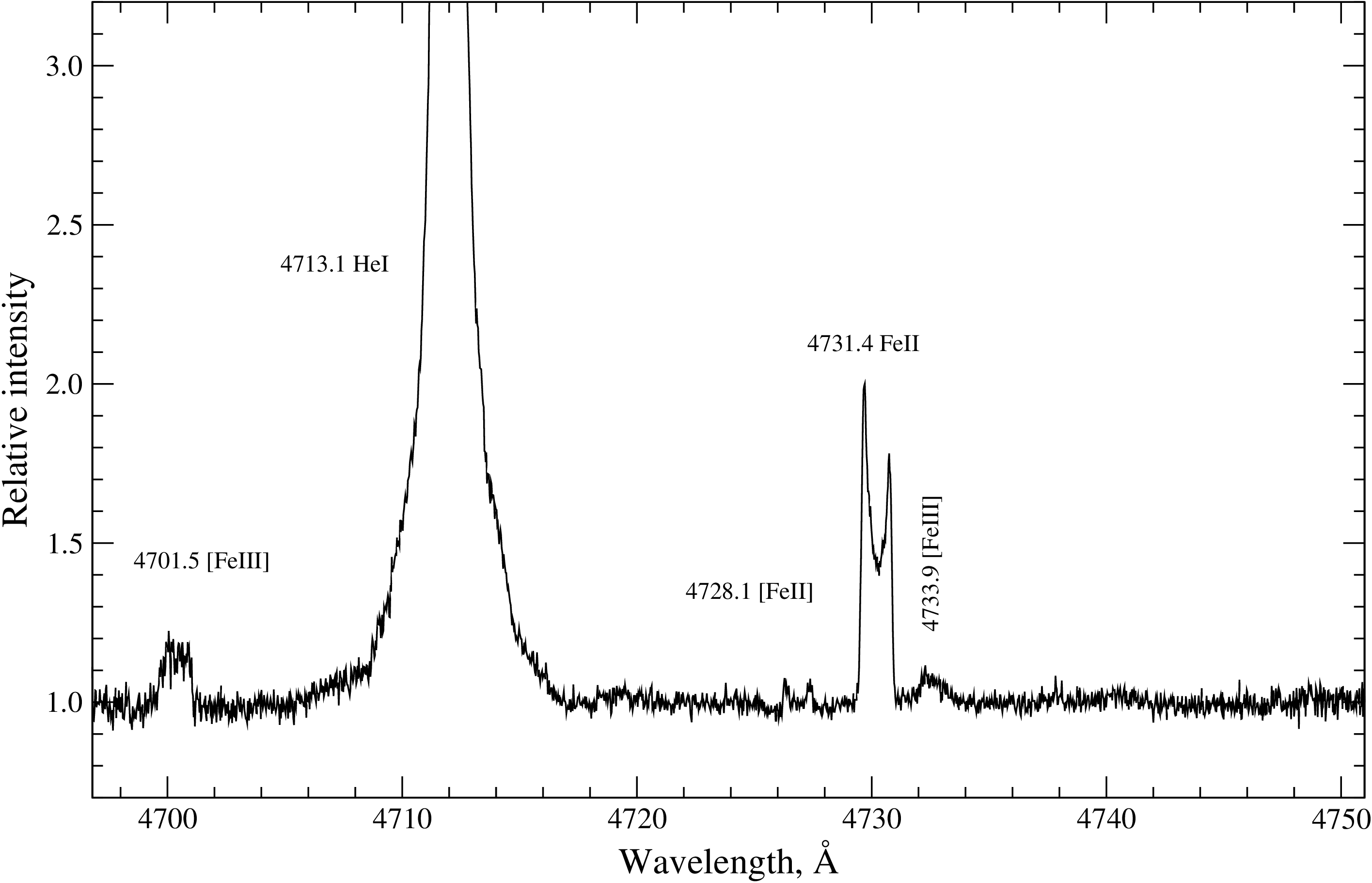}
\caption{CI\,Cam spectrum: fragments 18--20.}
\end{figure*}

\setcounter{figure}{0}
\begin{figure*}
\includegraphics[angle=0,width=0.8\textwidth,bb=0 40 1450 950,clip]{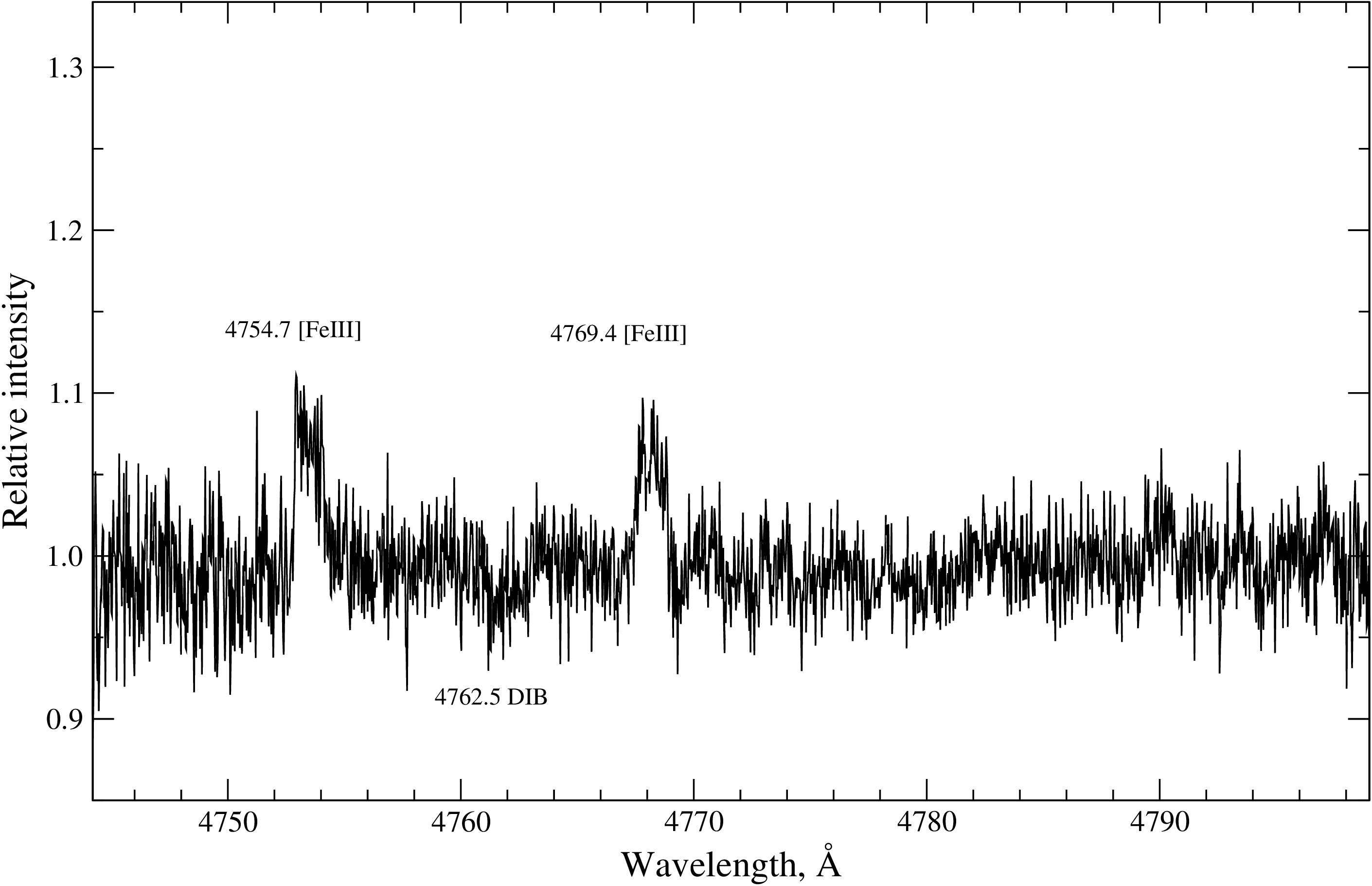}
\includegraphics[angle=0,width=0.8\textwidth,bb=0 40 1450 950,clip]{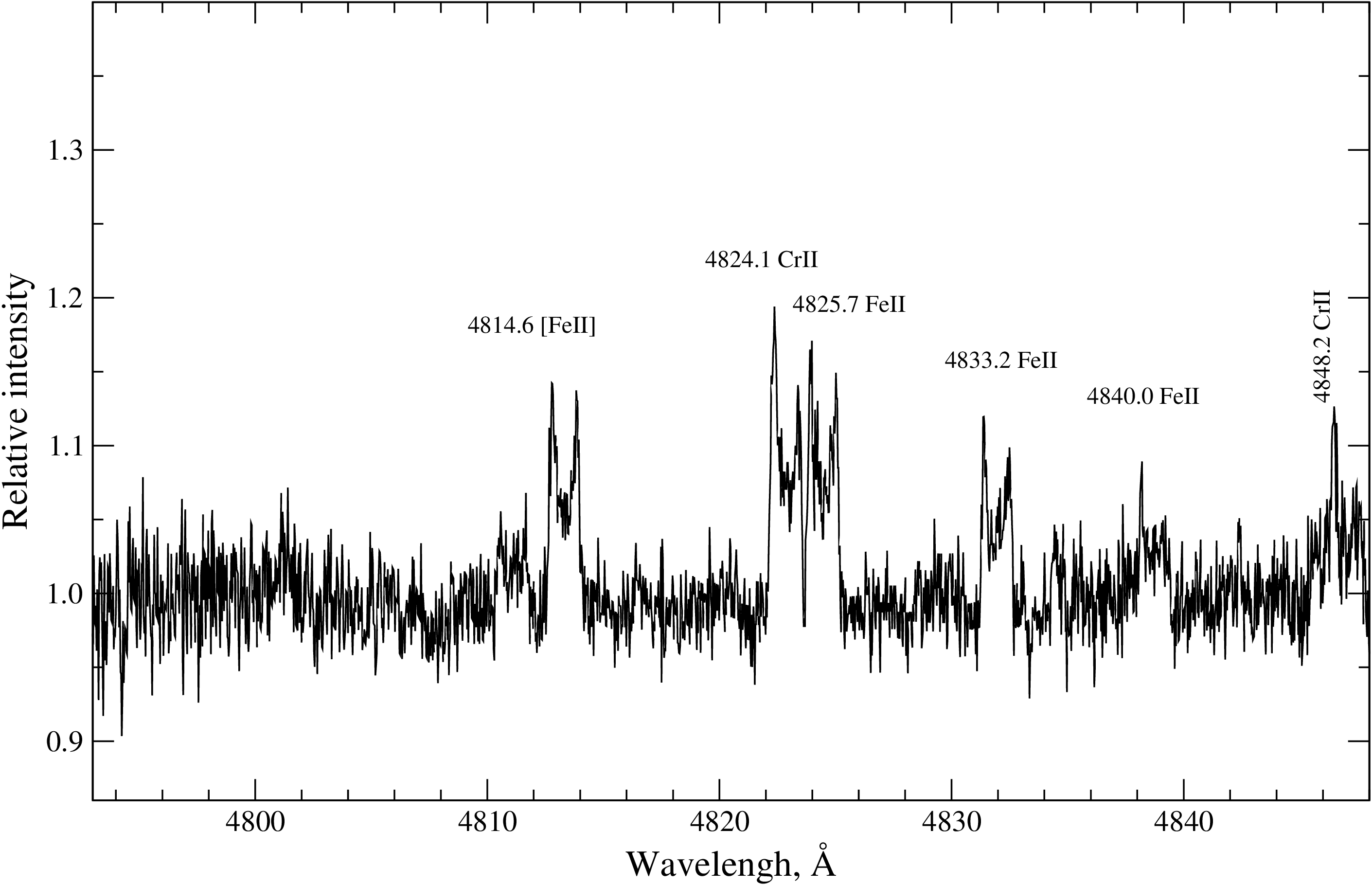}
\includegraphics[angle=0,width=0.8\textwidth,bb=0 40 1450 950,clip]{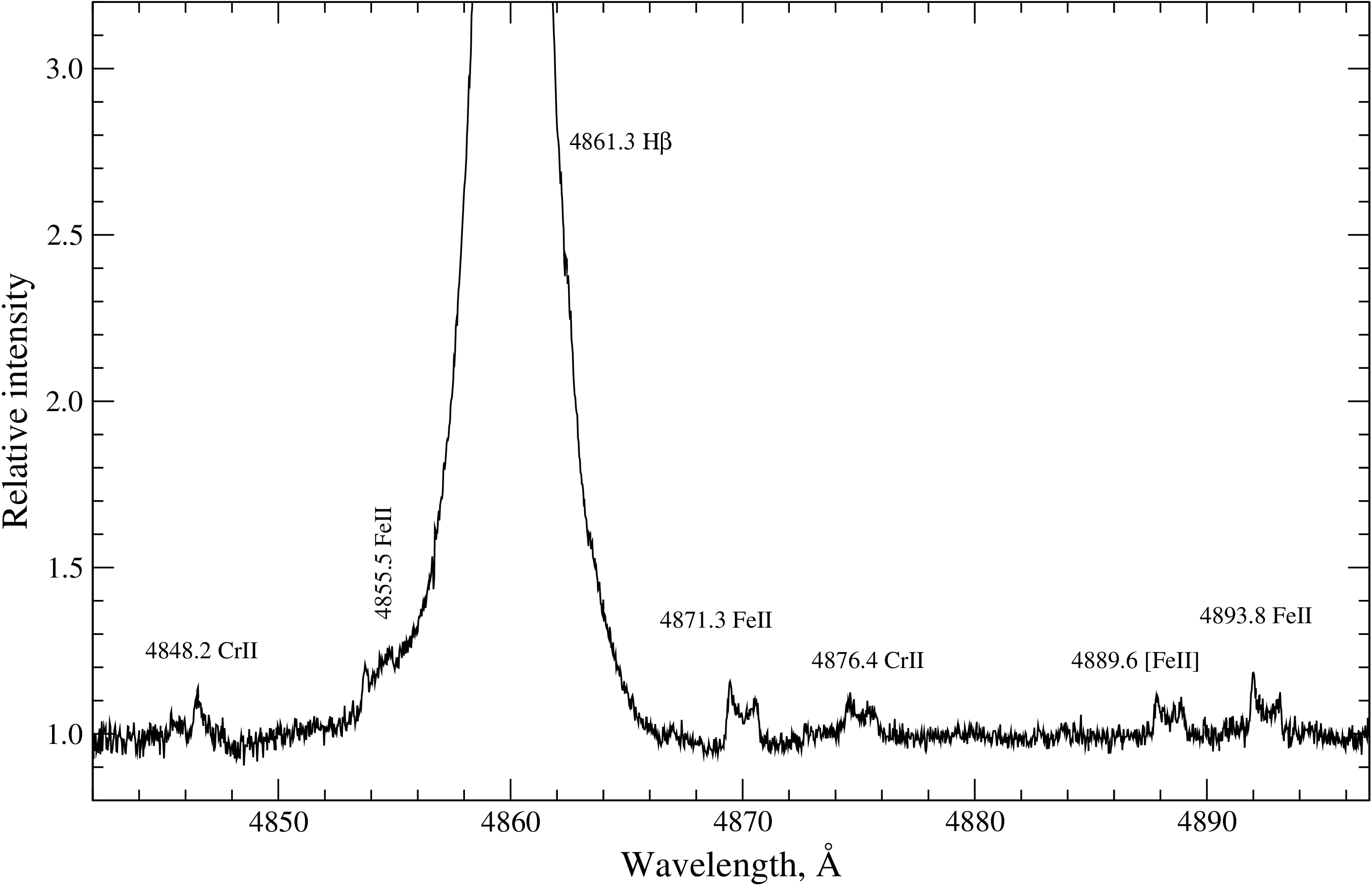}
\caption{CI\,Cam spectrum: fragments 21--23.}
\end{figure*}

\setcounter{figure}{0}
\begin{figure*}
\includegraphics[angle=0,width=0.8\textwidth,bb=0 40 1450 950,clip]{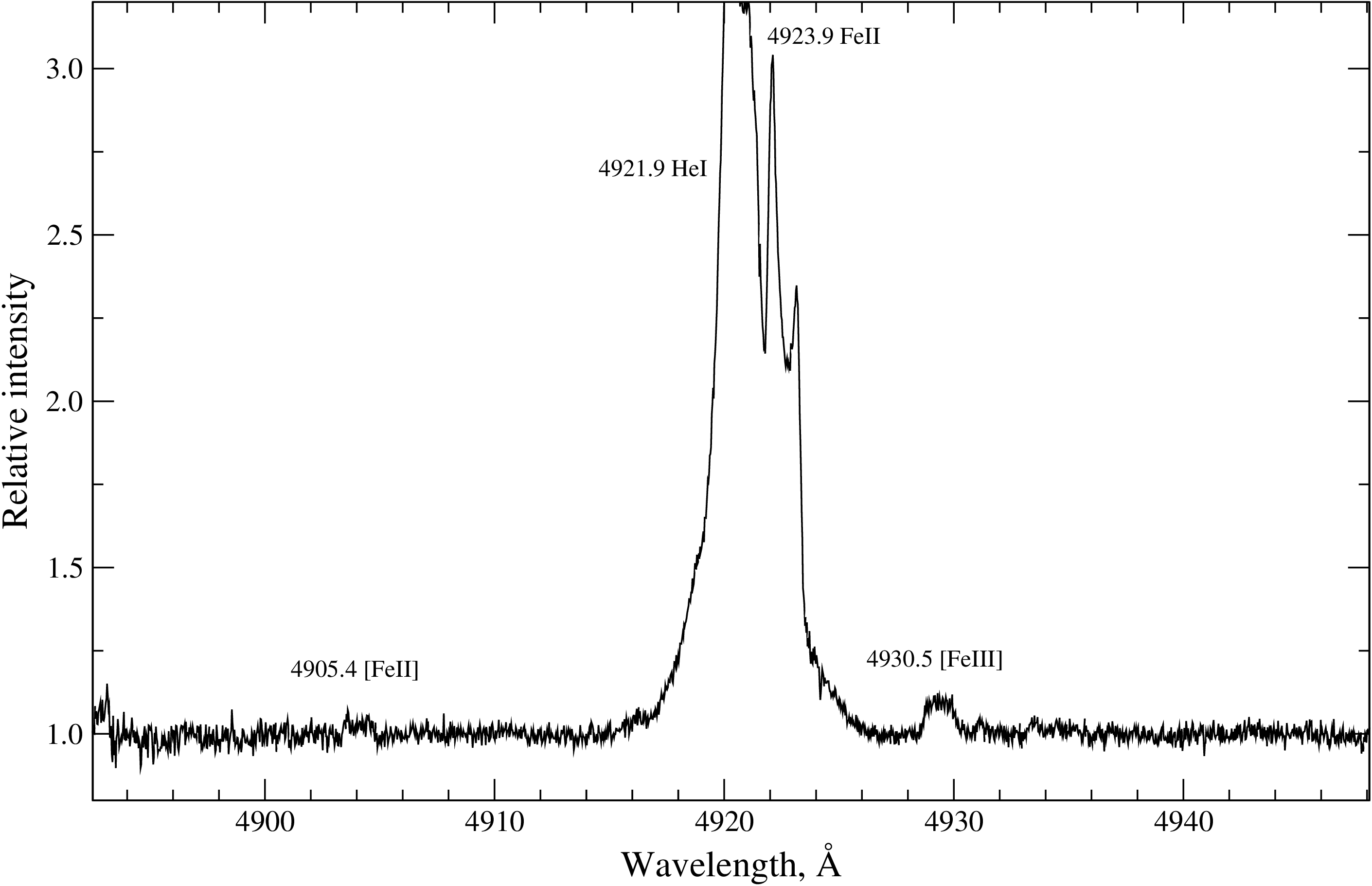}
\includegraphics[angle=0,width=0.8\textwidth,bb=0 40 1450 950,clip]{631_25k.eps}
\includegraphics[angle=0,width=0.8\textwidth,bb=0 40 1450 950,clip]{631_26k.eps}
\caption{CI\,Cam spectrum: fragments 24--26.}
\end{figure*}

\setcounter{figure}{0}
\begin{figure*}
\includegraphics[angle=0,width=0.8\textwidth,bb=0 40 1450 950,clip]{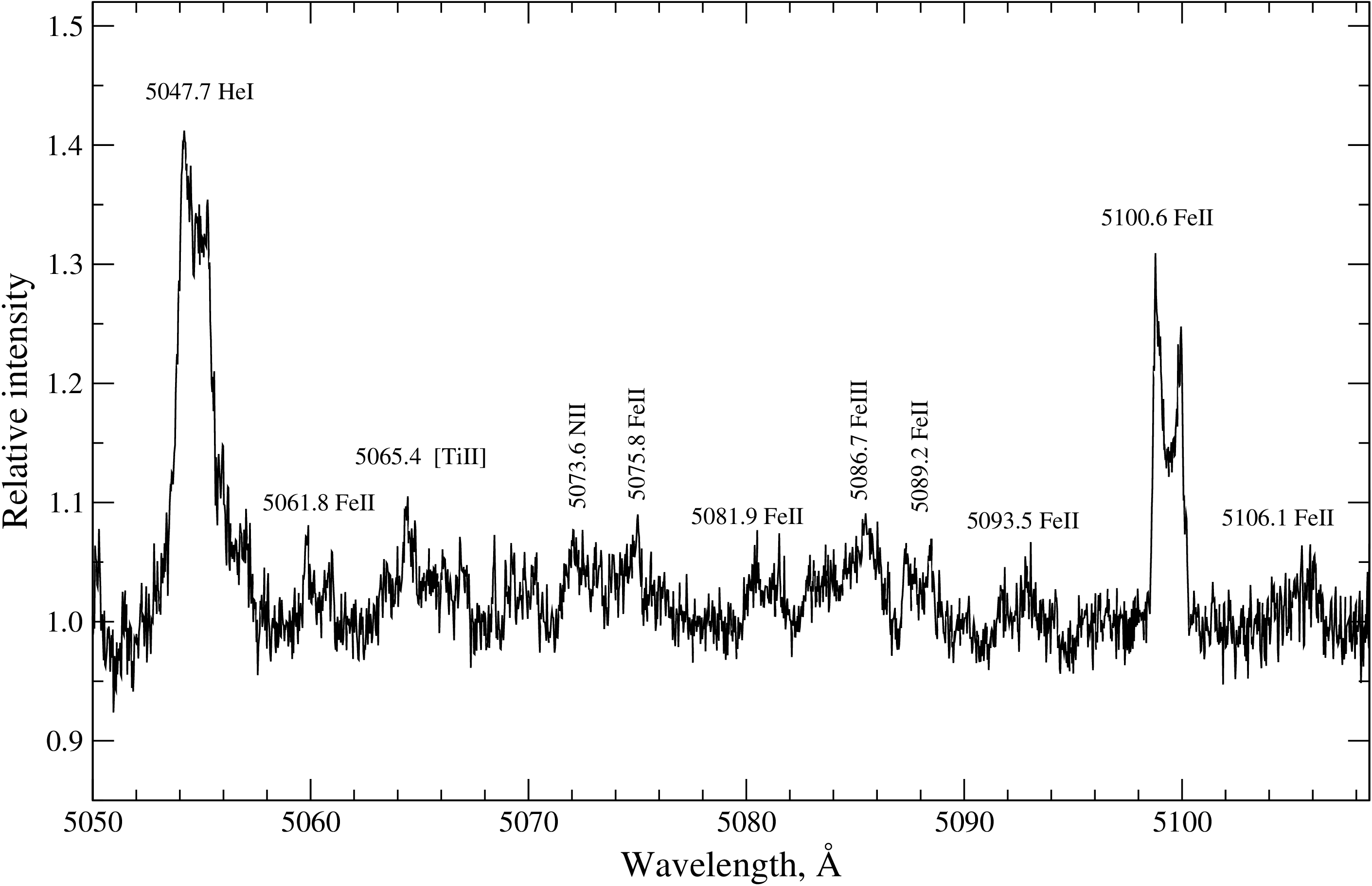}
\includegraphics[angle=0,width=0.8\textwidth,bb=0 40 1450 950,clip]{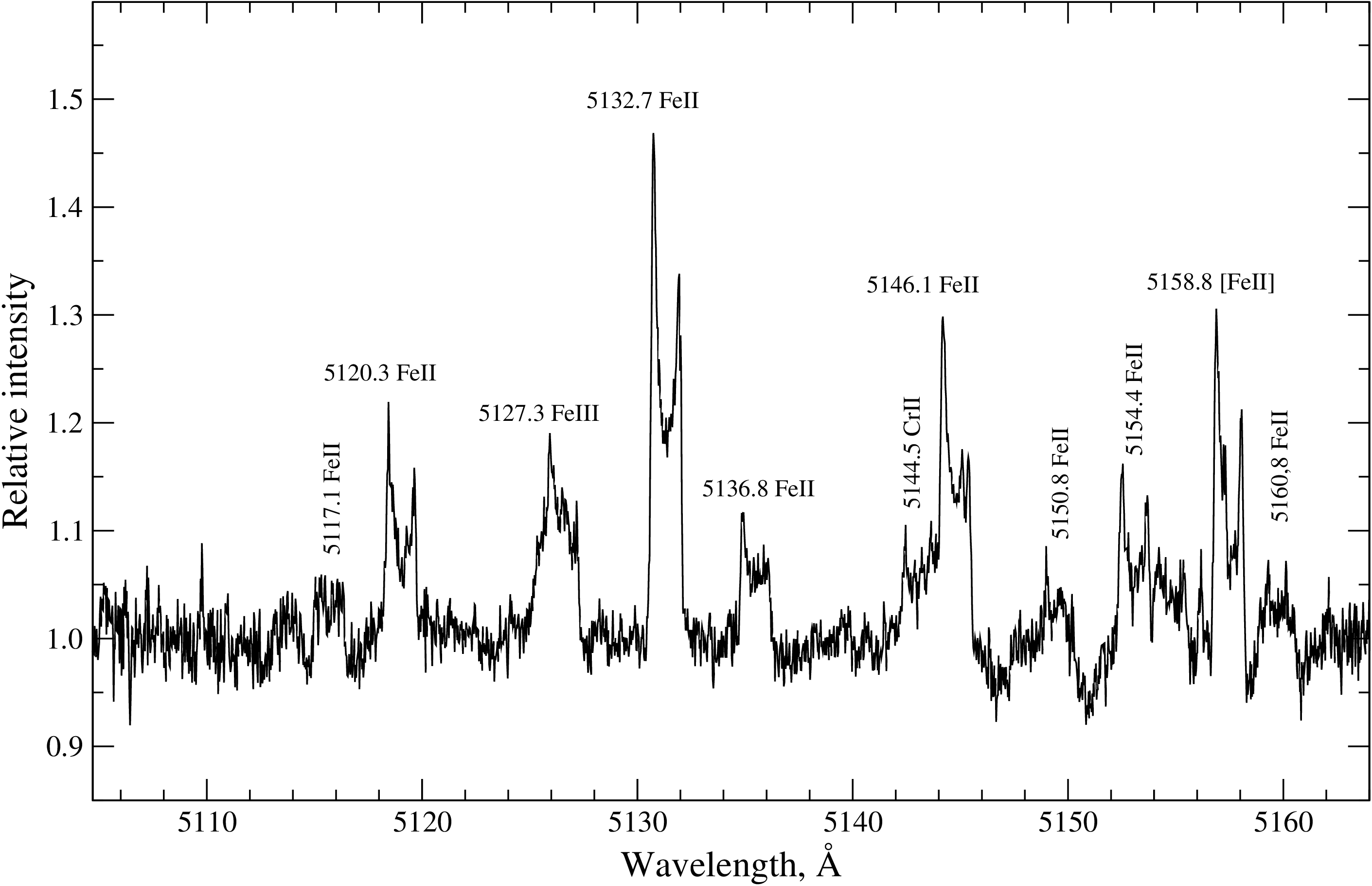}
\includegraphics[angle=0,width=0.8\textwidth,bb=0 40 1450 950,clip]{631_29k.eps}
\caption{CI\,Cam spectrum: fragments 27--29.}
\end{figure*}

\setcounter{figure}{0}
\begin{figure*}
\includegraphics[angle=0,width=0.8\textwidth,bb=0 40 1450 950,clip]{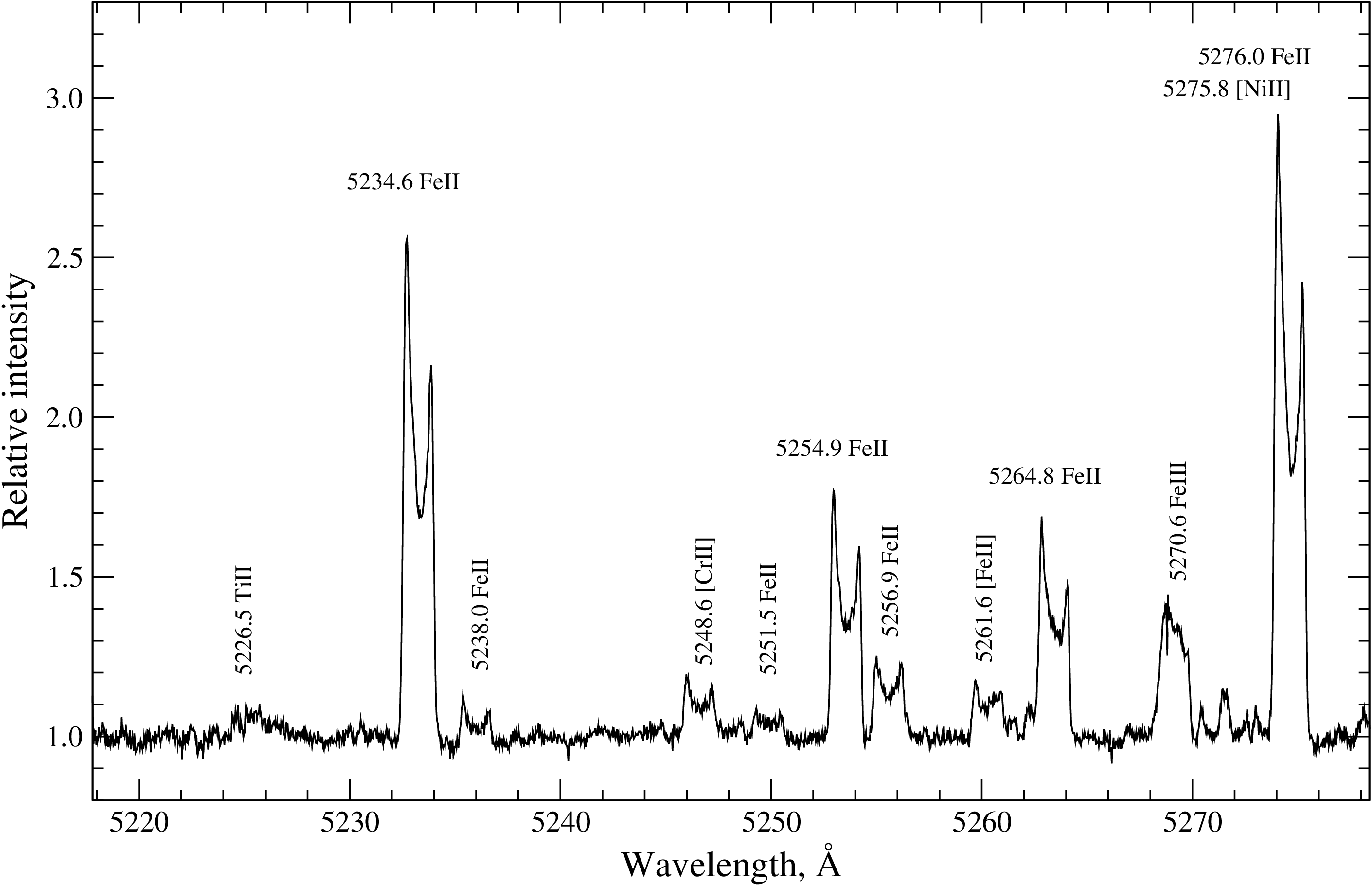}
\includegraphics[angle=0,width=0.8\textwidth,bb=0 40 1450 950,clip]{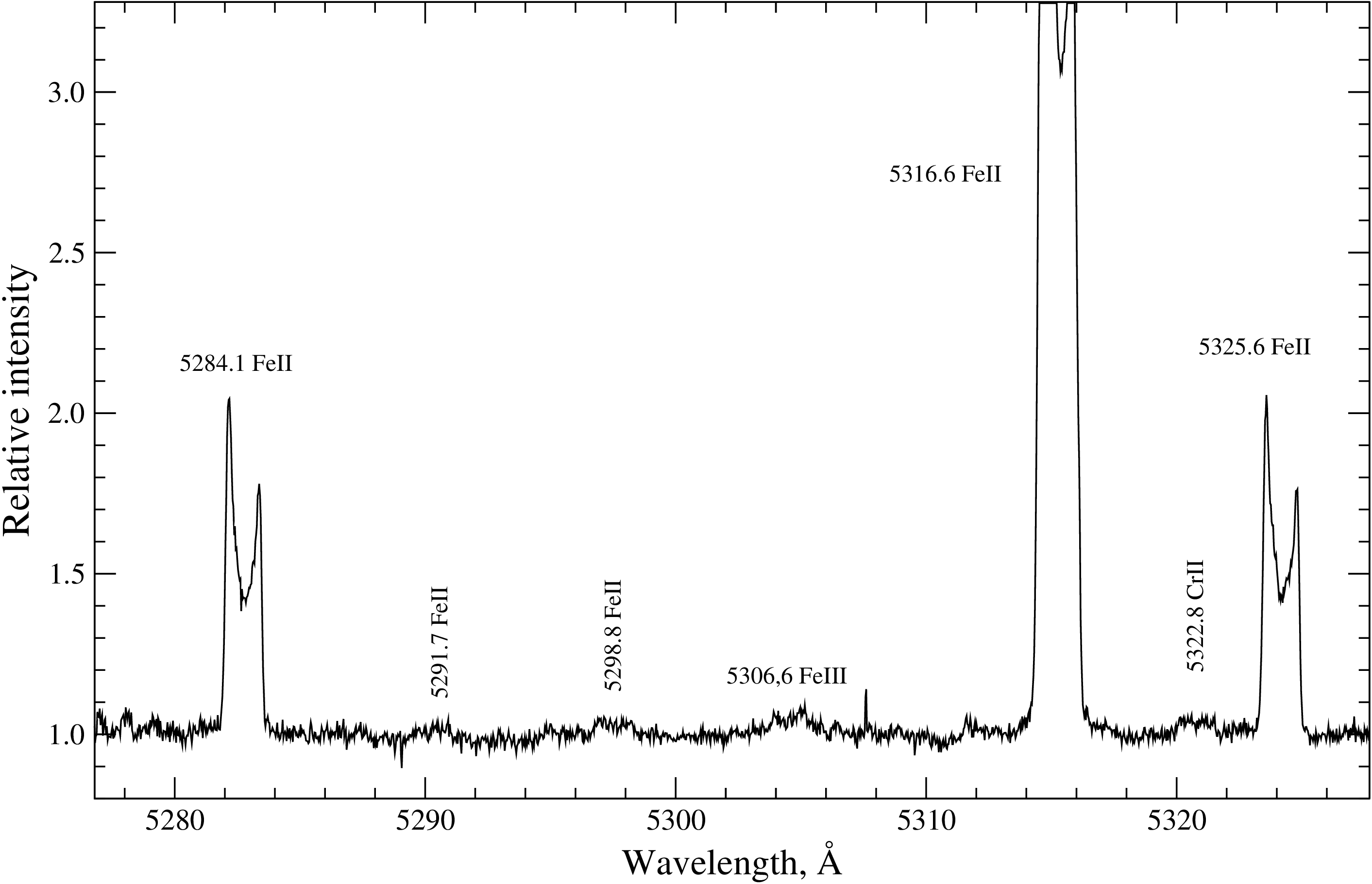}
\includegraphics[angle=0,width=0.8\textwidth,bb=0 40 1450 950,clip]{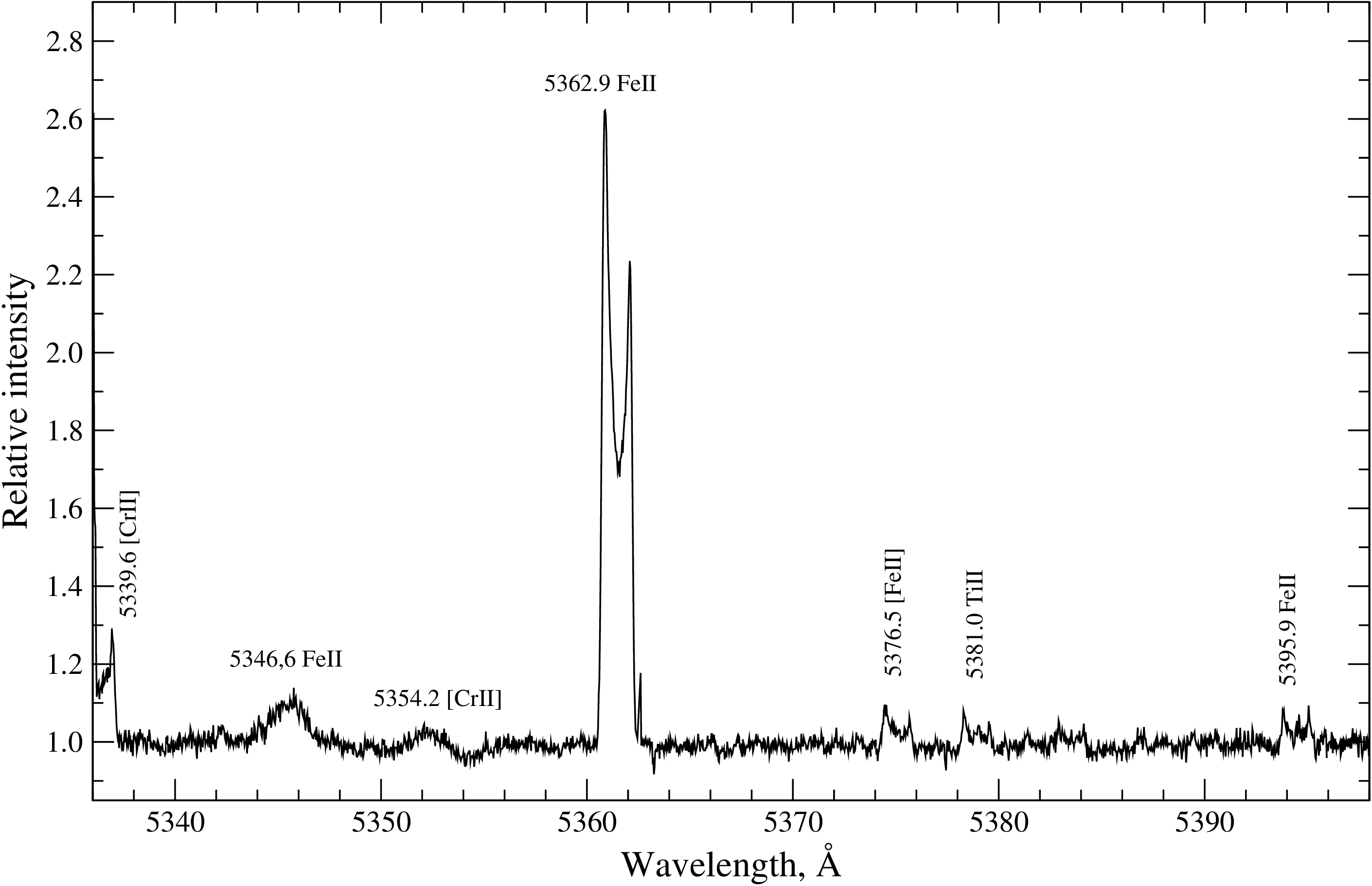}
\caption{CI\,Cam spectrum: fragments 30--32.}
\end{figure*}

\setcounter{figure}{0}
\begin{figure*}
\includegraphics[angle=0,width=0.8\textwidth,bb=0 40 1450 950,clip]{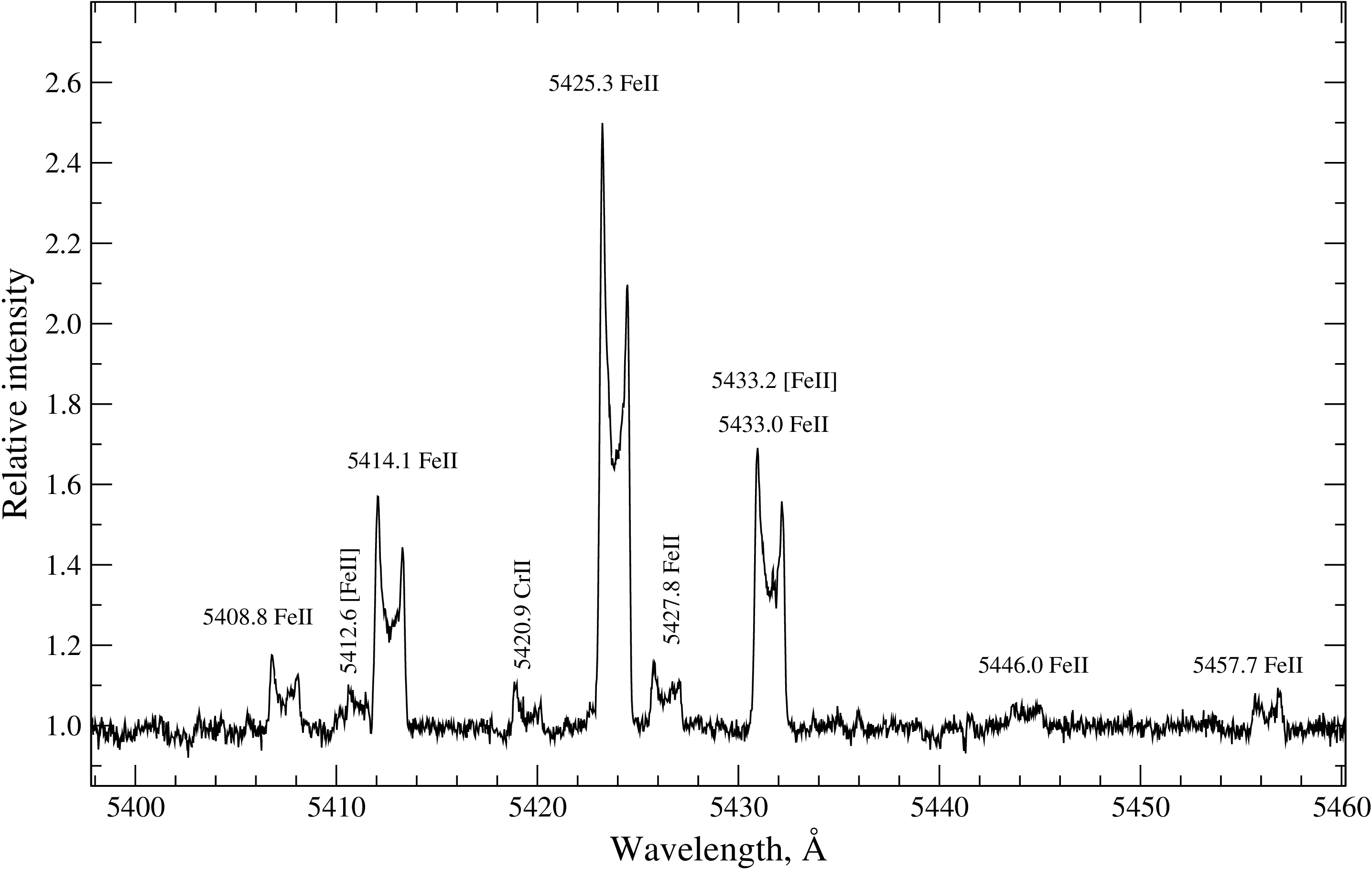}
\includegraphics[angle=0,width=0.8\textwidth,bb=0 40 1450 950,clip]{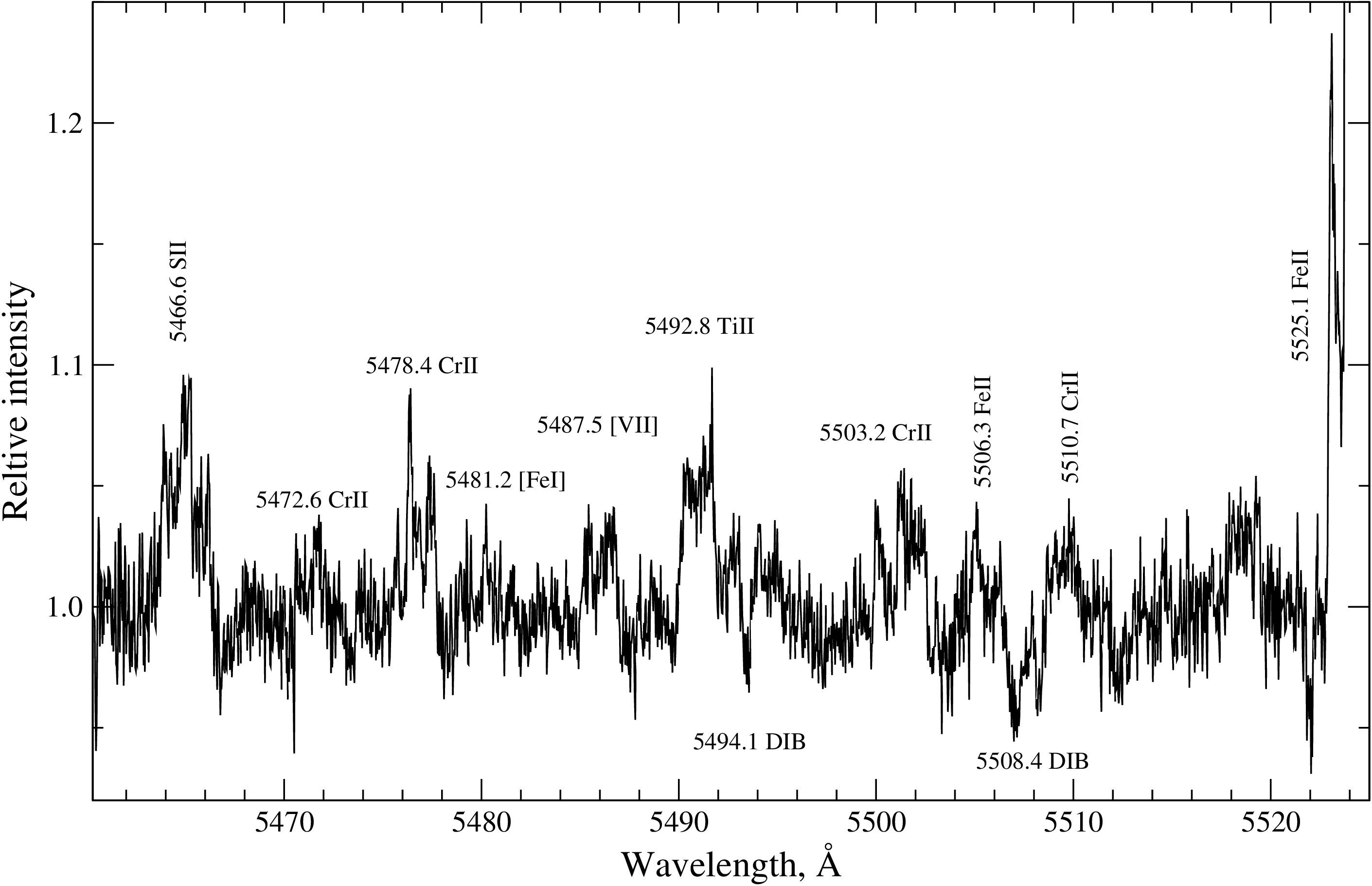}
\includegraphics[angle=0,width=0.8\textwidth,bb=0 40 1450 950,clip]{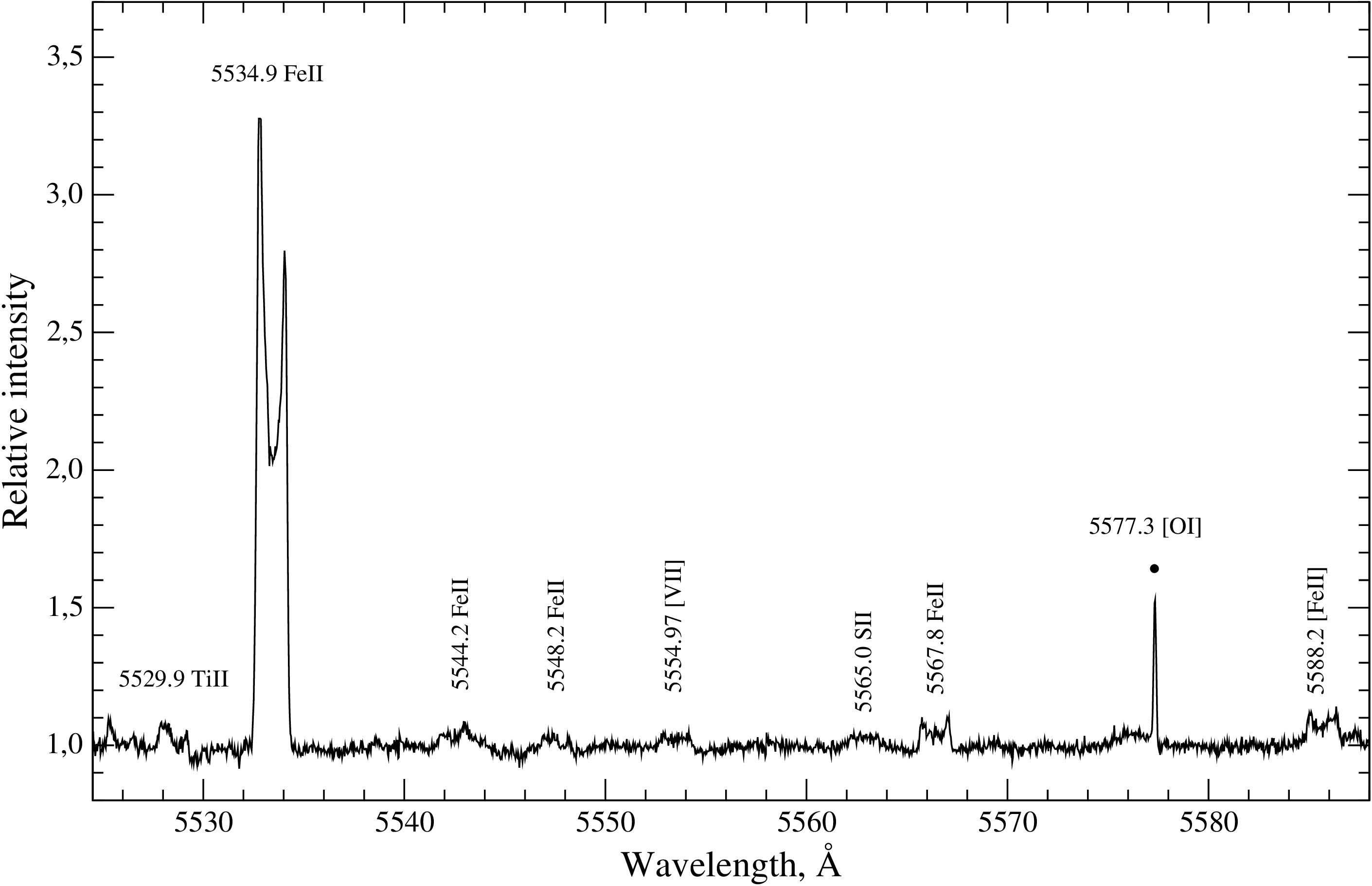}
\caption{CI\,Cam spectrum: fragments 33--35.}
\end{figure*}

\setcounter{figure}{0}
\begin{figure*}
\includegraphics[angle=0,width=0.8\textwidth,bb=0 40 1450 950,clip]{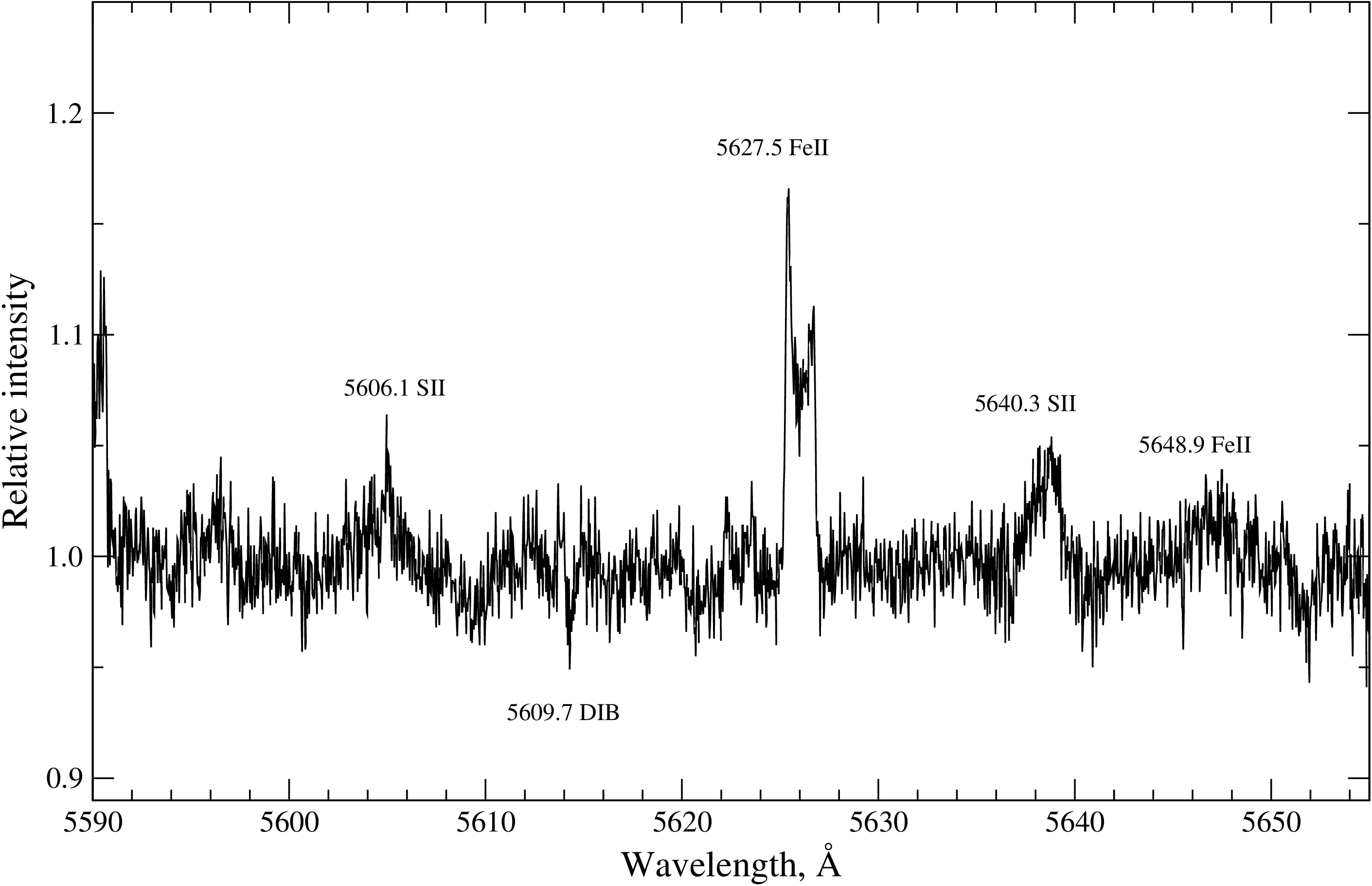}
\includegraphics[angle=0,width=0.8\textwidth,bb=0 40 1450 950,clip]{631_37k.eps}
\includegraphics[angle=0,width=0.8\textwidth,bb=0 40 1450 950,clip]{631_38mn.eps}
\caption{CI\,Cam spectrum: fragments 36--38.}
\end{figure*}

\setcounter{figure}{0}
\begin{figure*}
\includegraphics[angle=0,width=0.8\textwidth,bb=0 40 1450 950,clip]{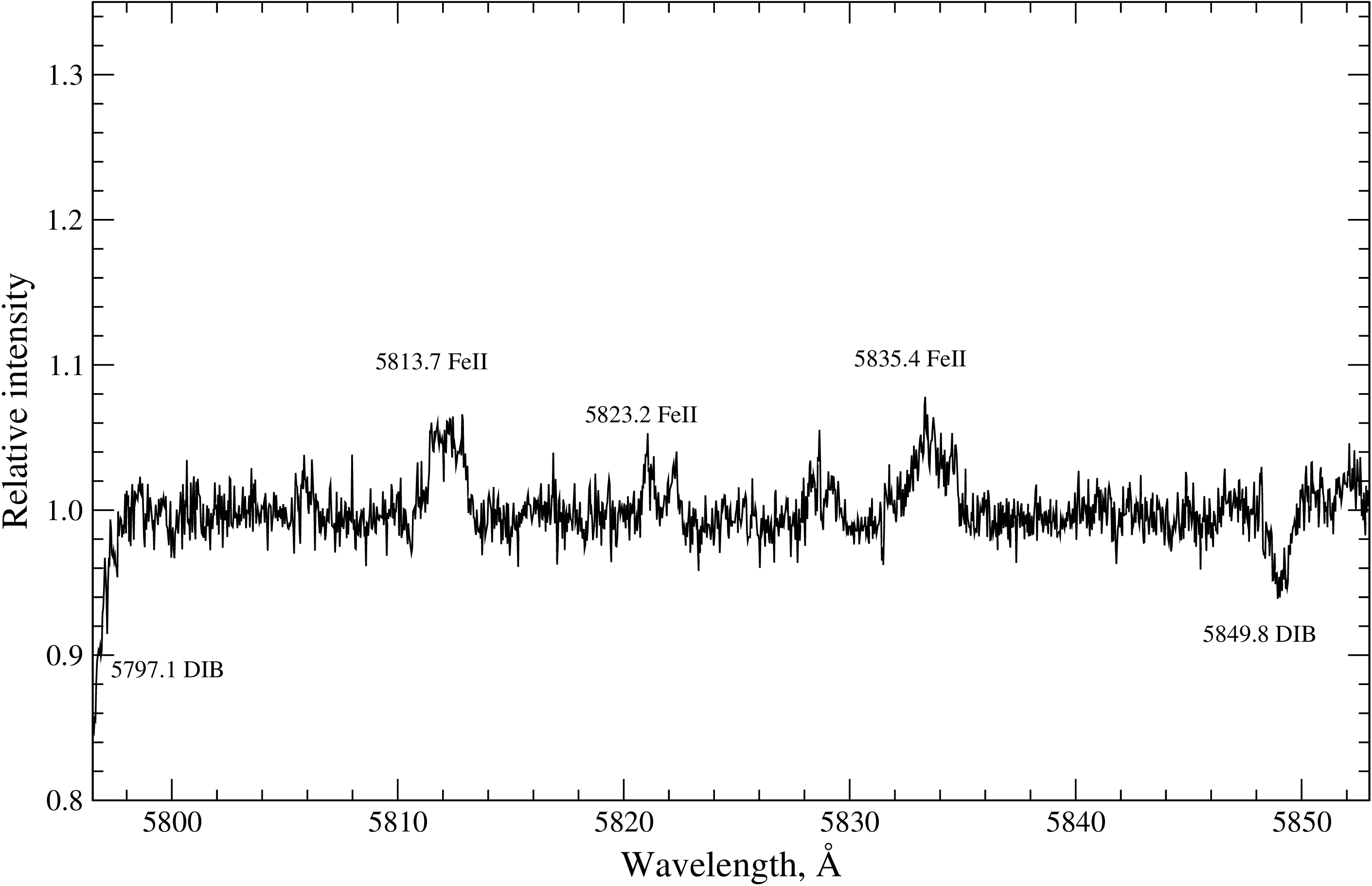}
\includegraphics[angle=0,width=0.8\textwidth,bb=0 40 1450 950,clip]{631_40k.eps}
\includegraphics[angle=0,width=0.8\textwidth,bb=0 40 1450 950,clip]{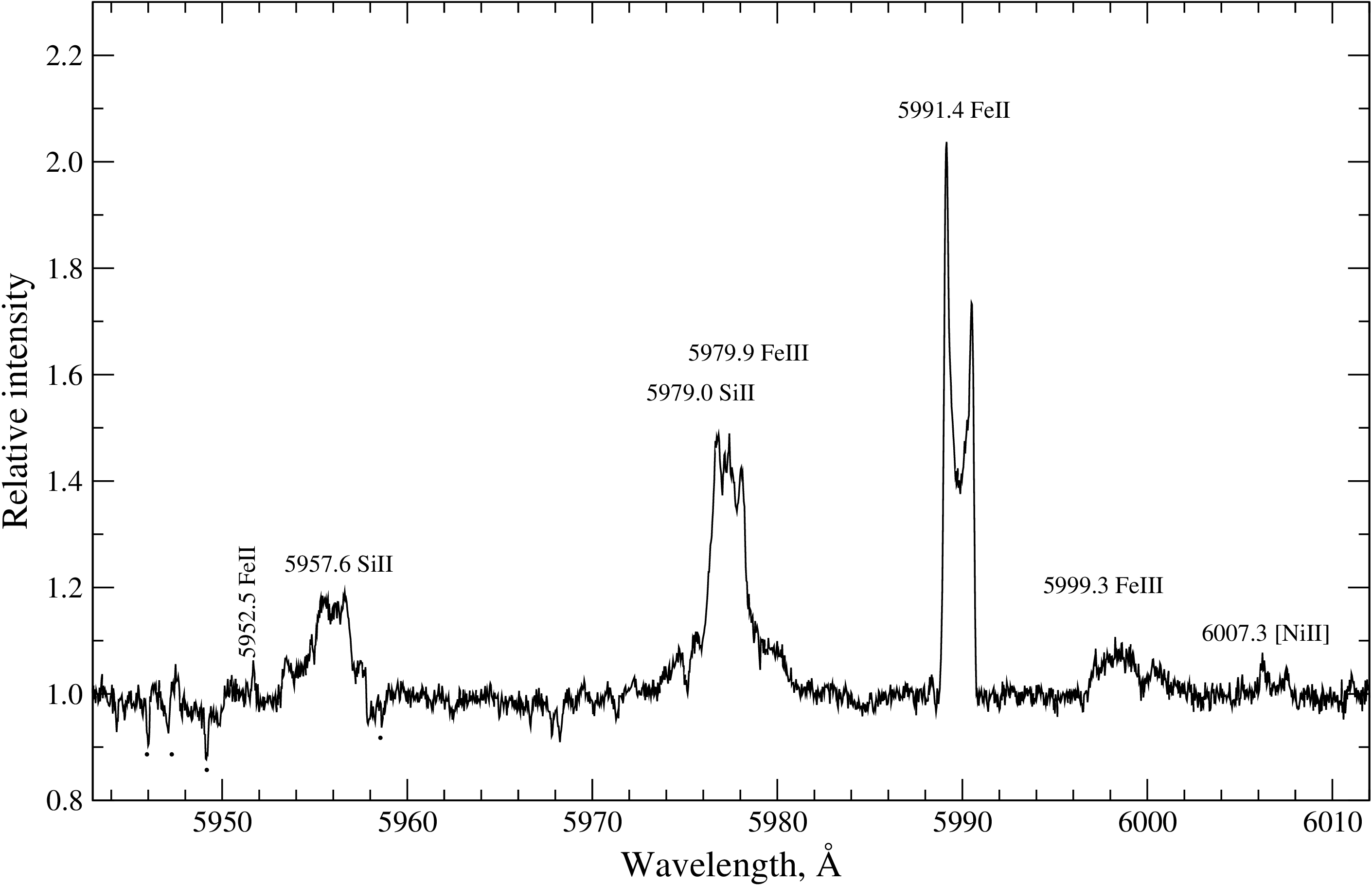}
\caption{CI\,Cam spectrum: fragments 39--41.}
\end{figure*}

\setcounter{figure}{0}
\begin{figure*}
\includegraphics[angle=0,width=0.8\textwidth,bb=0 40 1450 950,clip]{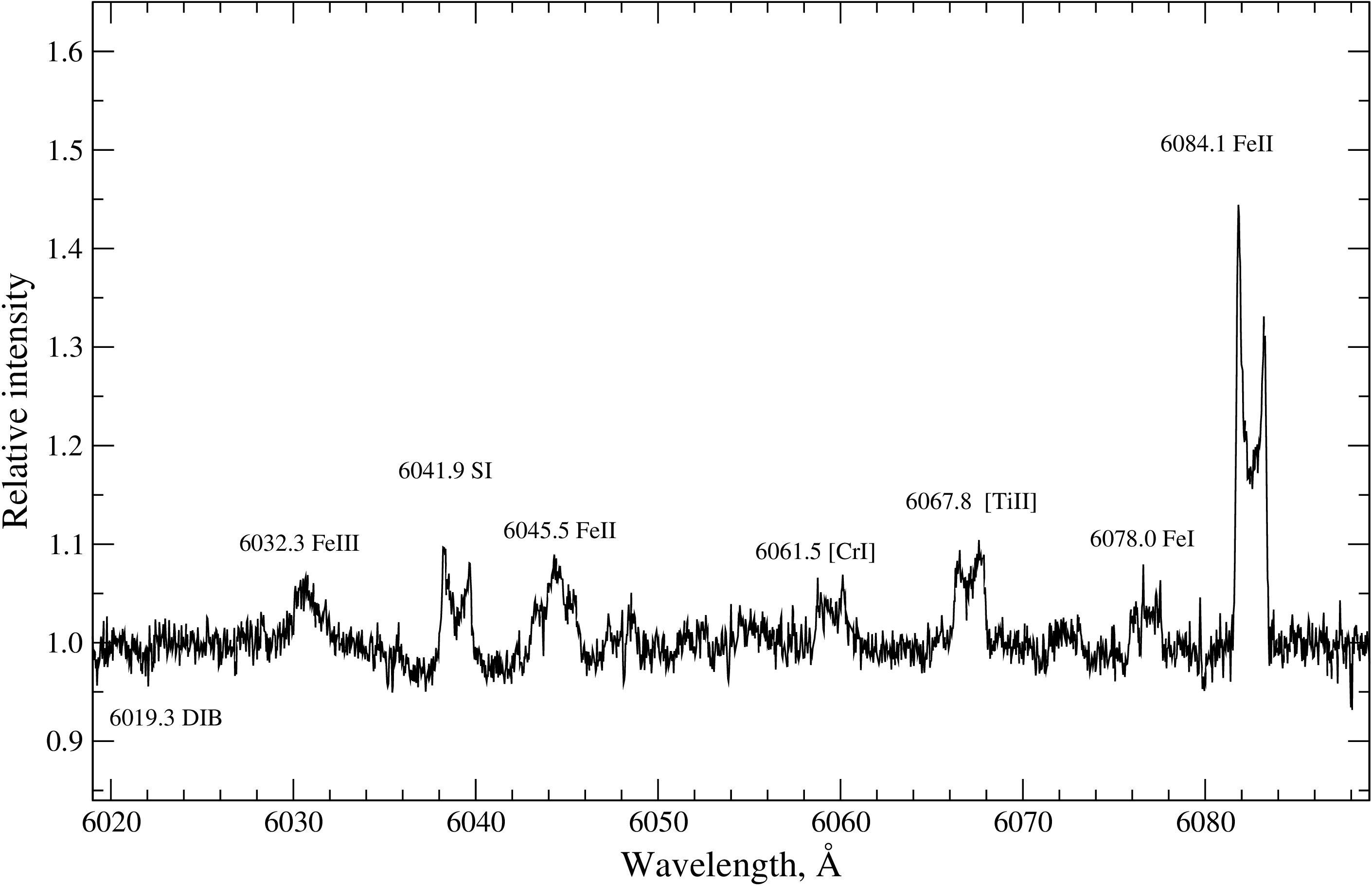}
\includegraphics[angle=0,width=0.8\textwidth,bb=0 40 1450 950,clip]{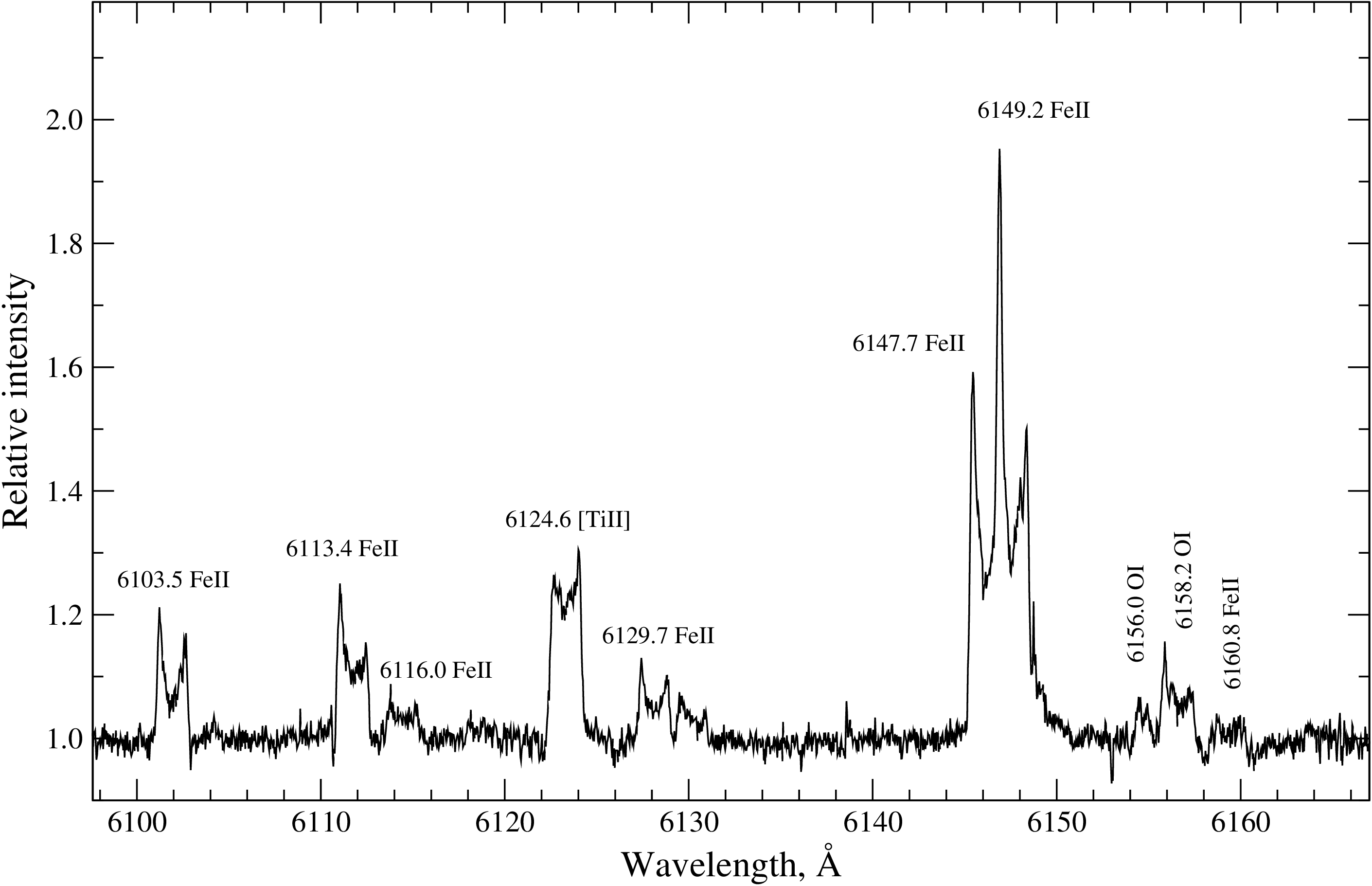}
\includegraphics[angle=0,width=0.8\textwidth,bb=0 40 1450 950,clip]{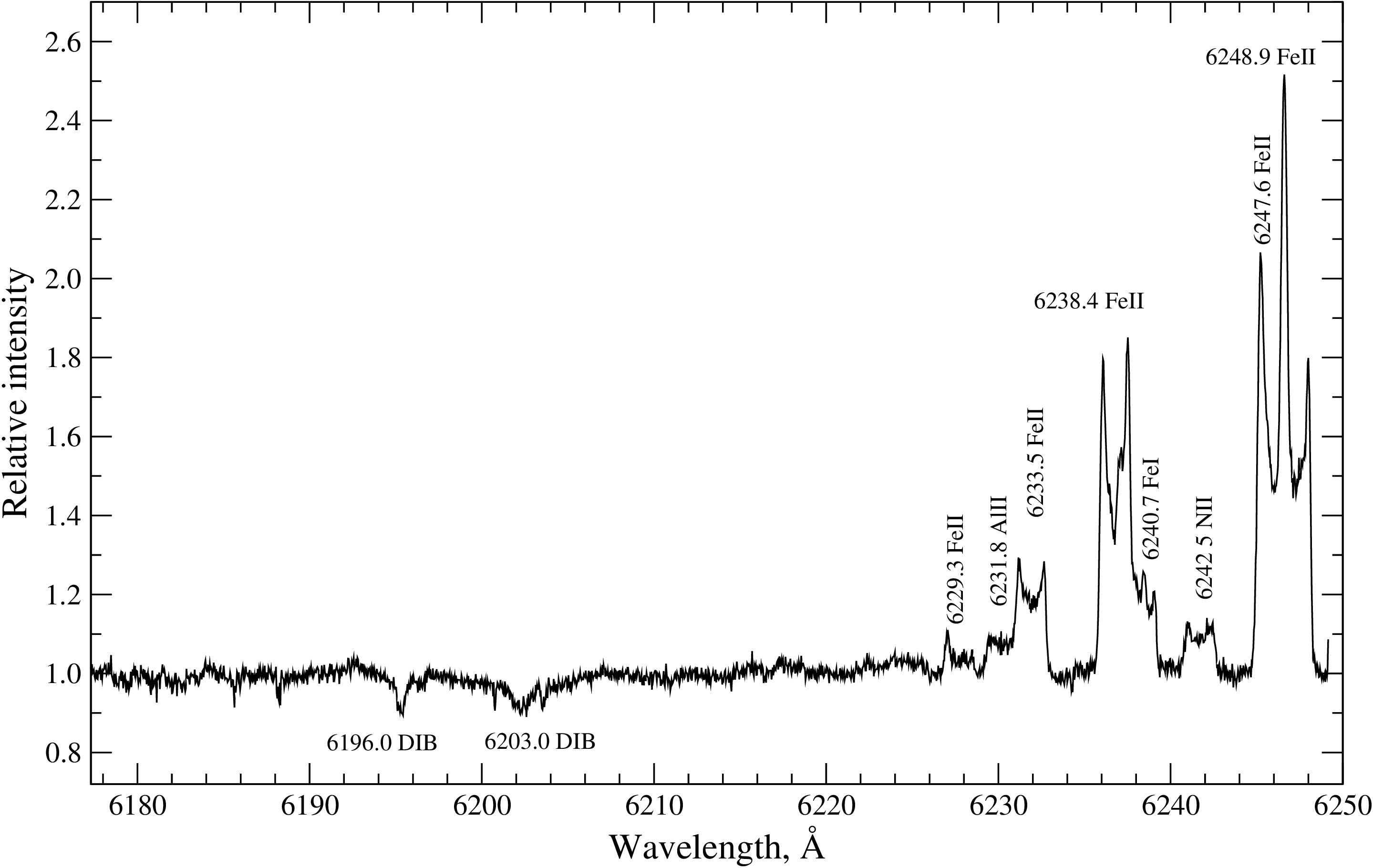}
\caption{CI\,Cam spectrum: fragments 42--44.}
\end{figure*}

\setcounter{figure}{0}
\begin{figure*}
\includegraphics[angle=0,width=0.8\textwidth,bb=0 40 1450 950,clip]{631_45k.eps}
\includegraphics[angle=0,width=0.8\textwidth,bb=0 40 1450 950,clip]{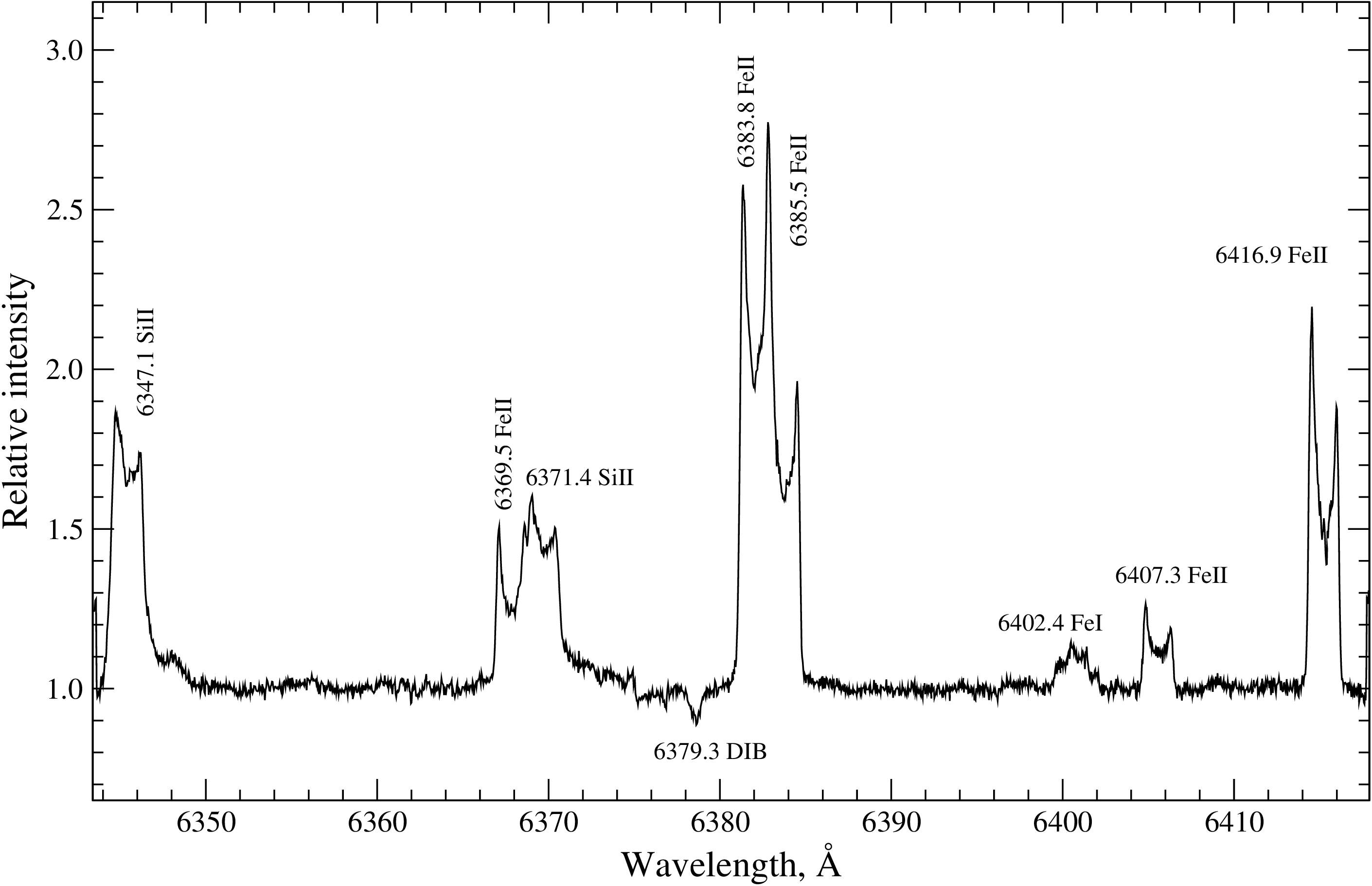}
\includegraphics[angle=0,width=0.8\textwidth,bb=0 40 1450 950,clip]{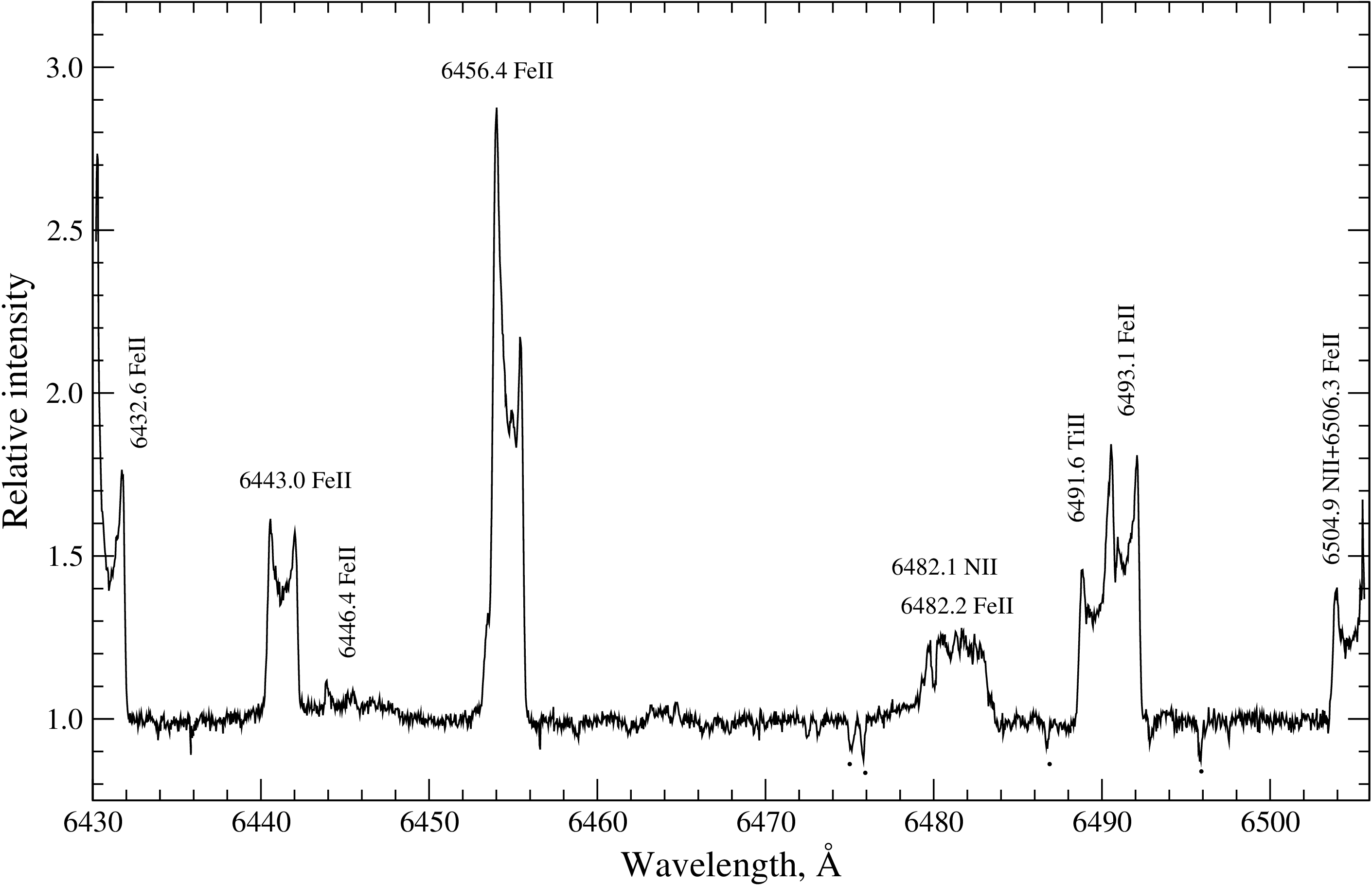}
\caption{CI\,Cam spectrum: fragments 45--47.}
\end{figure*}

\setcounter{figure}{0}
\begin{figure*}
\includegraphics[angle=0,width=0.8\textwidth,bb=0 40 1450 950,clip]{631_48k.eps}
\includegraphics[angle=0,width=0.8\textwidth,bb=0 40 1450 950,clip]{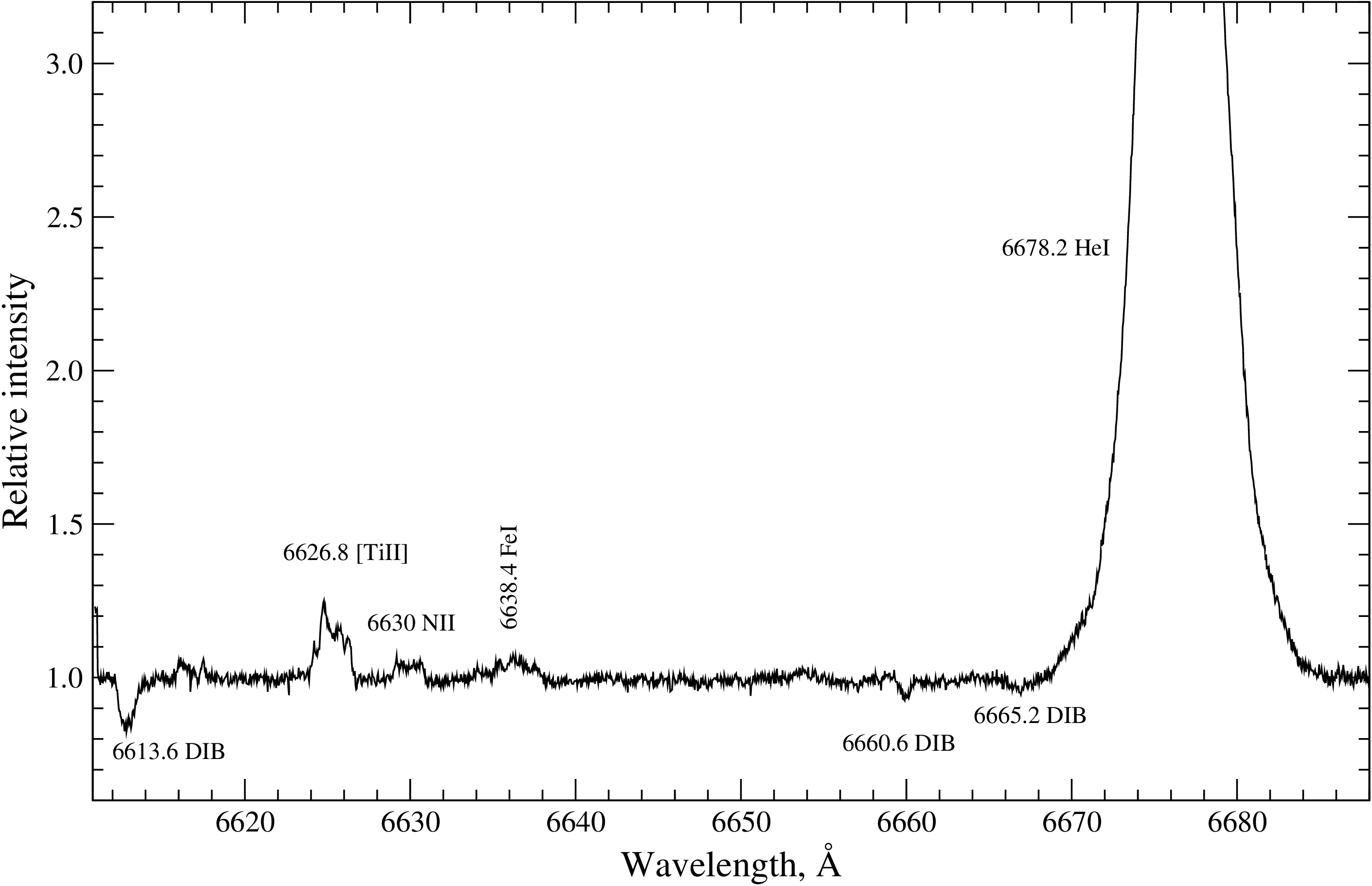}
\includegraphics[angle=0,width=0.8\textwidth,bb=0 40 1450 950,clip]{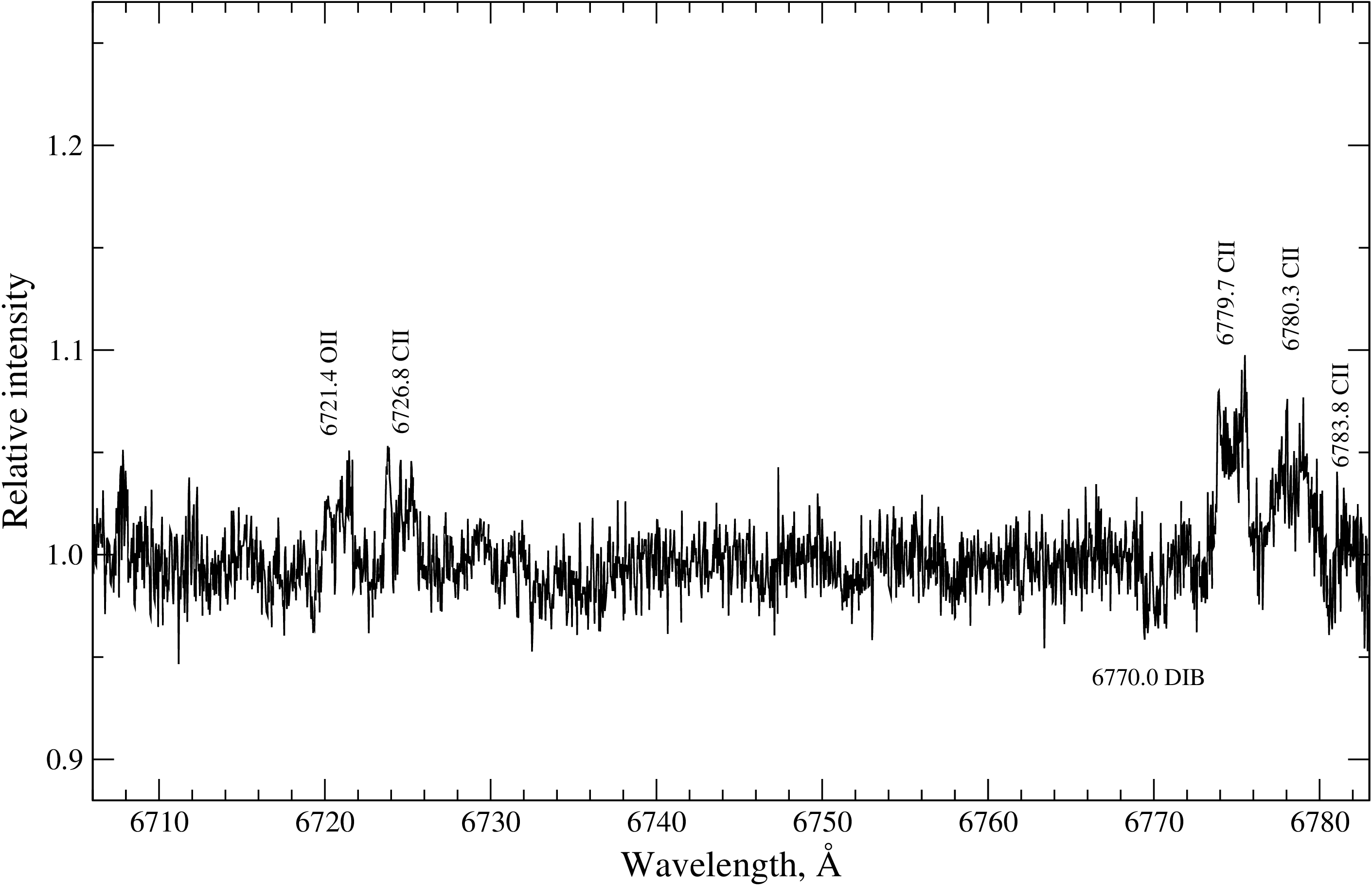}
\caption{CI\,Cam spectrum: fragments 48--50.}
\end{figure*}

\setcounter{figure}{0}
\begin{figure*}
\includegraphics[angle=0,width=0.8\textwidth,bb=0 40 1450 950,clip]{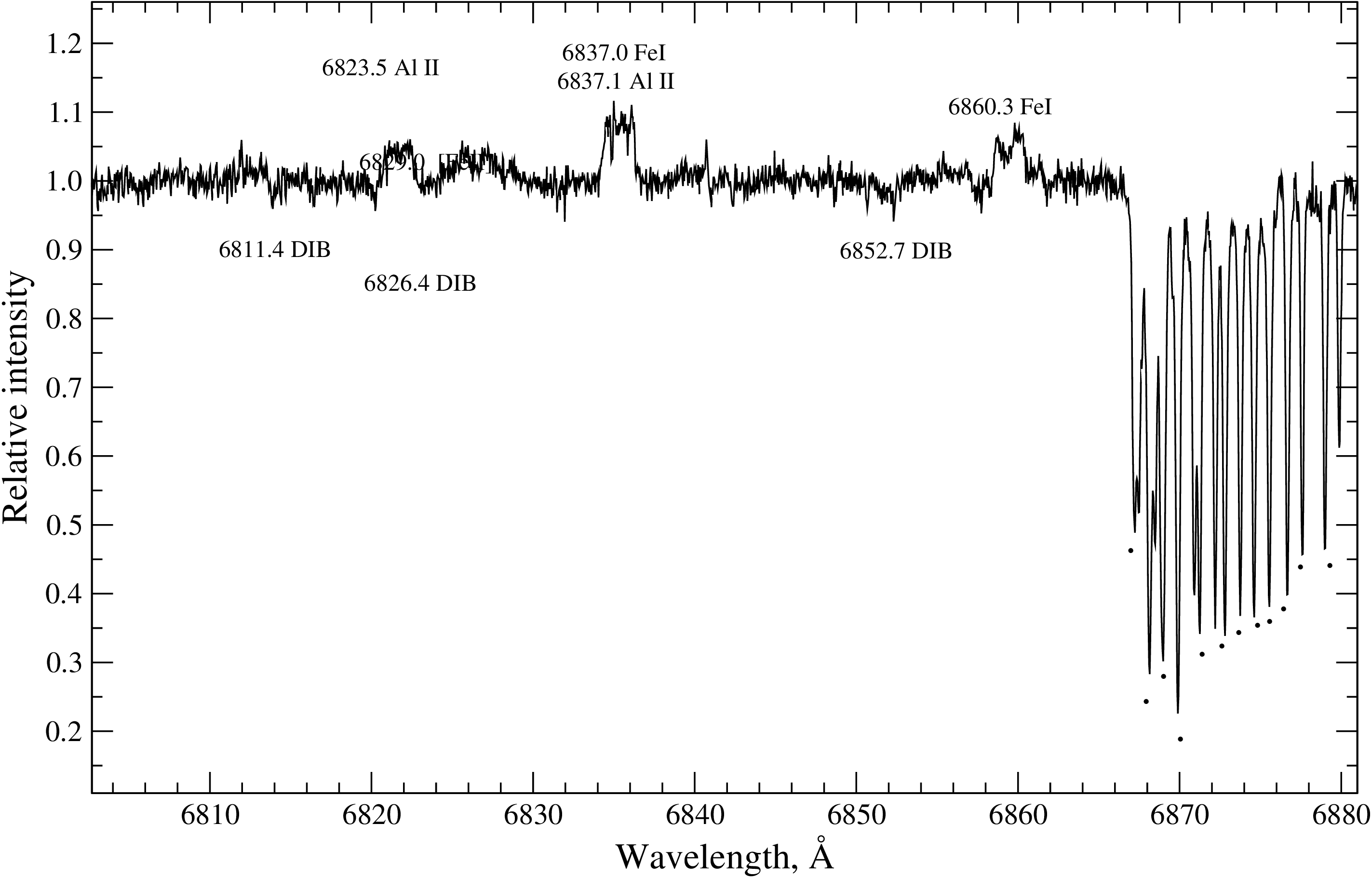}
\includegraphics[angle=0,width=0.8\textwidth,bb=0 40 1450 950,clip]{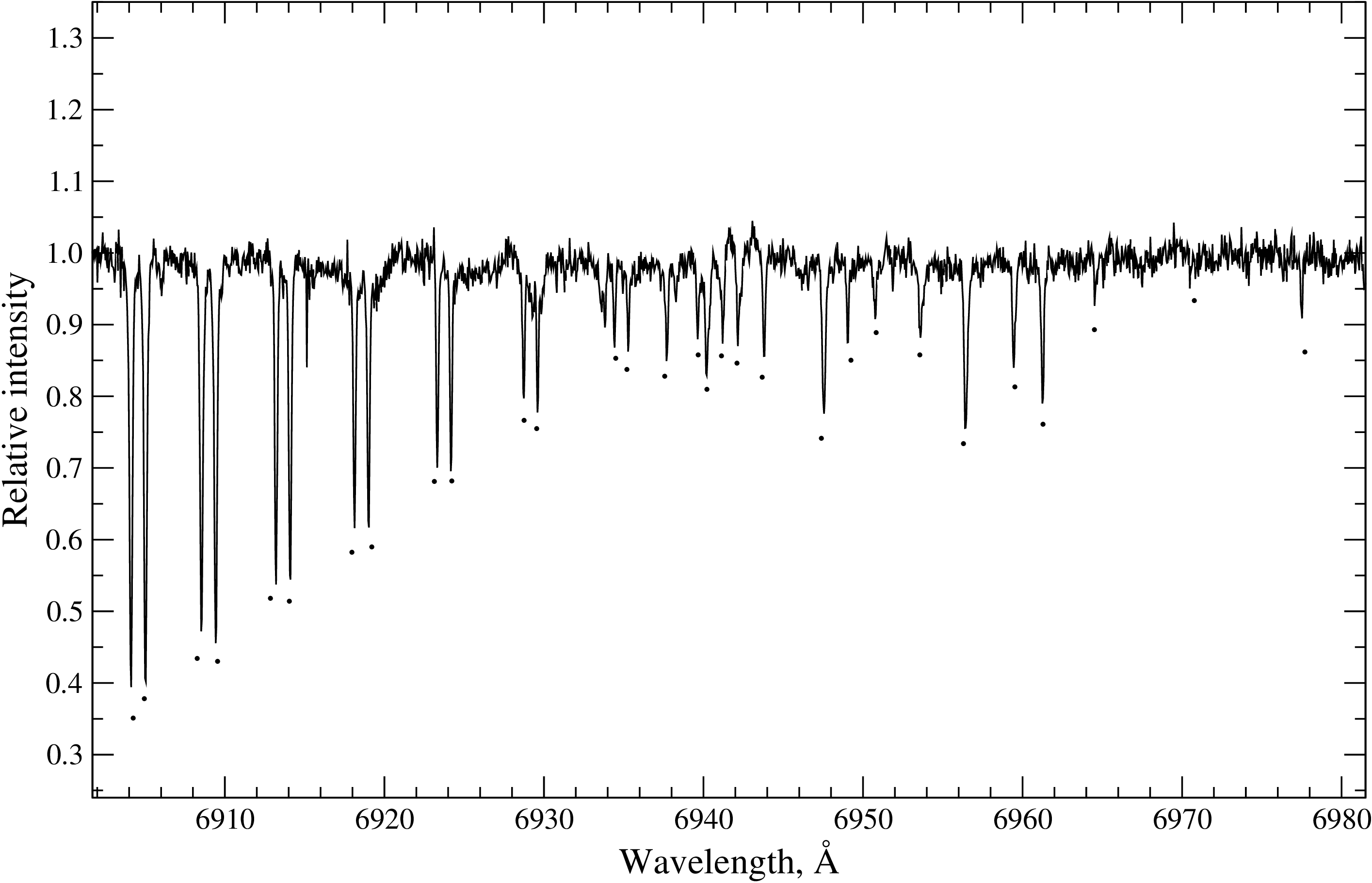}
\includegraphics[angle=0,width=0.8\textwidth,bb=0 40 1450 950,clip]{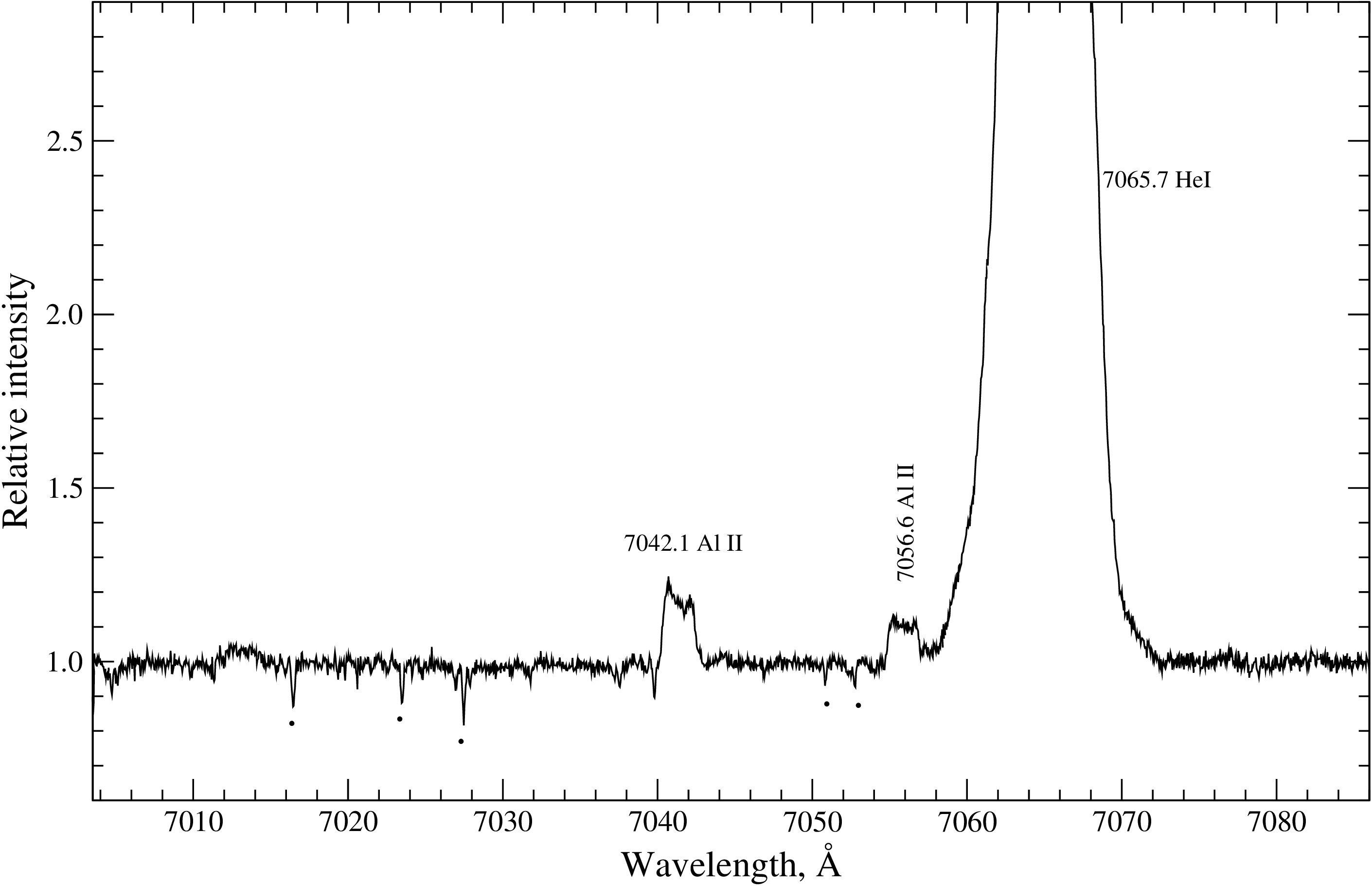}
\caption{CI\,Cam spectrum: fragments 51--53.}
\end{figure*}

\setcounter{figure}{0}
\begin{figure*}
\includegraphics[angle=0,width=0.8\textwidth,bb=0 40 1450 950,clip]{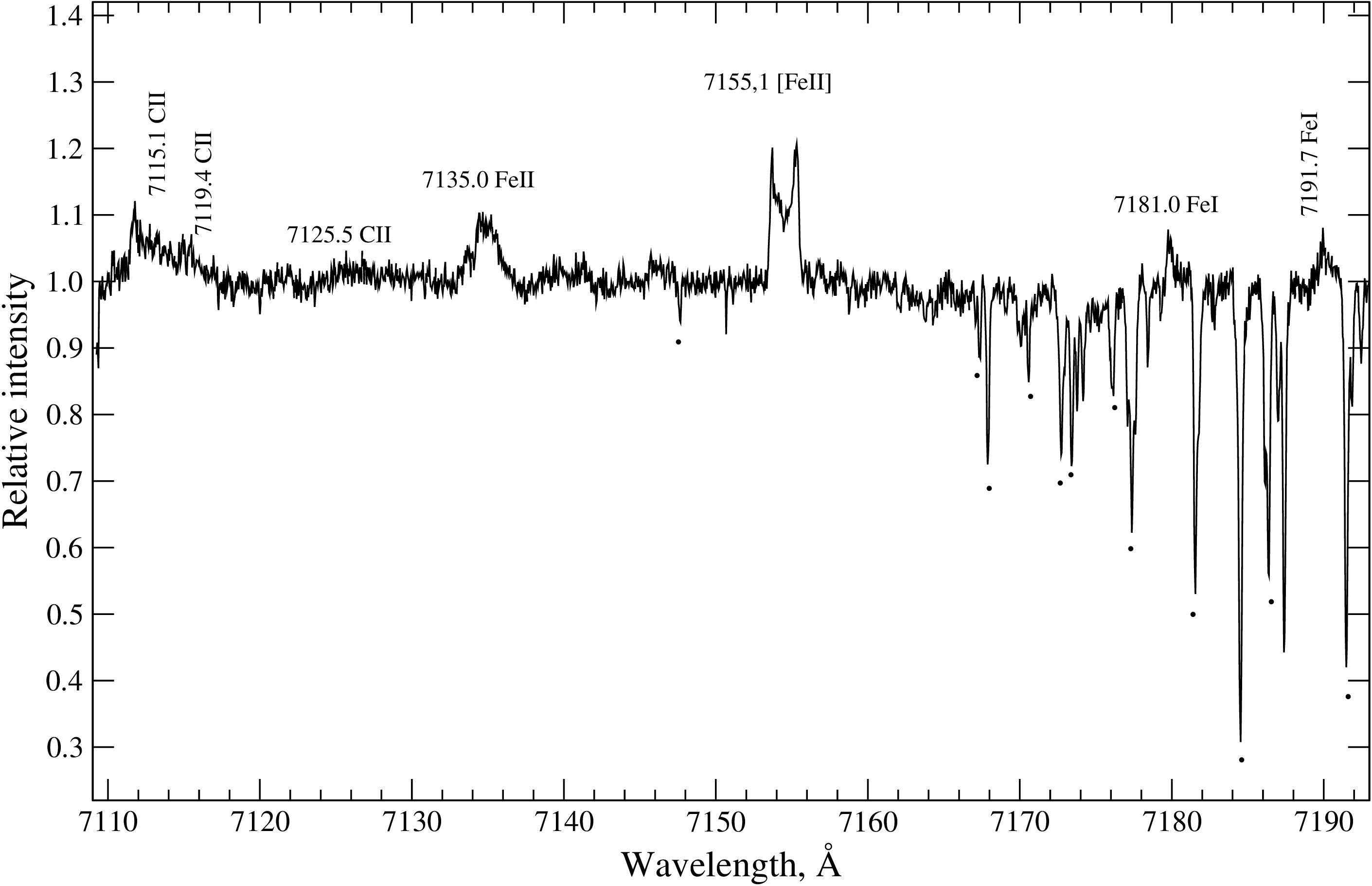}
\includegraphics[angle=0,width=0.8\textwidth,bb=0 40 1450 950,clip]{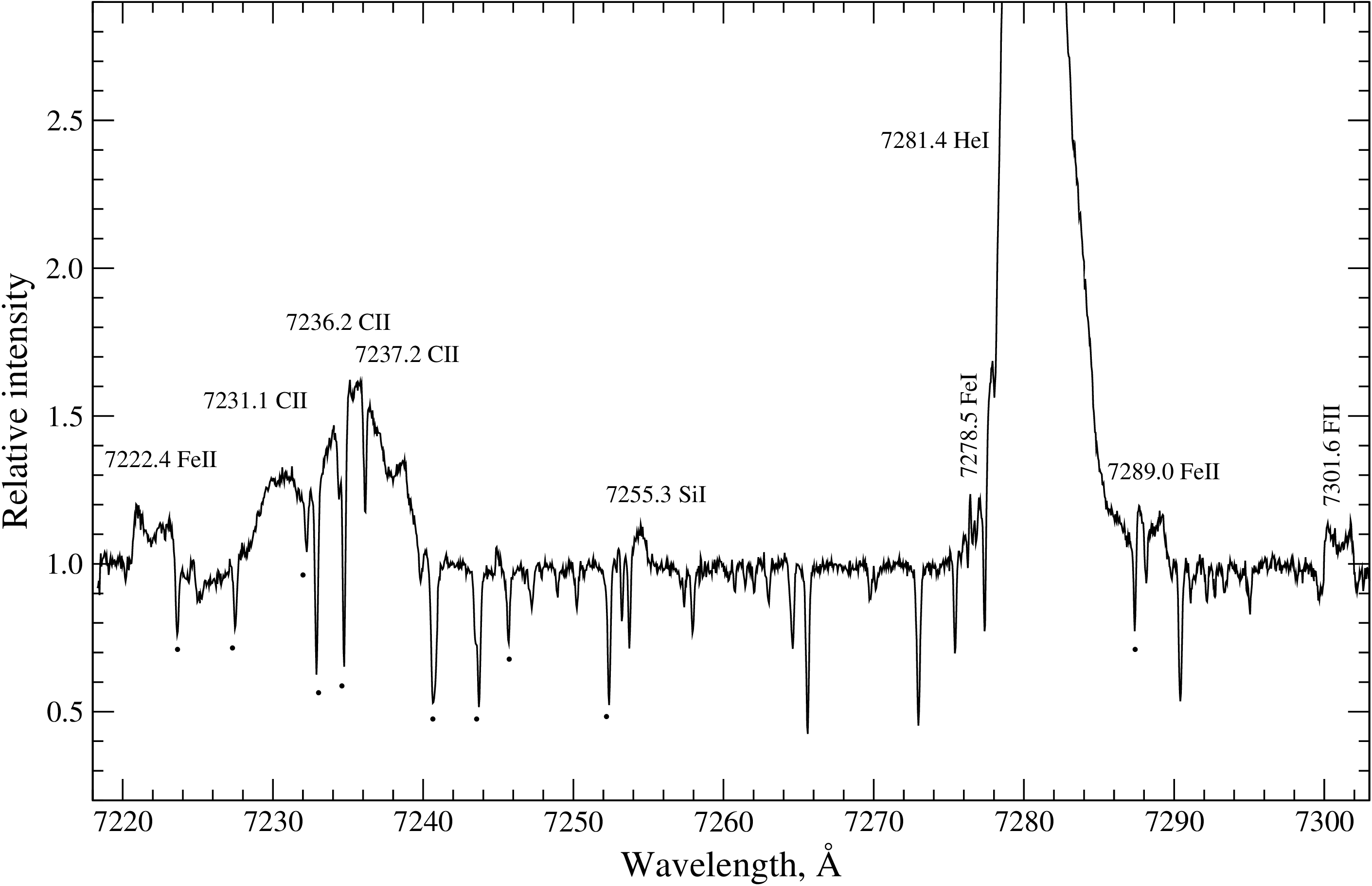}
\includegraphics[angle=0,width=0.8\textwidth,bb=0 40 1450 950,clip]{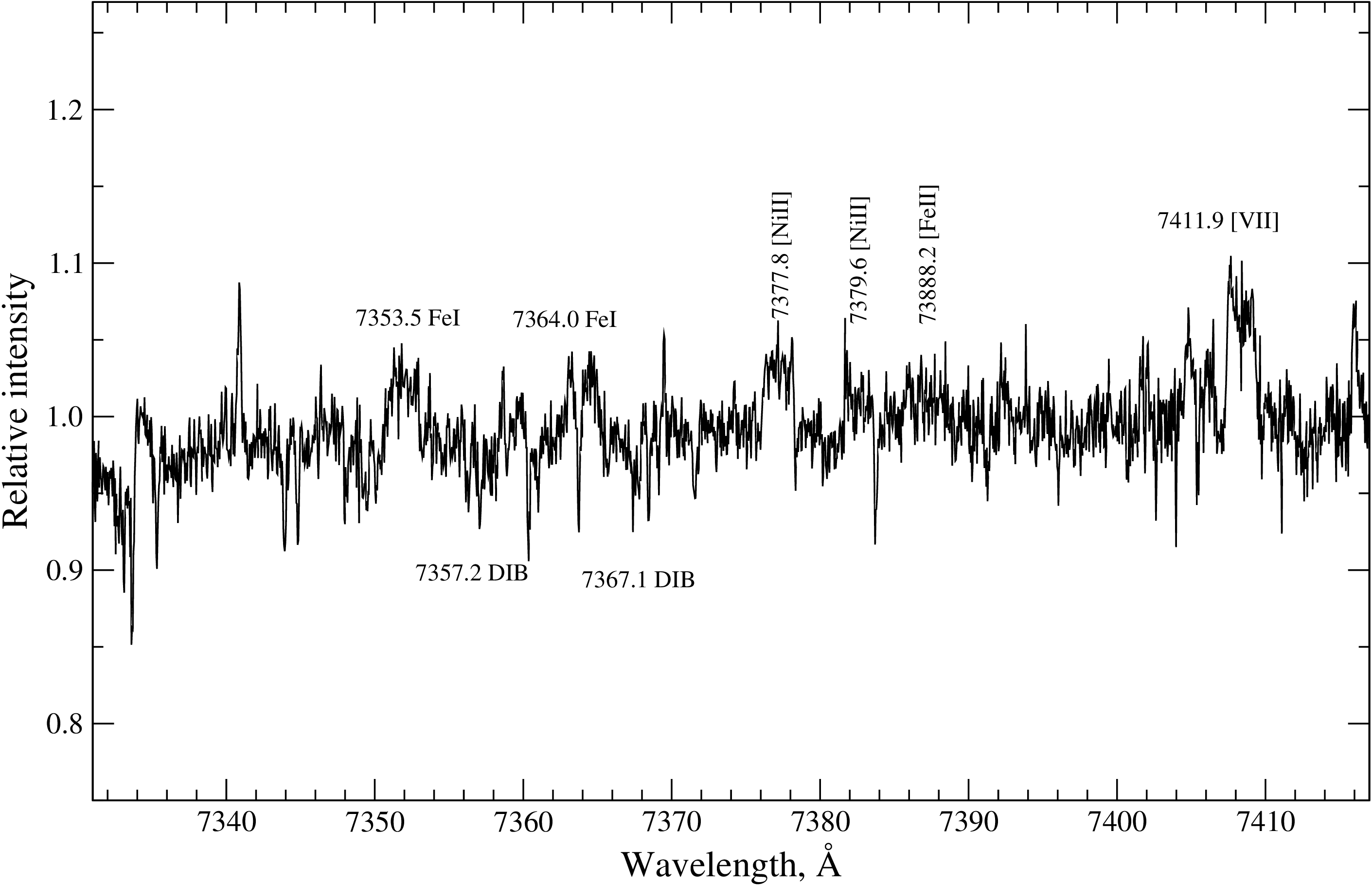}
\caption{CI\,Cam spectrum: fragments 54--56.}
\end{figure*}

\setcounter{figure}{0}
\begin{figure*}
\includegraphics[angle=0,width=0.8\textwidth,bb=0 40 1450 950,clip]{759_38k.eps}
\includegraphics[angle=0,width=0.8\textwidth,bb=0 40 1450 950,clip]{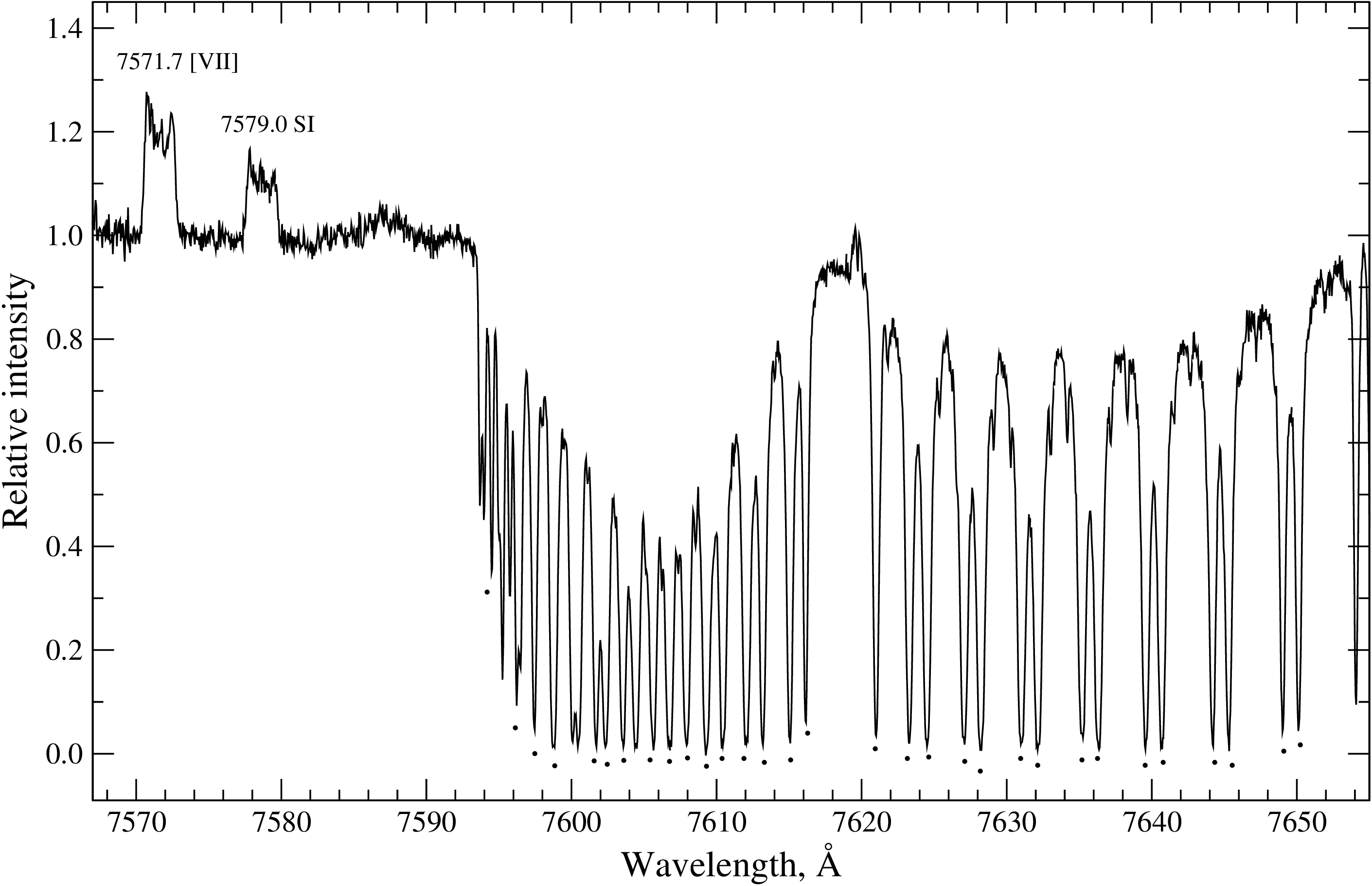}
\includegraphics[angle=0,width=0.8\textwidth,bb=0 40 1450 950,clip]{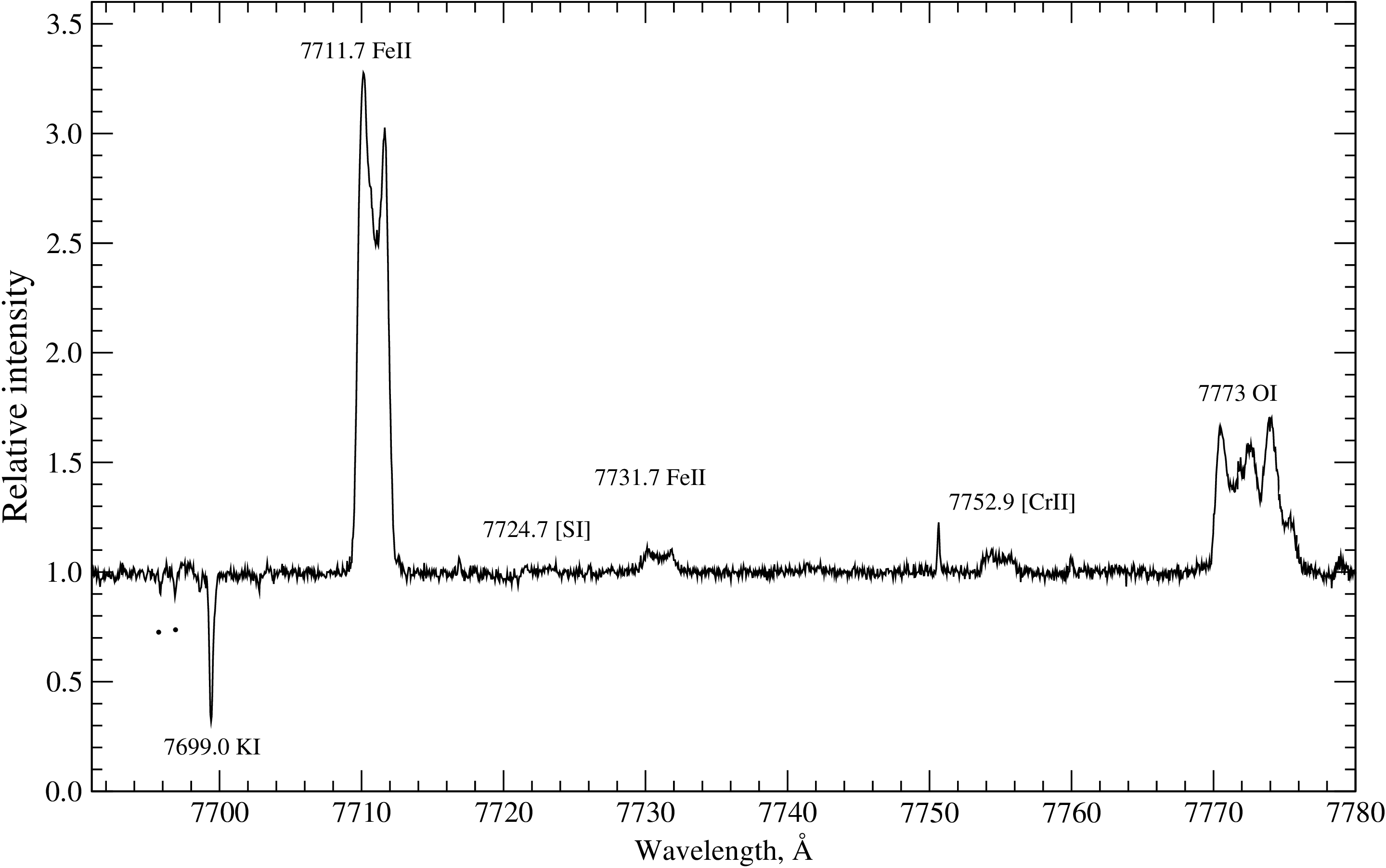}
\caption{CI\,Cam spectrum: fragments 57--59.}
\end{figure*}

\end{document}

%% file: sao_cmd_author.tex
\def\squareforqed{\hbox{\rlap{$\sqcap$}$\sqcup$}}

\def\sq{\ifmmode\squareforqed\else{\unskip\nobreak\hfil
\penalty50\hskip1em\null\nobreak\hfil\squareforqed
\parfillskip=0pt\finalhyphendemerits=0\endgraf}\fi}

\def\degr{\hbox{$^\circ$}}

\def\utw{\smash{\rlap{\lower5pt\hbox{$\sim$}}}}

\def\udtw{\smash{\rlap{\lower6pt\hbox{$\approx$}}}}

\def\diameter{{\ifmmode\mathchoice
{\ooalign{\hfil\hbox{$\displaystyle/$}\hfil\crcr
{\hbox{$\displaystyle\mathchar"20D$}}}}
{\ooalign{\hfil\hbox{$\textstyle/$}\hfil\crcr
{\hbox{$\textstyle\mathchar"20D$}}}}
{\ooalign{\hfil\hbox{$\scriptstyle/$}\hfil\crcr
{\hbox{$\scriptstyle\mathchar"20D$}}}}
{\ooalign{\hfil\hbox{$\scriptscriptstyle/$}\hfil\crcr
{\hbox{$\scriptscriptstyle\mathchar"20D$}}}}
\else{\ooalign{\hfil/\hfil\crcr\mathhexbox20D}}%
\fi}}